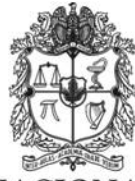

# Estudio de choques en plasmas astrofísicos ultrarelativistas como un posible escenario explicativo de Gamma Ray Bursts (GRBs)

## Maykol Camilo Delgado Correal

# Estudio de choques en plasmas astrofísicos ultrarelativistas como un posible escenario explicativo de Gamma Ray Bursts (GRBs)

## Maykol Camilo Delgado Correal

Tesis de grado presentada como requisito parcial para optar al título de:
**Magíster en Ciencias-Astronomía**

Director:
Prof. Benjamín Calvo-Mozo

Línea de Investigación:
Astrofísica
Universidad Nacional de Colombia
Facultad de Ciencias, Observatorio Astronómico Nacional
Bogotá, Colombia
Enero, 2012

**What is Science?**

"... I can also define science another way: Science is the belief in the ignorance of experts ..."
Dr. Feynman

Esta tesis es dedicada a a mi padre:
César Delgado Tamayo, un gran hombre de Ciencia y de lo mas inteligente que ha tenido el Puerto de Buenaventura; de gran resiliencia intelectual al máximo, según comunicación personal con Dr. Raúl Cuero.

# Agradecimientos

**Personales**

A mi madre Nina, por brindarme el apoyo moral y económico para el desarrollo de la tesis. Además es una persona muy valiosa, sin la cual mi trabajo de investigación no sería igual. Finalmente sin saber nada del tema me ha apoyado desde que comencé a estudiar física y posteriormente la maestría en astronomía.

A mi novia Alexandra por brindarme el apoyo sentimental que me sirvió de equilibrio en momentos de debilidad y también por brindarme momentos infinitos de felicidad.

A mi hermana Mónica por brindarme momentos de esparcimiento adecuados para liberarme de tensiones generadas en la realización de la tesis.

A mi hermano Andrés por las charlas de temas políticos que me ayudaban de vez en cuando a despejar mi mente.

A mi hermano Juan Carlos por sus valiosos concejos en la realización de la tesis.

**Académicos**

A Prof. Benjamín Calvo-Mozo. Como director de tesis me proporcionó la fuerza (en todo el sentido de Star Wars) necesaria para asumir el reto de hacer una tesis teórica en el tema de los GRBs.

A Andrés Castillo, por las fructíferas discusiones acerca de la posible conexión entre supernovas y GRBs, además de la física involucrada en ambos procesos.

A Dr. Sarira Sahu, por enseñarme los aspectos básicos de física de partículas involucrados en la descripción de la fase inicial de los GRBs. Además en Abril de 2010 en cada una de sus charlas acerca de los GRBs me presentó el estado actual de investigación en lo que yo llamo: "neutrinos + GRBs".

A los integrantes del GoSA (Group of Solar Astrophysics) por la discusión generada en su seminario semanal, en especial a: Julian, Juan Camilo, Mauricio, Manuel y al Prof. Benjamín.

# Resumen

Los *Gamma-Ray Bursts-GRBs* son unos de los eventos astronómicos mas energéticos del universo que aún no presentan una adecuada explicación del mecanismo que los origina. Este interrogante motiva a la comunidad astronómica a hacer observaciones en todas las longitudes de onda de emisiones de radiación asociadas con los *GRBs*. Además se hacen modelos físicos para explicar los comportamientos temporales de dichas observaciones. Todo con el ánimo principal de encontrar el escenario físico progenitor.

En esta tesis se presenta un modelo astrofísico que incluye aspectos hidrodinámicos y radiativos involucrados en la evolución temporal de una onda expansiva que lleva consigo un fluido adiabático ultra-relativista. Adicionalmente se realiza una comparación con el *GRB* 050525. Se encuentra que el radio de la onda expansiva $R$ presenta un incremento en el tiempo de la forma: $R \sim t^{\frac{1}{4}}$, el factor gamma de *Lorentz* de las partículas cargadas que componen el fluido ultra-relativista disminuye considerablemente al pasar el tiempo y el flujo de radiación teórico, de la forma $F_\nu \propto t^{-\beta}$ con $\beta$ del orden de 1 se ajusta muy bien a las dos pendientes de la curva de luz del *afterglow* en rayos x del *GRB 050525*.

# Abstract

The Gamma-Ray Bursts - GRBs are one of the most energetic astronomical events in the universe that have not yet an adequate explanation of what kind of mechanism is carried out. This question motivates the astronomical community to do radiation mesurements associated with the GRBs. They also are doing physical models to explain the temporal behavior of these observational results. The community are looking for an appropriate explanation of GRBs origin.

This thesis presents an astrophysical model that includes an hydrodynamic and radiative aspects. This model studies the temporal evolution of a wave carries an expansive ultra-relativistic adiabatic fluid. Additionally, this thesis present a comparison between model with GRB 050525 observational behavior. It is found that the blast wave radius $R$ shows an increase in the time: $R \sim t^{\frac{1}{4}}$, the Lorentz gamma factor of charged particles that are in ultra-relativistic fluid decreases significantly over time and the theoretical flow of radiation: $F_\nu \propto t^{-\beta}$ with $\beta$ around 1 fits very well with the two slopes of the x-ray afterglow light curve of GRB 050525.

# Contenido

# 1. Introducción

Los *Gamma-Ray Bursts-GRBs* son un fenómeno astrofísico que todavía no tiene una explicación concreta de su mecanismo progenitor, sabiéndose con certeza que son unas de las explosiones mas energéticas del universo con una energía total emitida alrededor de $10^{51}$ Erg, con una emisión de fotones en el rango de los 30keV hasta los 2MeV aproximadamente, y con una duración divida en dos grupos: menores de 2 s (Media entre 0.2s-1.3s) y mayores de 2s (Media entre 20s y 40 s). Además son eventos astrofísicos que ocurren espacialmente de forma isotrópica y del orden de 2 por semana. También se ha encontrado evidencia experimental que a cada GRBs le corresponde un rescoldo (*afterglow*), el cual es una emisión en las bandas de rayos x, uv, visible, infrarojo y/o radio que ocurre inmediatamente después de la ráfaga inicial de rayos gamma. El *afterglow* puede permanecer en el cielo desde unos pocos días hasta varios meses.

A continuación se muestra un resumen corto de la información que el lector va encontrar en los capítulos siguientes:

En el segundo capítulo se muestra una descripción detallada de las características observacionales de los Gamma-Ray Bursts-GRBs, y se hizo una clasificación bibliográfica de los tópicos actuales de investigación más representativos en este tema.

En el tercer capítulo se trabaja en el procedimiento físico-matemático detallado para encontrar las ecuaciones que describen una superficie de discontinuidad presente en el gas ideal relativista, también llamadas ecuaciones relativistas de *Rankine-Hugoniot*. Para ello primero se remplazó el tensor relativista *minkowskiano* de energía-momento en las ecuaciones que describen eulerianamente un fluido ideal relativista. Después generamos una superficie de discontinuidad en el interior de este fluido y mediante los principios de conservación de energía, número de partículas y momento encontramos las ecuaciones relativistas del choque adiabático.

En el cuarto capítulo se desarrolla con el mayor detalle posible el estudio de las características físicas de la radiación de sincrotrón, entre las que se cuentan: potencia total radiada, características geométricas, temporales y en frecuencia. Al final del capítulo derivamos una expresión que nos da cuenta de la potencia radiada por electrones relativistas en presencia de campos magnéticos transversales a su movimiento.

En el quinto capítulo en primera medida se muestra una revisión del modelo del *fireball* que es un tratamiento físico que busca explicar la tendencia de las curvas de luz observadas de los GRBs y de sus respectivos *afterglows*. Siguiente a esto en las otras secciones del mismo capítulo se muestra el estudio teórico de la evolución temporal observada en la curva de luz del *afterglow* en rayos x del GRB 050525 detectado por la sonda espacial *Swift* el 25 de Mayo de 2005. Para ello se muestran dos modelos complementarios: hidrodinámico y radiativo. Para el primero se estudio la variación temporal del factor gamma de *Lorentz* de un fluido adiabático ultra-relativista mediante el estudio de la onda de choque relativista producida por la interacción de este fluido (compuesto por electrones y contaminado de protones) con el Medio interestelar-ISM. Para el estudio radiativo se tuvo en cuenta que los primeros electrones por su movimiento ultra-relativista generan campos magnéticos transversales al movimiento de los electrones que vienen detrás de ellos y se produce emisión de radiación de sincrotrón. Se encontró que el factor gamma de *Lorentz* va decayendo a medida que pasa el tiempo, produciendo así que la flujo de radiación detectado por los sensores de *Swift* en la banda de los 2-10 keV también disminuya a medida que el tiempo transcurre.

# 2. Descripción fenomenológica de los Gamma-Ray Bursts (GRBs)

A mediados de los años setenta la guerra fría entre Estados Unidos de América (USA) y la antigua Unión de Repúblicas Socialistas Soviéticas (CCCP) desencadenó una carrera espacial seguida de una de espionaje donde cada país no sabía con certeza los adelantos tecnológicos de su par. Más específicamente los norteamericanos estaban muy preocupados por las explosiones nucleares que pudieran estar haciendo los rusos; para ello colocaron un grupo de satélites llamados Vela equipados con detectores de rayos gamma y rayos x, pues es bien sabido que durante una explosión nuclear se emiten rayos gamma. La sorpresa que se llevaron los investigadores que procesaban los datos de Vela 5a,5b,6a y 6b ocurrió entre julio de 1969 y julio de 1972 cuando detectaron (ver figura **2-1**) las primeros 16 *Gamma-Ray Bursts*-GRBs (Ráfagas de Rayos Gamma) de origen cósmico. El grupo de *Los Alamos National Lab* determinaron estimativos gruesos de las posiciones en la esfera celeste de las 16 ráfagas en rayos gamma, descartando así su origen tanto terrestre como del sistema solar [90]. Una vez desclasificado este tipo de eventos con la publicación de 1973 podemos afirmar que comienza la investigación en este tema por la comunidad astrofísica tanto teórica como observacionalmente, haciendo de este tema un tópico actual que ha presentado un crecimiento casi exponencial en publicaciones [79] (ver figura **2-2**).

Desde que comencé a explorar este evento astronómico me encontré que las publicaciones están distribuidas en los siguientes temas:

Modelo del *Fireball* y su relación con los GRBs [153, 140, 135, 134, 133, 84, 38, 52, 145, 114], observaciones en rayos gamma de los GRBs [68, 169], observaciones en rayos x, ultravioleta, el visible, infrarrojo y en radio del *afterglow* (rescoldo) de GRBs [26, 110, 88, 58, 53, 18, 69, 5, 65], clasificación temporal de los GRBs [122, 163], observaciones y modelos teóricos que muestran la posible conexión de GRBs con explosión de supernovas e hipernovas [71, 156, 37, 137, 109, 150], modelos teóricos y observaciones de posibles progenitores (motor engine) de GRBs [49, 118, 57, 152, 89, 99, 176, 102, 17, 124, 24, 123], búsqueda observacional y modelos teóricos de coincidencias entre GRBs y ondas gravitacionales [1, 91, 2, 73, 120], estudios numéricos MHD para la descripción de los GRBs [158, 159, 47], propiedades energéticas de los GRBs [54, 106, 32, 11], descripción teórica y observacional de los *afterglows* de GRBs [127, 117, 132, 34, 171, 43, 172, 173, 74, 162, 87, 113, 31] descripción teórica de los pulsos iniciales (*prompt emission*) de GRBs [85, 66, 12, 104], descripción teórica y observacional de los espectros de potencia (SED) de los GRBs [119, 10, 51, 170, 15, 16, 29, 166, 86], medición de *redshift* de GRBs [48, 60, 67, 43], alternativas al modelo del Fireball para explicar la evolución temporal de las curvas de luz de los GRBs [28, 35, 108, 107], inclusión teórica y observacional de la física de partículas en el estudio de los GRBs [4, 143, 141, 167, 93, 41, 80], estudio observacional de las galaxias huéspedes de GRBs [19, 6, 64, 138, 39, 147, 149], efectos bioquímicos de la interacción de los GRBs con la atmósfera terrestre [161, 72, 13, 112], lentes gravitacionales y GRBs [8, 125], usos de tipo cosmológico de los GRBs [9, 33, 40, 165, 100, 55], entre otros.

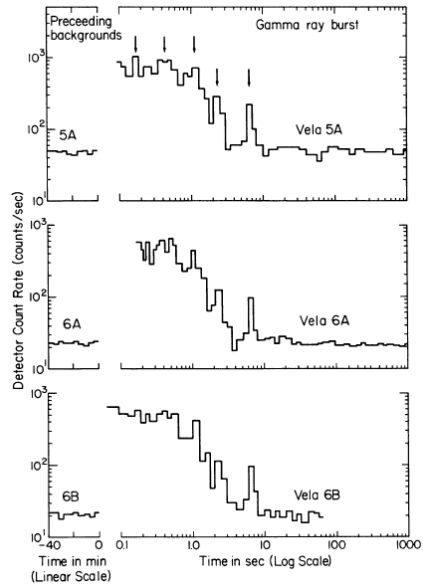

**Figura 2-1**.: Primera detección reportada de un *Gamma-Ray Burst* ocurrido el 22 de agosto de 1970 por los tres satélites Vela. [90].

Así, la primera parte de mi investigación fue realizar un filtro y una organización por temas de toda la información obtenida acerca de los GRBs. Después de ello se escogió estudiar con detalle los aspectos físicos involucrados en la explicación teórica de la fase del it afterglow de los GRBs. Para realizar esta descripción es necesario reproducir las ecuaciones involucradas en el modelo estándar del *Fireball* que contiene un aspecto hidrodinámico relativista (ondas de choque) y uno radiativo (emisión de radiación de sincrotrón). Por último y en búsqueda de resultados numéricos, esta tesis se basa en estudiar teóricamente el trasiente (*afterglow*) en rayos x del GRB050525.

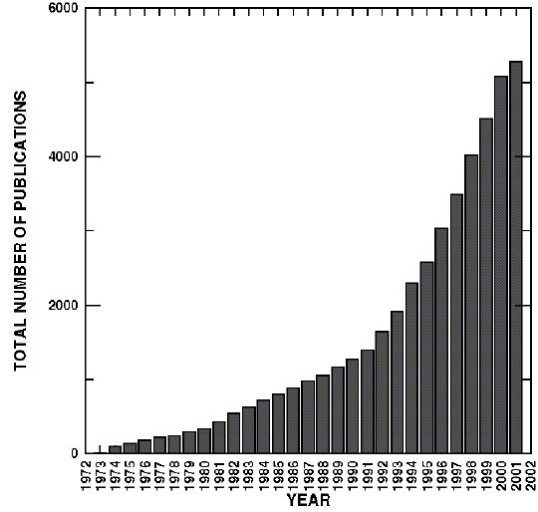

**Figura 2-2**.: Crecimiento exponencial del número de publicaciones relacionadas con GRBs hasta 2001. [79].

## 2.1. Características Observacionales de los Gamma-Ray Bursts-GRBs

Después de las primeras detecciones de GRBs, la comunidad astrofísica no se hizo esperar y con la ayuda de telescopios espaciales con detectores en la banda de rayos x y gamma encontraron resultados muy interesantes. Encontraron que la potencia emitida por cada GRB era del orden de $10^{52}$ erg por segundo (equivalente a la energía emitida por el Sol en un billón de años ($10^{12}$)) considerándose uno de los eventos más energéticos del universo e identificaron una distribución bi-modal de la duración de los GRBs a partir de la medición de 1234 GRBs que se detectaron entre Abril de 1991 y Agosto de 1996 tomados por los detectores de BATSE (*Burst and Transient Source Experiment*) a bordo del *Compton Gamma-Ray Observatory* (CGRO) [126], es decir que se clasificaron los GRBs en dos familias: *Long* GRBs ( $T_{90} > 2s$ ) y *Short* GRBs ($T_{90} < 2s$) (ver figura **2-3**), donde $T_{90}$ es el tiempo que duró en recibirse el 90 % del burst. Desde esa época se han hecho modelos para explicar el origen de esas dos familias de GRBs, el más importante de ellos el conocido como el *modelo del Fireball*; sin embargo, la comunidad sigue investigando en ello concentrada específicamente en los *Short* GRBs.

Otra característica observacional de los GRBs de acuerdo a los resultados acumulados por la misión *Swift Gamma-Ray Burst Mission* durante 5 años es que se detectan alrededor de 100 GRB por año, lo cual nos da un promedio de casi dos GRBs por semana. Ver por ejemplo: http://www.nasa.gov/mission_pages/swift/bursts/500th.html. Resaltamos que su distribución en coordenadas galácticas es isotrópica, dándonos a entender que muy probablemente

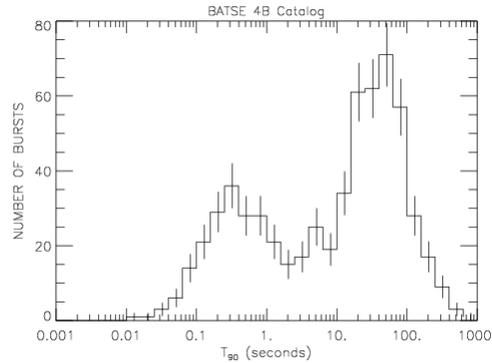

**Figura 2-3**.: Distribución temporal de los GRBs generada a partir de los datos de la duración del *bursts*. [126].

su origen es extragaláctico, como se muestra en la figura **2-4**, extraída de:
http://gammaray.msfc.nasa.gov/batse/grb/duration/.

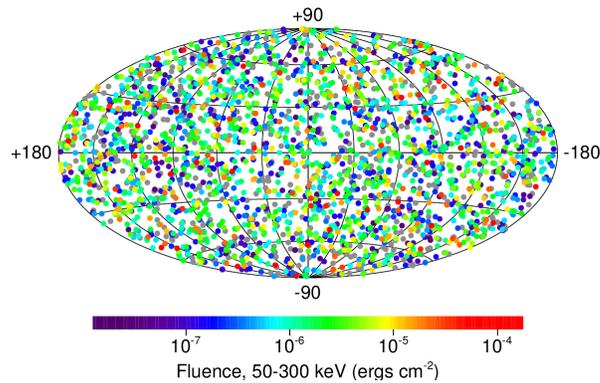

**Figura 2-4**.: Distribución espacial de 2704 GRBs registrados por BATSE.

Las detecciones observadas en rayos gamma de estos eventos astronómicos mostraron un espectro de energía no termal (ver figura **2-5**) lo que llevó a realizar observaciones de sus debidas contrapartes en otras bandas del espectro electromagnético que ayudaran a esclarecer el proceso físico que originaba dichas explosiones.

La primera detección de una contraparte en rayos x de un GRB sucedió el 28 de Febrero de 1997 y se denominó *afterglow* (rescoldo). Seguido de esto se observó la disminución de su flujo de radiación en los días siguientes (ver figura **2-6**).

A partir de este punto la comunidad se concentró en caracterizar las curvas de luz de los afterglows para darse cuenta que se podían ajustar con una ley inversa de potencias que al paso del tiempo cambiaba su índice espectral (ver figura **2-7**). Cuando se realizaron medidas del *afterglow* en el óptico, se encontró también un cambio del índice espectral (llamado jet break), pero en este caso este fue acromático (ver figura **2-8**).

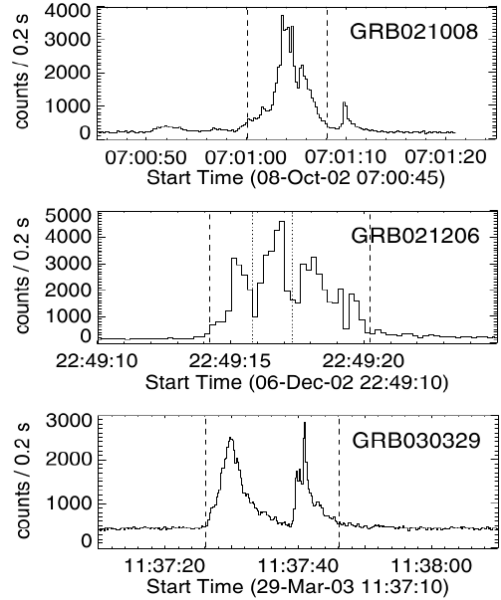

**Figura 2-5**.: Curvas de luz de 3 Gamma-Ray Bursts observados con RHESSI en la banda de los 20-1000KeV. [169]

Suponiendo que le podemos asignar una galaxia huésped a cada GRB, entonces determinando el corrimiento al rojo, z, de las líneas de emisión de esta galaxia ($z = \frac{\lambda - \lambda_0}{\lambda_0}$), en principio podemos asociarle este z al GRB (ver figura **2-9**). Otra forma de medir el z de un GRB es utilizando el espectro de su transiente óptico e identificar líneas de absorción por su presencia en la galaxia huésped (ver figura **2-10** ).

Actualmente existe una base de datos de las propiedades mas representativas de las galaxias huésped de GRBs llamada GHostS [148], entre las cuales se encuentra la información de su z, de su distancia de luminosidad, de la extinción de su polvo y la duración de la ráfaga de rayos gamma (http://www.grbhosts.org/). Los datos son de uso libre para la comunidad astronómica y es permitido hacer estudio de correlación de las variables allí presentadas (ver figura **2-11**). Por ejemplo, el equipo de GHostS encontró que de una muestra de 39 GRBs, alrededor de 2/5 del número total de GRBs (a 2007) con medición de su z, la mayoría de la población tenía z menor a dos (ver figura **2-12**). Esta información registrada en esta database GHostS es fundamental para la descripción teórica de las curvas de luz de los afterglows de GRBs. Cabe resaltar que la base GHostS tiene 128 GRBs con galaxia huésped a la fecha de hoy. Además algunos GRBs como GRB990123 no presentan contraparte extragalactica.

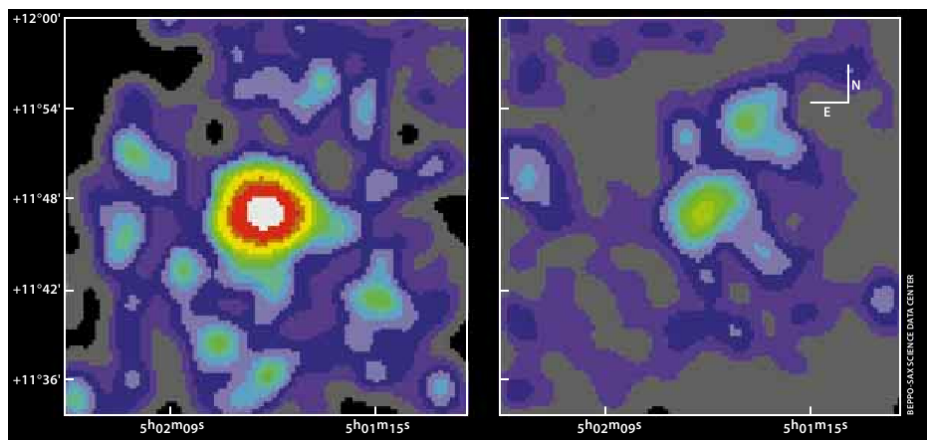

**Figura 2-6**.: Imagen en rayos x tomada por el satélite Beppo-SAX. La imagen de la izquierda corresponde al 28 de febrero de 1997 y la imagen de la derecha al 3 de Marzo [50]

En los últimos años se ha empezado a realizar distribuciones espectrales de energía-SED en la banda de los rayos gamma en búsqueda de ajustar observacionalmente las predicciones de los posibles mecanismos centrales (*central engine*) de producción de los primeros GRBs. En la figura **2-13** se muestra una medición hecha con un arreglo de 9 detectores de Germanio de *RHESSI* que trabajan en la banda de los 20keV hasta los 17 MeV, con una resolución espectral alrededor de los 4keV y una temporal de $1\mu s$ [169].

En conclusión podemos decir que las observaciones detectadas de los GRBs nos están mostrando características extremas presentes en los procesos físicos más energéticos que ocurren en diferentes regiones del universo.

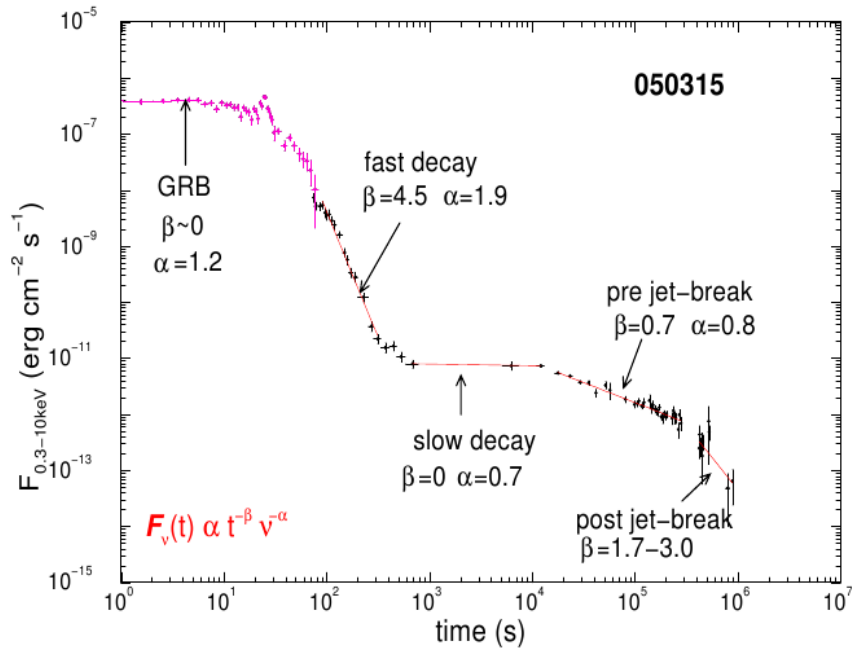

**Figura 2-7**.: Fases del afterglow del GRB050315 y el ajuste a una ley de potencia adecuada para cada fase de decaimiento. Los puntos rojos indican medidas de Burst Alert Telescope-BAT (15-150keV) extrapoladas a la banda de medida de X-ray Telescope-XRT (0.3 a 10keV). Los puntos negros indican medidas XRT. Ambos son instrumentos de SWIFT. [130]

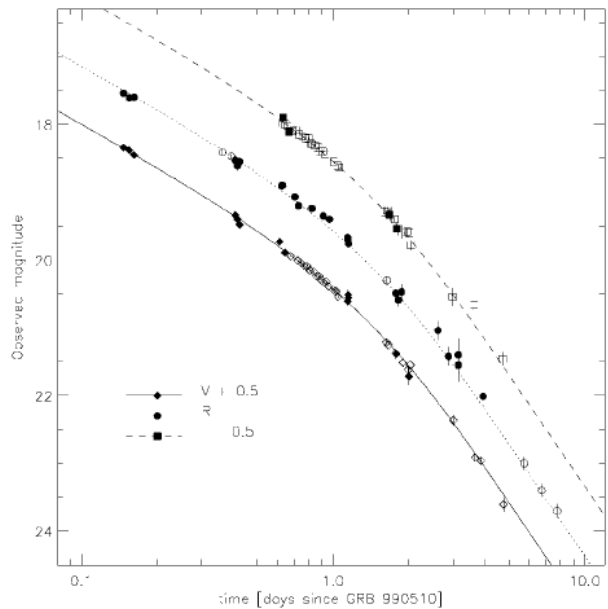

**Figura 2-8**.: Curvas de Luz ópticas del afterglow del GRB 990510. [70]

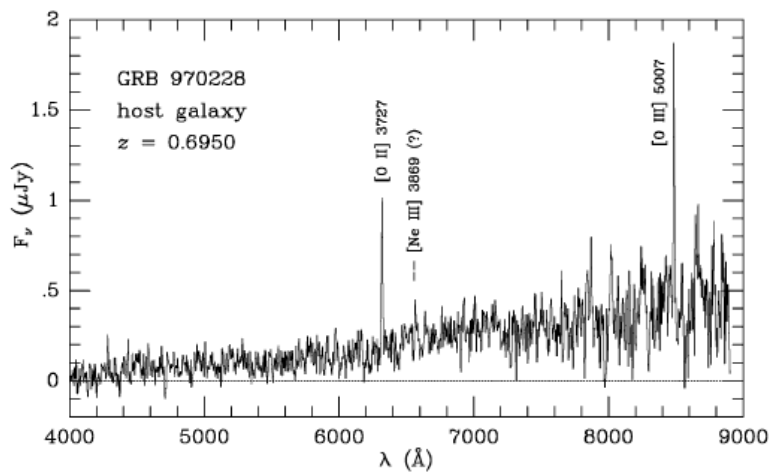

**Figura 2-9**.: Espectro de la galaxia huésped del GRB 970228 obtenido con le telescopio Keck II. Las líneas prominentes[O II],[O III] y posiblemente [Ne III] son colocadas asumiendo que las líneas se originan desde la galaxia huésped con z= 0.695. Modificado de: [21]

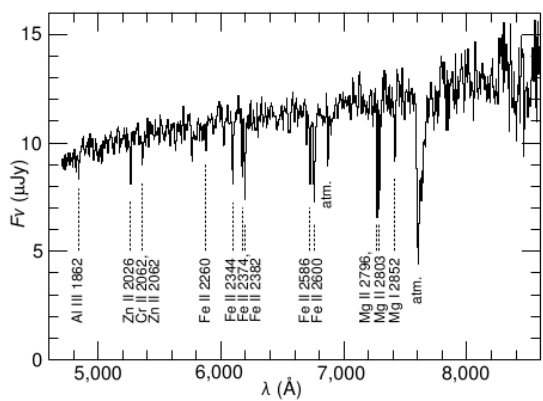

**Figura 2-10**.: Espectro del transiente óptico del GRB 990123. [96]

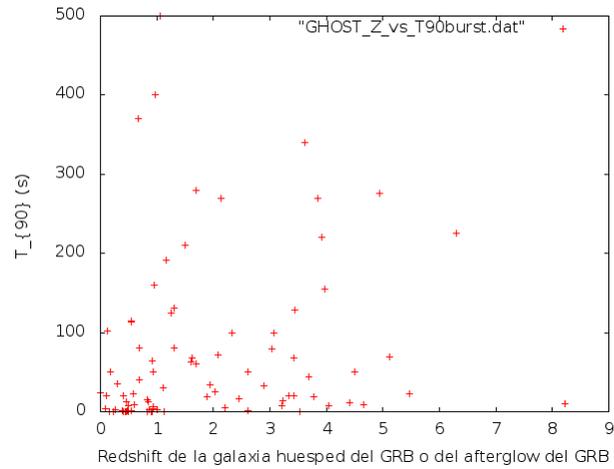

**Figura 2-11**.: Relación entre la duración del Burst: $T_{90}$ y su corrimiento al rojo z.

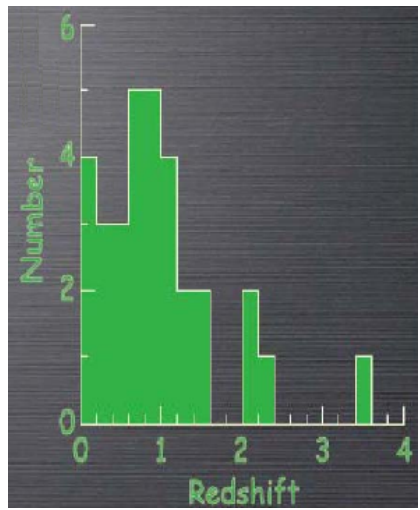

**Figura 2-12**.: Histograma de Distribución de los valores de z asociados con 37 GRBs presentes en GHostS [148]

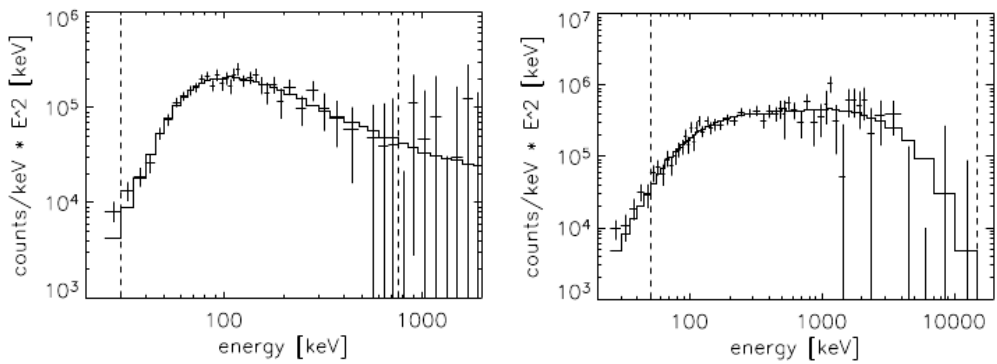

**Figura 2-13**.: Espectro de distribución de energía del GRB 050525(izquierda) y GRB 050717(derecha). Los datos que fueron tomados por detectores de *RHESSI* son mostrados por las cruces. [170]

# 3. Revisión detallada de la deducción de las Ecuaciones Relativistas de las Ondas de Choque: Ecuaciones Relativistas de Rankine-Hugoniot

En este capitulo se muestra el procedimiento físico-matemático detallado para encontrar las ecuaciones que describen una superficie de discontinuidad presente en el gas ideal relativista, también llamadas ecuaciones relativistas de *Rankine-Hugoniot*. Para ello primero se reemplazó el tensor relativista *minkowskiano* de energía-momento en las ecuaciones que describen *eulerianamente* un fluido ideal relativista. Después generamos una superficie de discontinuidad en el interior de este fluido y mediante los principios de la conservación de la energía, del número de partículas y del momento, encontramos las ecuaciones relativistas del choque asumiendo que el mismo se lleva a cabo en forma adiabática.

La importancia de entender las suposiciones físicas detrás de la demostración de estas ecuaciones radica en que son muy utilizadas en el modelo estándar del *fireball* [1] para hacer la descripción de las curvas de luz de los eventos astrofísicos llamados *Gamma-Ray Bursts*. Estas ecuaciones fueron obtenidas por primera vez por *A. H. Taub* en 1948 [157] en un marco general y adaptadas como las "*jump conditions*" en el contexto de ondas expansivas relativistas involucradas en procesos astrofísicos por *R. D. Blandford* y *C. F. McKee* en 1976 [20]. En el presente capítulo haremos entonces la deducción de las ecuaciones de choque relativista.

En la descripción física de los fluidos Newtonianos tenemos usualmente dos opciones. Una de ellas es la descripción de Euler, en la cual para un instante dado de tiempo asociamos campos escalares, vectoriales o tensoriales a cada punto del espacio por donde se mueve el fluido, en donde tales campos representan propiedades físicas del fluido en cada localidad por donde fluye, para cada instante de tiempo. La otra descripción es la de Lagrange, la que conlleva a un seguimiento de un elemento dado del fluido en su movimiento. Nosotros utilizaremos la primera de ellas, pero en un contexto relativista, en donde el espacio en el que se describen los campos del fluido es el espacio-tiempo de Minkoswki.

---

[1] Explicado al inicio del capítulo 5

Las ecuaciones que gobiernan el movimiento relativista de un gas ideal que no está sujeto a fuerzas externas pueden ser escritas como [98, 157]:

$$\left(\rho^0 u^i\right)_{,i} = 0 \tag{3-1}$$

$$\left(T^{ik}\right)_{,k} = 0 \tag{3-2}$$

en donde los índices griegos varían de 1 a 3 y los índices latinos recorren los valores de 0 a 3. Además, el subíndice precedido por una coma indica la derivación con respecto a la respectiva variable. Así por ejemplo, $A^i_{,k}$ representa la derivada parcial $\partial A^i/\partial x^k$.

La cuadrivelocidad $u^i$ de cada una de las partículas del fluido esta definida como,

$$u^i = (\gamma, \xi^\alpha) \tag{3-3}$$

$$\xi^\alpha = \frac{\frac{v^\alpha}{c}}{\left(1 - \frac{v^2}{c^2}\right)^{\frac{1}{2}}} \tag{3-4}$$

$$v^2 = \sum_{\alpha=1}^{3} (v^\alpha)^2 \tag{3-5}$$

Donde $v^\alpha$ son las componentes de la velocidad de cada partícula en un referencial dado.

La ecuación $(3-1)$ representa la continuidad del fluido ideal relativista, mientras que la ecuación $(3-2)$ la conservación del momento y la energía. En nuestro desarrollo resulta conveniente tener en cuenta la densidad numérica de partículas, $n$, la cual expresa el número de partículas del fluido por unidad de volumen en un referencial dado. Cuando el fluido está en reposo tenemos que $n = n_0$ y está relacionado con la densidad $\rho^0$ del elemento de volumen del fluido en reposo de la siguiente forma,

$$\rho^0 = n_0 m_0 \tag{3-6}$$

en donde $m_0$ es la masa en reposo de las partículas constituyentes del fluido, en el caso en el que se trate de una sola especie de partículas. Si llevamos $(3-6)$ a $(3-1)$ obtenemos,

$$\left(n_0 u^i\right)_{,i} = 0 \tag{3-7}$$

la cual representa la conservación del número de partículas.

Al cuadrivector:

$$n^i = n_0 u^i \tag{3-8}$$

se le denomina cuadrivector de flujo de partículas. Su componente 0, $n^0 = n_0\gamma$, es la densidad numérica de partículas (*number density*), i.e., el número de partículas por unidad de volumen.

El tensor de Energía-Momento, $T^{ik}$, se define a partir de integrar en todo el volumen el cambio en el tiempo de la función de distribución de las partículas [2] que componen el fluido. La expresión de $T^{ik}$ que se obtiene se muestra a continuación ([157], ver Parte II),

$$T^{ik} = \rho^0 \left(c^2 + \epsilon + \frac{P}{\rho^0}\right) u^i u^k + P g^{ik} \tag{3-9}$$

$$g_{11} = g_{22} = g_{33} = 1, g_{00} = -1 \tag{3-10}$$

La convención anterior de la métrica tiene signatura $+2$. Para el intervalo espacio-temporal escogemos:

$$-ds^2 = g_{ik} dx^i dx^k \tag{3-11}$$

Donde $\epsilon$ en las ecs.$(3-9)$ y $(3-12)$ representa [3] la energía interna $U$ por unidad de masa del fluido en reposo, tal como es medida por un observador en reposo con respecto al elemento de fluido (referencial propio[4], $U = U_0$) y P es la presión en el elemento de volumen del fluido ($V_0$):

$$\epsilon = \frac{U_0}{\rho_0 V_0} \qquad \Rightarrow \rho_0 \epsilon = \frac{U_0}{V_0} \tag{3-12}$$

---

[2]En este análisis se considera que el número de partículas del fluido no cambia, las colisiones ocurren entre ellas y se asume conservación de la masa, energía y del momento en todo el sistema ( Caso 3, Cap. 6, "Ecuación de Transporte de Boltzman", [95] )

[3]Como $[\rho^0] = \frac{Kg}{m^3}$ entonces $[\epsilon] = \frac{J}{Kg}$

[4]En el caso general desde cualquier otro sistema de referencia la energía interna del fluido es: $U = U_0 + n_0 m_0 c^2$ y su entalpía por unidad de volumen propio es: $\omega = (\rho^0 c^2 + \rho^0)\epsilon + P = e + P$

A veces se emplean las componentes covariantes del tensor de energía-momento es tal que $T^{ik}$ las cuales se pueden inferir del modo normal "subiendo" y "bajando" índices a partir del $T^{ik}$ dado en la ecuación $(3-9)$ y de la métrica dada en $(3-10)$ de acuerdo con: $T_{ik} = g_{il}g_{km}T^{lm}$. De este modo las componentes covariantes $T_{ik}$ quedan determinadas por,

$$T_{00} = T^{00}; T_{\alpha\beta} = T^{\alpha\beta}; T_{0\alpha} = -T^{0\alpha} \tag{3-13}$$

La interpretación usual de las componentes contravariantes del tensor de energía-momento es tal que $T^{00}$ es la densidad de energía del fluido, $\frac{T^{0\alpha}}{c}$ es la densidad de las componentes de momento, y las magnitudes $T^{\alpha\beta}$ forman el tensor tridimensional de densidad de flujo de momento ([98], Cap. 15).

En el sistema de referencia propio un elemento del fluido en reposo allí, es tal que la presión ejercida por una porción determinada del fluido es la misma en todas las direcciones (Principio de Pascal), siendo esta perpendicular a la superficie sobre la cual actúa. Así, de la ec. $(3-9)$ podemos ver que en este caso los $T^{\alpha\beta}$ y el $T^{00}$ en el referencial propio son:

$$T^{00}_{ReferencialPropio} = \rho^0 c^2 + \rho^0 \epsilon = \frac{U}{V_0} = e \tag{3-14}$$

$$T^{\alpha\beta}_{ReferencialPropio} = P g^{\alpha\beta} \tag{3-15}$$

Donde $e$ representa la densidad de energía propia del fluido, es decir la energía interna por unidad de volumen propio.

El caso no relativista del $T^{ik}$ se obtiene al considerar un movimiento de las partículas del fluido con bajas velocidades ( $v << c$ ) en la ec. $(3-9)$. En este caso los $T^{\alpha\beta}$ y el $T^{00}$ son de la forma,

$$T^{00}_{NoRelativista} = \rho^0 c^2 + \rho^0 \epsilon \tag{3-16}$$

$$T^{\alpha\beta}_{NoRelativista} = \rho^0 v^\alpha v^\beta + P \delta^{\alpha\beta} \tag{3-17}$$

Donde la ec. $(3-17)$ es el tensor de densidad de flujo de momento no relativista de nuestro fluido ideal ([98], Cap. 1 ). [5]

---

[5] Para una descripción más detallada ver sección: anexo.

Ahora substituyendo la ec. $(3-9)$ en la ec. $(3-2)$ tenemos que se satisface,

$$\left(T^{ik}\right), k = \frac{\partial T^{ik}}{\partial x^k} = 0 \tag{3-18}$$

$$\rho^0 \left(c^2 + \epsilon + \frac{P}{\rho^0}\right) \frac{\partial (u^i u^k)}{\partial x^k} + \frac{\partial P g^{ik}}{\partial x^k} = 0 \tag{3-19}$$

$$\rho^0 \left(c^2 + \epsilon + \frac{P}{\rho^0}\right) u^k \frac{\partial (u^i)}{\partial x^k} + \left(c^2 + \epsilon + \frac{P}{\rho^0}\right) u^i \frac{\partial (\rho^0 u^k)}{\partial x^k} + \frac{\partial P}{\partial x^k} g^{ik} = 0 \tag{3-20}$$

En los desarrollos anteriores tuvimos en cuenta que $\frac{\partial (\rho^0 c^2 + \rho^0 \epsilon + P)}{\partial x^k} = 0$ dado que la entalpía específica se conserva a lo largo de las líneas de flujo. Teniendo en cuenta la ec. $(3-1)$ en el anterior resultado obtenemos,

$$\rho^0 \left(c^2 + \epsilon + \frac{P}{\rho^0}\right) u^k \frac{\partial (u^i)}{\partial x^k} + \frac{\partial P}{\partial x^k} g^{ik} = 0 \tag{3-21}$$

$$\Rightarrow \rho^0 \left(c^2 + \epsilon + \frac{P}{\rho^0}\right) u^k \frac{\partial (u^i)}{\partial x^k} + \frac{\partial P}{\partial x^k} g^{ik} = 0 \tag{3-22}$$

$$\Rightarrow \rho^0 u^k \frac{\partial (c^2 \mu u^i)}{\partial x^k} + \frac{\partial P}{\partial x^k} g^{ik} = 0 \tag{3-23}$$

Donde,

$$\mu = 1 + (\frac{1}{c^2})(\epsilon + \frac{P}{\rho^0}) = 1 + \frac{\omega}{c^2} \tag{3-24}$$

siendo $\omega = \epsilon + \frac{P}{\rho^0}$ la entalpía por unidad de masa ( "entalpía específica") del fluido. Es claro que la variable $\mu$ introducida arriba es adimensional.

Multiplicando $(3-23)$ por $-u_i$ y sumando índices tenemos que se cumple,

$$-\rho^0 u_i u^k \frac{\partial (c^2 \mu u^i)}{\partial x^k} - u_i \frac{\partial P}{\partial x^k} g^{ik} = 0 \tag{3-25}$$

$$-\rho^0 u^k c^2 \mu u_i \frac{\partial (u^i)}{\partial x^k} - \rho^0 u^k u_i u^i c^2 \frac{\partial (\mu)}{\partial x^k} - u_i g^{ik} \frac{\partial P}{\partial x^k} = 0 \tag{3-26}$$

De acuerdo con la definición del cuadrivector de velocidad,$(3-3)$, y teniendo en cuenta la métrica $(3-10)$ vemos que,

$$u_i u^i = -1 \tag{3-27}$$

y por tanto se cumple que [44]:

$$u_i \left( \frac{\partial u^i}{\partial x^k} \right) = 0 \tag{3-28}$$

Teniendo en cuenta esta última expresión vemos que $(3-26)$ se transforma en,

$$\rho^0 u^k \frac{\partial \left( 1 + (\frac{1}{c^2})(\epsilon + \frac{P}{\rho^0}) \right)}{\partial x^k} - u^k \frac{\partial P}{\partial x^k} = 0 \tag{3-29}$$

$$\rho^0 \left( \frac{\partial \epsilon}{\partial x^k} u^k + \left( \frac{\partial \frac{P}{\rho^0}}{\partial x^k} \right) u^k \right) = 0 \tag{3-30}$$

La ec. $(3-30)$ representa la conservación de energía presente en nuestro gas ideal relativista.

Ahora nuestro propósito es encontrar las ecuaciones que describen una superficie de discontinuidad presente en el gas ideal relativista, también llamadas ecuaciones relativistas de *Rankine-Hugoniot*. Para ello debemos comenzar integrando sobre un volumen en el espacio-tiempo las ecuaciones $(3-1)$ y $(3-2)$,

$$\int \left( \left( \rho^0 u^i \right), i \right) d^4 x = \int \left( \rho^0 u^i \lambda_i \right) d^3 x = 0 \tag{3-31}$$

$$\int \left( \left( T^{ik} \right), k \right) d^4 x = \int \left( T^{ik} \lambda_k \right) d^3 x = 0 \tag{3-32}$$

Donde las ecuaciones de la derecha son tomadas sobre la hipersuperficie tridimensional que está acotando al volumen de fluido en estudio. A su vez $\lambda_i$ son las componentes covariantes del vector normal (en todas las direcciones) a la hipersuperficie.

Suponiendo que el volumen tridimensional es una capa (*shell*) de espesor $\sigma$ que hace parte (incrustada) de una superficie de discontinuidad $\Sigma$, donde el vector normal tridimensional a dicha superficie es $\Lambda_\alpha$ (Ver Fígura **3-1**).

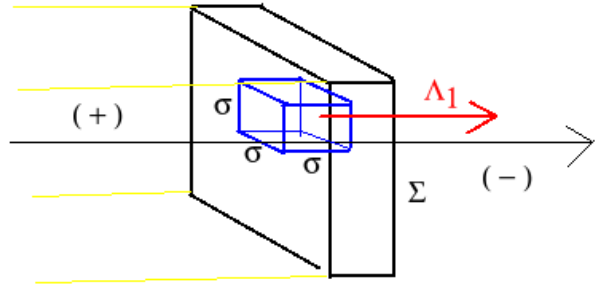

**Figura 3-1**.: Esquema pictórico de la superficie de discontinuidad, que separa dos regiones del fluido con diferente $\rho^0$, $P$, y $u^\alpha$ .

Escogiendo nuestro sistema coordenado tal que la superficie de discontinuidad esté en reposo con respecto a él [6] , obtenemos que,

$$\lambda_\alpha = \Lambda_\alpha \tag{3-33}$$

$$\lambda_0 = 0 \tag{3-34}$$

debido a que: $\lambda_i \lambda^i = 1$ y $\sum_{\alpha=1}^{3}(\Lambda_\alpha)^2 = 1$ .

Reemplazando $(3-33)$ en $(3-31)$ obtenemos,

$$\int \left(\rho^0 u^i \lambda_i\right) d^3x = \int \left(\rho^0 u^\alpha \Lambda_\alpha\right) d^3x = 0 \tag{3-35}$$

$$\int_{-\infty}^{+\infty} \int_{-\infty}^{+\infty} \int_{-\infty}^{+\infty} \left(\rho^0 \left(u^1 \Lambda_1 + u^2 \Lambda_2 + u^3 \Lambda_3\right)\right) dx^1 dx^2 dx^3 = 0 \tag{3-36}$$

Observando la figura **3-1** nos damos cuenta que la integral anterior toma la forma,

$$\int_{-\infty}^{-\frac{\sigma}{2}} \int_{-\infty}^{-\frac{\sigma}{2}} \int_{-\infty}^{-\frac{\sigma}{2}} \left(\rho^0 \left(u^1 \Lambda_1 + u^2 \Lambda_2 + u^3 \Lambda_3\right)\right) dx^1 dx^2 dx^3 + \tag{3-37}$$

---

[6]Suposición hecha también por ([98], $LL$84, Cap. IX) para derivar las ecuaciones estándar (no relativistas) de *Rankine-Hugoniot*

$$\int_{-\frac{\sigma}{2}}^{\frac{\sigma}{2}} \int_{-\frac{\sigma}{2}}^{\frac{\sigma}{2}} \int_{-\frac{\sigma}{2}}^{\frac{\sigma}{2}} \left( \rho^0 \left( u^1 \Lambda_1 + u^2 \Lambda_2 + u^3 \Lambda_3 \right) \right) dx^1 dx^2 dx^3 + \tag{3-38}$$

$$\int_{\frac{\sigma}{2}}^{\infty} \int_{\frac{\sigma}{2}}^{\infty} \int_{\frac{\sigma}{2}}^{\infty} \left( \rho^0 \left( u^1 \Lambda_1 + u^2 \Lambda_2 + u^3 \Lambda_3 \right) \right) dx^1 dx^2 dx^3 = 0 \tag{3-39}$$

Denotando como $f_+$ al valor de la función $f$ (ya sea $\rho^0$, $u^i$, $P$, $V$, $\omega$, $\epsilon$, $T$, o $S$) sobre un lado de la superficie de discontinuidad $\Sigma$ (en la vecindad de $\Sigma$) y a $f_-$ como el valor de la función $f$ en el lado opuesto [7], (ver figura **3-1**), tenemos que las anteriores integrales se re-escriben como,

$$\int_{-\infty}^{-\frac{\sigma}{2}} \int_{-\infty}^{-\frac{\sigma}{2}} \int_{-\infty}^{-\frac{\sigma}{2}} \left( \rho_+^0 \left( u_+^1 \Lambda_1 + u_+^2 \Lambda_2 + u_+^3 \Lambda_3 \right) \right) dx^1 dx^2 dx^3 + \tag{3-40}$$

$$\int_{-\frac{\sigma}{2}}^{0} \int_{-\frac{\sigma}{2}}^{0} \int_{-\frac{\sigma}{2}}^{0} \left( \rho_+^0 \left( u_+^1 \Lambda_1 + u_+^2 \Lambda_2 + u_+^3 \Lambda_3 \right) \right) dx^1 dx^2 dx^3 + \tag{3-41}$$

$$\int_{0}^{\frac{\sigma}{2}} \int_{0}^{\frac{\sigma}{2}} \int_{0}^{\frac{\sigma}{2}} \left( \rho_-^0 \left( u_-^1 \Lambda_1 + u_-^2 \Lambda_2 + u_-^3 \Lambda_3 \right) \right) dx^1 dx^2 dx^3 + \tag{3-42}$$

$$\int_{\frac{\sigma}{2}}^{\infty} \int_{\frac{\sigma}{2}}^{\infty} \int_{\frac{\sigma}{2}}^{\infty} \left[ \rho_-^0 \left( u_-^1 \Lambda_1 + u_-^2 \Lambda_2 + u_-^3 \Lambda_3 \right) \right] dx^1 dx^2 dx^3 = 0 \tag{3-43}$$

La anterior ecuación integral se satisface cuando (por la conservación del número de partículas) al $\sigma \longmapsto 0$,

$$\rho_+^0 u_+^1 \Lambda_1 = \rho_-^0 u_-^1 \Lambda_1 \tag{3-44}$$

$$\rho_+^0 u_+^2 \Lambda_2 = \rho_-^0 u_-^2 \Lambda_2 \tag{3-45}$$

---

[7]Donde $f_+$ se encuentra en la dirección opuesta a la dirección del vector normal a $\Sigma$ y $f_-$ en la misma dirección de $\Lambda_1$. En el estudio de las ondas de choque no relativistas hecho en ([98], *LL*84, Cap. IX) y en [160] se toma $f_+ = f_2$ y $f_- = f_1$

$$\rho_+^0 u_+^3 \Lambda_3 = \rho_-^0 u_-^1 \Lambda_3 \tag{3-46}$$

$$\Rightarrow \rho_+^0 u_+^\alpha \Lambda_\alpha = \rho_-^0 u_-^\alpha \Lambda_\alpha \tag{3-47}$$

o bien, escrito en otra notación,

$$\left[\rho^0 u^\alpha \Lambda_\alpha\right] = 0 \tag{3-48}$$

Donde hemos definido $[f]$ como: $[f] = f_+ - f_-$ .

Haciendo un análisis similar al hecho con $(3-31)$ , pero ahora partiendo desde $(3-32)$ obtenemos que cuando al $\sigma \longmapsto 0$

$$T_+^{i\alpha} \Lambda_\alpha = T_-^{i\alpha} \Lambda_\alpha \tag{3-49}$$

$$\Rightarrow \left[T^{i\alpha} \Lambda_\alpha\right] = 0 \tag{3-50}$$

Donde hemos tenido en cuenta la conservación de la energía y del momento a través del fluido, lo cual se representa mediante la nulidad de la cuadri-divergencia del tensor de energía-momento ec.(3-2). Finalmente como pasa en ([98], $LL84$, Cap. IX) nos damos cuenta que las relaciones de discontinuidad $(3-48)$ y $(3-50)$ son válidas en la vecindad de la superficie de discontinuidad, es decir cuando $\sigma \longmapsto 0$).

Considerando el movimiento del gas ideal relativista como unidimensional (a lo largo del eje $x$ como se muestra en la Figura **3-1** ), entonces la superficie de discontinuidad $\Sigma$ es perpendicular al eje $x$, es decir que las siguientes relaciones son válidas,

$$\Lambda_\alpha = \delta_\alpha^1 \tag{3-51}$$

$$u^0 = \frac{d(ct)}{ds} = \frac{d(ct)}{cd\tau} = \frac{dt}{d\tau} = \gamma = \frac{1}{(1-v^2)^{\frac{1}{2}}} \tag{3-52}$$

$$u^1 = \frac{dx}{ds} = \frac{dx}{cd\tau} = \frac{1}{c}\frac{dx}{dt}\frac{dt}{d\tau} = \frac{1}{c}v\gamma = \frac{v}{(1-v^2)^{\frac{1}{2}}} \tag{3-53}$$

$$u^2 = u^3 = 0 \tag{3-54}$$

Donde $v = \frac{dx}{dt}$ se considera como la velocidad del gas en unidades en las cuales $c = 1$.

Reemplazando $(3-9)$, $(3-52)$, $(3-53)$ en $(3-48)$ y en $(3-50)$ podemos encontrar las ecuaciones relativistas de *Rankine-Hugoniot* (*jump conditions*).

A partir de $(3-51)$ y $(3-53)$ en $(3-48)$ obtenemos,

$$\rho_+^0 u_+^\alpha \delta_\alpha^1 = \rho_-^0 u_-^\alpha \delta_\alpha^1 \tag{3-55}$$

$$\rho_+^0 u_+^1 = \rho_-^0 u_-^1 \tag{3-56}$$

$$\rho_+^0 \frac{v_+}{(1-v_+^2)^{\frac{1}{2}}} = \rho_-^0 \frac{v_-}{(1-v_-^2)^{\frac{1}{2}}} = j \tag{3-57}$$

Donde $j$ es la densidad de flujo másico relativista del fluido [98, 160] y $(3-57)$ es la primera ecuación relativista de *Rankine-Hugoniot*.

Ahora, a partir de reemplazar $(3-51)$ en $(3-50)$ tenemos que se satisface,

$$T_+^{i\alpha} \delta_\alpha^1 = T_-^{i\alpha} \delta_\alpha^1 \tag{3-58}$$

$$T_+^{i1} = T_-^{i1} \tag{3-59}$$

Si recordamos que el tensor de energía-momento viene dado por la ecuación (3-2), es decir, $T^{ik} = \rho^0 \left(c^2 + \epsilon + \frac{P}{\rho^0}\right) u^i u^k + P g^{ik}$, es fácil darnos cuenta que cuando $i = 2, 3$ tenemos que $T^{i1} = 0$ puesto que $u^2 = u^3 = 0$ (problema unidimensional) y $g^{21} = g^{31} = 0$ dado que la métrica es diagonal. Así solamente sobreviven los términos $T^{01}$ y $T^{11}$ para ser tenidos en cuenta en $(3-59)$, es decir que $(3-59)$ se reduce al siguiente par de ecuaciones,

$$T_+^{01} = T_-^{01} \tag{3-60}$$

$$T_+^{11} = T_-^{11} \tag{3-61}$$

Reemplazando $(3-9)$, $(3-52)$ y $(3-53)$ en $(3-60)$ obtenemos,

$$\rho_+^0 \left(c^2 + \epsilon_+ + \frac{P_+}{\rho_+^0}\right) u_+^0 u_+^1 = \rho_-^0 \left(c^2 + \epsilon_- + \frac{P_-}{\rho_-^0}\right) u_-^0 u_-^1 \tag{3-62}$$

$$\Rightarrow \frac{\rho_+^0 u_+^0 \left(c^2 + \epsilon_+ + \frac{P_+}{\rho_+^0}\right) v_+}{(1-v_+^2)^{\frac{1}{2}}} = \frac{\rho_-^0 u_-^0 \left(c^2 + \epsilon_- + \frac{P_-}{\rho_-^0}\right) v_-}{(1-v_-^2)^{\frac{1}{2}}} \tag{3-63}$$

Recordando que en $(3-24)$ habíamos denotado que $\mu = 1 + (\frac{1}{c^2})(\epsilon + \frac{P}{\rho^0})$ entonces tenemos que la siguiente ecuación es válida,

$$\Rightarrow c^2 + \epsilon + \frac{P}{\rho^0} = c^2 \left(1 + \left(\frac{1}{c^2}\right)\epsilon + \frac{1}{c^2}\left(\frac{P}{\rho^0}\right)\right) = c^2 \mu \tag{3-64}$$

Reemplazando $(3-64)$ en $(3-63)$ tenemos que,

$$\Rightarrow \frac{\rho_+^0 u_+^0 c^2 \mu_+ v_+}{(1-v_+^2)^{\frac{1}{2}}} = \frac{\rho_-^0 u_-^0 c^2 \mu_- v_-}{(1-v_-^2)^{\frac{1}{2}}} \tag{3-65}$$

Recordando que $u^0 = \frac{1}{(1-v^2)^{\frac{1}{2}}}$ y que $\rho_+^0 \frac{v_+}{(1-v_+^2)^{\frac{1}{2}}} = \rho_-^0 \frac{v_-}{(1-v_-^2)^{\frac{1}{2}}} = j$, tenemos que la ecuación $(3-65)$ se transforma en,

$$\frac{c^2 \mu_+ \rho_+^0 v_+}{(1-v_+^2)^{\frac{1}{2}}(1-v_+^2)^{\frac{1}{2}}} = \frac{c^2 \mu_- \rho_-^0 v_-}{(1-v_-^2)^{\frac{1}{2}}(1-v_-^2)^{\frac{1}{2}}} \tag{3-66}$$

$$\Rightarrow \frac{c^2 \mu_+ j}{(1-v_+^2)^{\frac{1}{2}}} = \frac{c^2 \mu_- j}{(1-v_-^2)^{\frac{1}{2}}} \Rightarrow \mu_+ \gamma_+ = \mu_- \gamma_- \tag{3-67}$$

Ahora partiendo de la otra condición de choque $(3-61)$ tenemos que se cumple,

$$T_+^{11} = T_-^{11} \tag{3-68}$$

$$\Rightarrow \rho_+^0 \left(c^2 + \epsilon_+ + \frac{P_+}{\rho_+^0}\right) u_+^1 u_+^1 + P_+ = \rho_-^0 \left(c^2 + \epsilon_- + \frac{P_-}{\rho_-^0}\right) u_-^1 u_-^1 + P_- \tag{3-69}$$

Reemplazando $(3-64)$ en $(3-69)$ tenemos que,

$$\left(\rho_+^0 c^2 \mu_+ u_+^1 u_+^1\right) + P_+ = \left(\rho_-^0 c^2 \mu_- u_-^1 u_-^1\right) + P_- \tag{3-70}$$

y teniendo en cuenta $(3-53)$ y $(3-57)$ en $(3-70)$ obtenemos,

$$\frac{\rho_+^0 v_+ c^2 \mu_+ v_+}{(1-v_+^2)^{\frac{1}{2}}(1-v_+^2)^{\frac{1}{2}}} + P_+ = \frac{\rho_-^0 v_- c^2 \mu_- v_-}{(1-v_-^2)^{\frac{1}{2}}(1-v_-^2)^{\frac{1}{2}}} + P_- \tag{3-71}$$

$$\Rightarrow \frac{jc^2 \mu_+ v_+}{(1-v_+^2)^{\frac{1}{2}}} + P_+ = \frac{jc^2 \mu_- v_-}{(1-v_-^2)^{\frac{1}{2}}} + P_- \tag{3-72}$$

y por tanto se tiene:

$$jc^2 \left(\frac{\mu_+ v_+}{(1-v_+^2)^{\frac{1}{2}}} - \frac{\mu_- v_-}{(1-v_-^2)^{\frac{1}{2}}}\right) = P_- - P_+ \tag{3-73}$$

Como $\rho_+^0 \frac{v_+}{(1-v_+^2)^{\frac{1}{2}}} = \rho_-^0 \frac{v_-}{(1-v_-^2)^{\frac{1}{2}}} = j$ entonces podemos escribir $\frac{v_+}{(1-v_+^2)^{\frac{1}{2}}} = \frac{j}{\rho_+^0}$ y $\frac{v_-}{(1-v_-^2)^{\frac{1}{2}}} = \frac{j}{\rho_-^0}$. Utilizando este resultado en $(3-73)$ obtenemos,

$$jc^2 \left(\frac{\mu_+ j}{\rho_+^0} - \frac{\mu_- j}{\rho_-^0}\right) = P_- - P_+ \tag{3-74}$$

$$\Rightarrow j^2 c^2 \left(\frac{\mu_+}{\rho_+^0} - \frac{\mu_-}{\rho_-^0}\right) = P_- - P_+ \tag{3-75}$$

$$\Rightarrow j^2 = \frac{1}{c^2}\left(\frac{P_- - P_+}{\frac{\mu_+}{\rho_+^0} - \frac{\mu_-}{\rho_-^0}}\right) = \frac{1}{c^2}\left(\frac{P_+ - P_-}{\frac{\mu_-}{\rho_-^0} - \frac{\mu_+}{\rho_+^0}}\right) = -\frac{1}{c^2}\left(\frac{P_+ - P_-}{\frac{\mu_+}{\rho_+^0} - \frac{\mu_-}{\rho_-^0}}\right) = -\frac{1}{c^2}\left(\frac{P_+ - P_-}{\mu_+ V_+ - \mu_- V_-}\right) \tag{3-76}$$

$$\Rightarrow j = \frac{1}{c}\left(\frac{P_+ - P_-}{\frac{\mu_-}{\rho_-^0} - \frac{\mu_+}{\rho_+^0}}\right)^{\frac{1}{2}} \tag{3-77}$$

La ec. $(3-76)$ representa la segunda condición relativista del choque.

Ahora para encontrar la tercera ecuación relativista del choque debemos utilizar $(3-72)$, resultado de hacer $T_+^{11} = T_-^{11}$ de la siguiente manera:

$$\frac{jc^2\mu_+ v_+}{(1-v_+^2)^{\frac{1}{2}}} - \frac{jc^2\mu_- v_-}{(1-v_-^2)^{\frac{1}{2}}} = P_- - P_+ \tag{3-78}$$

Elevando al cuadrado a ambos lados de la igualdad anterior tenemos que se satisface,

$$\frac{j^2c^4\mu_+^2 v_+^2}{(1-v_+^2)} - \frac{2j^2c^4\mu_+ v_+ \mu_- v_-}{(1-v_+^2)^{\frac{1}{2}}(1-v_-^2)^{\frac{1}{2}}} + \frac{j^2c^4\mu_-^2 v_-^2}{(1-v_-^2)} = (P_- - P_+)^2 \tag{3-79}$$

Reemplazando $(3-67)$ y $(3-75)$ en $(3-79)$ obtenemos la siguiente relación,

$$\frac{j^2c^4\mu_+^2 v_+^2}{(1-v_+^2)} - \frac{j^2c^4\mu_-^2 v_-^2}{(1-v_-^2)} = (P_- - P_+)(P_- - P_+) \tag{3-80}$$

$$\Rightarrow \frac{j^2c^4\mu_+^2 v_+^2}{(1-v_+^2)} - \frac{j^2c^4\mu_-^2 v_-^2}{(1-v_-^2)} = (P_- - P_+)j^2c^2\left(\frac{\mu_+}{\rho_+^0} - \frac{\mu_-}{\rho_-^0}\right) \tag{3-81}$$

de donde obtenemos:

$$j^2c^4\left(\frac{\mu_+^2 v_+^2}{(1-v_+^2)} - \frac{\mu_-^2 v_-^2}{(1-v_-^2)}\right) = (P_- - P_+)j^2c^2\left(\frac{\mu_+}{\rho_+^0} - \frac{\mu_-}{\rho_-^0}\right) \tag{3-82}$$

Recordando que $(3-67)$

$$\frac{jc^2\mu_+}{(1-v_+^2)^{\frac{1}{2}}} = \frac{jc^2\mu_-}{(1-v_-^2)^{\frac{1}{2}}} \tag{3-83}$$

Entonces al elevar al cuadrado a ambos lados de la anterior ecuación tenemos que la siguiente relación es válida,

$$j^2c^4\left(\frac{\mu_+^2}{(1-v_+^2)} - \frac{\mu_-^2}{(1-v_-^2)}\right) = 0 \tag{3-84}$$

Finalmente restando $(3-82)$ a $(3-84)$ (es decir haciendo $(3-84)$ - $(3-82)$ ) obtenemos la tercera ecuación relativista de *Rankine-Hugoniot* $(3-86)$ que en su versión no relativista representa la ecuación adiabática del choque [8] ([98] *LL85*).

---

[8]Haciendo $1 - \frac{P}{c^2\rho} \approx 1$ en: $(3-76)$, $(3-86)$ y $(3-57)$, se obtiene las tres relaciones que describen un choque adiabático no relativista

$$\Rightarrow j^2 c^4 \left( \frac{\mu_+^2 (1 - v_+^2)}{(1 - v_+^2)} - \frac{\mu_-^2 (1 - v_-^2)}{(1 - v_-^2)} \right) = - (P_- - P_+) j^2 c^2 \left( \frac{\mu_+}{\rho_+^0} - \frac{\mu_-}{\rho_-^0} \right) \tag{3-85}$$

$$\Rightarrow j^2 c^2 \left( \mu_+^2 - \mu_-^2 \right) = j^2 (P_+ - P_-) \left( \frac{\mu_+}{\rho_+^0} - \frac{\mu_-}{\rho_-^0} \right) \tag{3-86}$$

La anterior ecuación también se puede escribir como,

$$c^2 \left( \mu_+^2 - \mu_-^2 \right) = (P_+ - P_-)(\mu_+ V_+ - \mu_- V_-) \tag{3-87}$$

# 4. La Radiación de Sincrotrón

A mediados de los años cincuenta, específicamente en Junio de 1947 se reportó en la comunidad física la detección de una radiación intensa visible proveniente de un haz de electrones ultraenergéticos que eran acelerados en un sincrotrón *General Electric* de 70MeV [45] y por esta razón a esta radiación se le acuñó el nombre de sincrotrón. Desde ese momento se recurrió a predicciones teóricas hechas con anterioridad que pudieran explicar dicho fenómeno. Estas inicialmente fueron hechas por *A. Liénard*, *E. Wiechert* y *J. Larmor* de manera independiente donde mostraron que electrones moviéndose en forma circular deberían radiar energía electromagnética [164]. Hacia 1912 *G. Schott* incluyó a los estudios anteriores la distribución de frecuencia de la radiación emitida y escribió sobre sus propiedades de polarización [154].

Con los años esos modelos primigenios se complementaron con nociones de la naciente relatividad especial y lograron su máxima expresión con el trabajo titulado: "*On the Classical Radiation of Accelerated Electrons*" [81], en el cual *J. Schwinger* encuentra expresiones para calcular la potencia radiada instantánea por electrones moviéndose de forma relativista en trayectorias arbitrarias, y de modo especial en movimiento circular uniforme. En este mismo trabajo obtiene la distribución angular y espectral de la radiación emitida. Con ayuda de estas ecuaciones fue posible explicar cuantitativamente la medición reportada en 1947 [45].

Con ayuda de los resultados analíticos encontrados por *Schwinger* la comunidad astronómica de esa época empezó a describir la emisión en radio proveniente de las estrellas sugiriendo que esta era producida por rayos cósmicos (electrones relativistas) que están atrapados en el campo magnético de la estrella [3, 61]. Después de ello la radiación de sincrotrón se convirtió en un caballo de batalla para explicar las emisiones detectadas desde ondas de radio hasta rayos x, donde los únicos prerrequisitos necesarios era la presencia de partículas de alta energía y campos magnéticos [62, 25, 168, 121].

La emisión de sincrotrón es un fenómeno que puede ser observado en muchas clases de objetos y contextos astrofísicos tales como [78]:

- Jets provenientes de radio-galaxias.

- Supernovas y remanentes de supernovas.

- Radiación no termal proveniente de la estrellas.
- *afterglow* de GRBs, entre otros.

En muchos de estos tópicos, por ejemplo en los modelos de fuentes de radiación cósmica, los problemas que involucran la interacción de electrones ultrarelativistas con un medio a su alrededor son muy comunes. Este "medio" es usualmente un gas parcialmente ionizado de baja densidad con una abundancia de elementos cósmicos (en su mayoría hidrógeno y helio), y este gas es permeado por un campo de radiación y un campo magnético. Los electrones interactúan en este medio mediante esencialmente cuatro procesos [22],

**i)** Haciendo colisiones elásticas e inelásticas con los átomos e iones del gas.

**ii)** Emitiendo un fotón *bremsstrahlung* (Radiación de frenado) durante estos procesos de dispersión.

**iii)** Realizando procesos de dispersión *Compton* con los fotones del campo de radiación.

**iv)** Siendo desviados por el campo magnético, emitiendo radiación de sincrotrón o *magneto-Bremsstrahlung* (Radiación de frenado magnética)

El primer proceso es importante únicamente a bajas energías $\gamma_e = \frac{E_e}{mc^2} < 1000$. Los tres procesos restantes son procesos de producción de fotones. Es necesario aclarar que el proceso de dispersión *Compton* (ítem iii) no produce un nuevo fotón; sin embargo, en colisiones con electrones ultrarelativistas, los fotones de baja energía provenientes del campo de radiación tienen un incremento de energía por un gran factor, así que se podría decir que un nuevo fotón de alta energía es producido. Detectando estos fotones producidos en estos procesos se dar razón del tipo de interacción de los electrones con el medio. Todos los tres procesos son esencialmente un caso especial de un proceso básico (dispersión *Compton*). Por ejemplo, el *Bremsstrahlang* (proceso ii) puede ser considerado como una dispersión *Compton* de un fotón virtual de los campos *Coulumbianos* de tres partículas en el sistema de dispersión. Por su parte la radiación de sincrotrón (proceso iv) en este contexto puede ser visto como una dispersión *Compton* de fotones virtuales de campos magnéticos estáticos [22].

Después de años de investigación se encontró que la radiación de sincrotrón no es radiación termal, es decir que las partículas que están emitiendo no se encuentran en equilibrio térmico. Su más destacada propiedad es que este tipo de radiación es fuertemente polarizada, un rasgo que la hace distinguible de la radiación térmica (tipo cuerpo negro) o de otros mecanismos no termales y permite al observador deducir la configuración tanto de los campos magnéticos como de la fuente.

Es objetivo de este capítulo describir matemáticamente las características más relevantes de la emisión de sincrotrón. Comenzaremos en la siguiente sección derivando la potencia radiada por una carga moviéndose de una forma ultra-relativista en un campo magnético homogéneo, teniendo de base el planteamiento mostrado por *T. Huelge* y *H. Falcke* [78], que a su vez toma muy de cerca el tratamiento matemático publicado por *G. Rybicki* y *R. Lightman* [142] .

## 4.1. Potencia Total Radiada

La potencia total $P$ radiada por una carga acelerada $q$ con velocidad $\vec{v}$ está dada por la fórmula de *Larmor*, la cual es hallada integrando el vector de *Pointing* (calculado a partir del producto cruz entre el campo eléctrico y el magnético retardados) sobre un ángulo sólido (ver deducción en Cap. 5,[42]). Esta fórmula tiene la siguiente expresión,

$$P = \frac{2q^2}{3c^3}|\dot{\vec{v}}|^2 \tag{4-1}$$

La aceleración de cargas moviéndose es gobernada por la fuerza de *Lorentz*,

$$\vec{F} = q\left(\vec{E} + \frac{\vec{v}}{c} \times \vec{B}\right) \tag{4-2}$$

En el caso de emisión de sincrotrón podemos asumir que la partícula está sometida solo a campos magnéticos, es decir que $\vec{E} = 0$, tal que se satisface,

$$\vec{F} = \frac{d\vec{p}}{dt} = \frac{d(\gamma m \vec{v})}{dt} \Rightarrow \frac{d(\gamma m \vec{v})}{dt} = \frac{q}{c}\vec{v} \times \vec{B} \tag{4-3}$$

donde $\gamma$ es el factor gamma de *Lorentz*.

Debido a la forma de la fuerza de *Lorentz*, vemos que se tiene que $\vec{v} \cdot \vec{F} = \vec{v} \cdot \left(\frac{q}{c}\vec{v} \times \vec{B}\right)$. Gracias a ello tendremos que la fuerza de *Lorentz* será perpendicular al vector velocidad (instantáneamente), es decir que (Ver figura **4-1** ): $\vec{v} \cdot \left(\vec{F}\right) = 0$ .

Esto implica que $|\vec{v}|$ es constante y entonces $\gamma$ es constante, mientras que la dirección de $\vec{v}$ cambia, asumiendo que el campo magnético $\vec{B}$ sea constante.
En otras palabras tenemos conservación de la energía en la partícula puesto que,

$$\frac{dW}{dt} = \frac{d\left(\gamma m c^2\right)}{dt} = \vec{v} \cdot \vec{F} = 0 \tag{4-4}$$

Y la ecuación de movimiento queda de la siguiente forma.

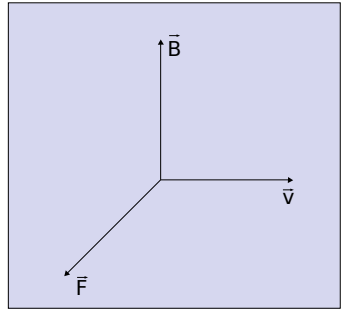

**Figura 4-1**.: Esquema pictórico de la ortogonalidad entre $\vec{B}$, $\vec{F}$ y $\vec{v}$

$$\gamma m \frac{d\vec{v}}{dt} = \frac{q}{c} \vec{v} \times \vec{B} \tag{4-5}$$

Como la aceleración $\vec{a} = d\vec{v}/dt$ es también perpendicular a $\vec{B}$, sus componentes paralela y perpendicular a las líneas de campo magnético son:

$$\vec{a_\parallel} = \frac{d\vec{v_\parallel}}{dt} = 0 \qquad \vec{a_\perp} = \frac{d\vec{v_\perp}}{dt} = \frac{d\vec{v_\perp}}{dt} = \frac{d\vec{v}}{dt} = \frac{q}{mc} \vec{v} \times \vec{B} \tag{4-6}$$

Donde $\vec{v}_\parallel$ es la componente paralela y $\vec{v_\perp}$ es la componente perpendicular de la velocidad.

Asumiendo que la partícula está en presencia de un campo magnético homogéneo, es decir que, $B = \left|\vec{B}\right|$ es constante en el espacio, y observando la figura **4-2**, es válido afirmar que,

$$\vec{v} = \vec{v_\perp} + \vec{v_\parallel} \;\; \Rightarrow |\vec{v}|^2 = |\vec{v_\perp}|^2 + |\vec{v_\parallel}|^2 \;\; \Rightarrow v_\perp = |\vec{v_\perp}| = \sqrt{|\vec{v}|^2 - |\vec{v_\parallel}|^2} = |\vec{v}| \sin\alpha = Cte \tag{4-7}$$

donde $\alpha$ (*pitch angle*) se define como el ángulo entre la dirección de $\vec{B}$ y $\vec{v}$.

Ahora a partir de,

$$\gamma m \frac{d\vec{v}}{dt} = \frac{q}{c} \vec{v} \times \vec{B}$$

y con ayuda de,

$$\left|\dot{\vec{v}}\right| = \left|\frac{d\vec{v}}{dt}\right| = \left|\frac{\vec{v_\perp}}{dt} + \frac{\vec{v_\parallel}}{dt}\right| ; \left|\vec{A} \times \vec{C}\right| = |A||C|\sin\alpha$$

obtenemos que,

$$\left|\frac{d\vec{v}}{dt}\right| = \left|\frac{d\vec{v_\perp}}{dt}\right| = \frac{q}{\gamma mc} \left|\vec{v_\perp} \times \vec{B}\right|$$

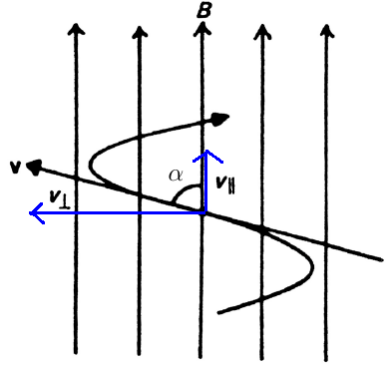

**Figura 4-2**.: Esquema pictórico de la trayectoria seguida por una partícula cargada en la presencia de un campo magnético transversal a su trayectoria inicial. Gráfico adaptado de ([105], fig.8.1)

$$\Rightarrow \left|\frac{d\vec{v}}{dt}\right| = \frac{q}{\gamma mc}|\vec{v_\perp}||\vec{B}| = \left|\dot{\vec{v}}\right| = \frac{q}{\gamma mc}v_\perp B$$

$$\left|\dot{\vec{v}}\right| = w_B v_\perp$$

en donde $w_B = \frac{qB}{\gamma mc}$ es la frecuencia sincrotrónica.

Asumiendo que $w_B$ sea constante, tenemos que $\frac{dw_B}{dt} = 0$, y por tanto,

$$a_\theta = \left|\dot{\vec{v}}\right| = w_B v_\perp$$

en donde $\theta$ es la coordenada angular polar perpendicular al campo magnético $\vec{B}$.

Este resultado [1] corresponde a un movimiento uniforme líneal a lo largo de las líneas del campo magnético $\left(\frac{dv_\parallel}{dt} = 0 \Rightarrow v_\parallel = cte\right)$ superpuesto por un movimiento circular ($w_B = cte$) en el plano perpendicular a $\vec{B}$ con frecuencia angular $w_B = \frac{qB}{\gamma mc}$, lo cual implica que el movimiento de la partícula alrededor del campo magnético sea helicoidal.

Transformaciones desde un sistema de referencia co-móvil [2] a la partícula, al sistema de referencia del observador introducen un factor adicional $\gamma^2$ en $\dot{v}_\perp$ ([142] Ec. 4-91b) tal que en el sistema de referencia del observador [3] tenemos que: $|\dot{\vec{v}}|' = \gamma^2 w_B v_\perp$.

En el sistema de referencia co-móvil tenemos que la potencia total radiada por un electrón

---

[1] Esta es la magnitud de la aceleración tangencial en coordenadas polares $a_r = \ddot{r} + 2r\dot{\theta}^2$ y $a_\theta = r\ddot{\theta} + \dot{r}\dot{\theta}$
[2] También llamado *Instantaneous Rest Frame* ([142],pp. 138)
[3] Ver detalles de la demostración: $a'_\perp = \gamma a_\perp$ en ([175], Apéndice B.2)

vía emisión de sincrotrón, usando la formula de *Larmor* es,

$$P' = \frac{2q^2}{2c^3}|\dot{\vec{v}}|'^2$$

$$P' = \frac{2q^2}{2c^3}\gamma^4 w_B^2 v_\perp^2$$

$$P' = \frac{2q^2}{2c^3}\gamma^4 \frac{q^2 B^2 v_\perp^2}{\gamma^2 m^2 c^2} \equiv cte$$

Ahora vamos a mostrar que $\boxed{P = P'}$ ([142],pp. 139) donde $P$ es la potencia radiada medida por el observador.

Por la invarianza del cuadrimomento al cuadrado tenemos, respectivamente para el sistema de referencia propio y para el otro sistema,

$$p^\mu p_\mu = p'^\mu p'_\mu = p^2$$

donde

$$p^\mu = \gamma m (c, \vec{v}) = \left(\frac{E}{c}, \vec{p}\right)$$

Teniendo en cuenta la métrica de Minkowski diag(-1,1,1,1), tenemos, para el marco de referencia propio ($\gamma = 1$):

$$p^\mu p_\mu = -p^0 p^0 = -m_0^2 c^2$$

Mientras que en cualquier sistema de referencia que se mueve con velocidad $v$ con respecto al sistema en reposo:

$$p'^\mu p'_\mu = -p^0 p^0 + p^1 p^1 + p^2 p^2 + p^3 p^3 = -\frac{E^2}{c^2} + p^2$$

en donde $p^2 = p_x^2 + p_y^2 + p_z^2$.

Por la invarianza del cuadrimomento al cuadrado tenemos,

$$-m_0^2 c^2 = -\frac{E^2}{c^2} + p^2 \Rightarrow$$

$$\boxed{E^2 = m_0^2 c^4 + p^2 c^2}$$

Para el caso de los fotones (radiación emitida vía emisión de sincrotrón) tenemos que,

$$m_0 = 0 \Rightarrow E^2 = p^2 c^2 \Rightarrow E = |\vec{p}|c$$

En el proceso anterior, se consideró al momentum-energía como un cuadrivector. Como tal, la componente cero se transforma como,

$$E = \gamma(E' + p'_{x'}c)$$

en donde las variables primadas corresponden a las magnitudes físicas observadas en cierto sistema de referencia inercial K' que se mueve uniformemente en el sentido positivo del eje-x con respecto a cierto sistema de referencia inercial K (variables no primadas). Si el referencial K' es el sistema de referencia co-móvil de los electrones tenemos que $dE = \gamma dE'$ y $dt = \gamma dt'$, en donde esta última expresión proviene de la componente cero de las transformaciones de *Lorentz* en forma diferencial. De estas dos expresiones y teniendo en cuenta que,

$$P = \frac{dE}{dt}, \quad P' = \frac{dE'}{dt'}$$

obtenemos:

$$P_{rad} = P'_{rad}$$

es decir, la potencia total emitida es un invariante de *Lorentz*. Ahora bien, en el sistema de referencia del observador tenemos que,

$$P = \frac{2q^2}{3c^3}\gamma^4 \frac{q^2 B^2}{\gamma^2 m^2 c^2} v_\perp^2 \equiv cte$$

Note que la radiación emitida por protones relativistas es menor que la emitida por electrones o positrones debido a que los protones tienen una relación de *carga/masa* menor que la de los electrones y los positrones, pues $m_p = 1836 m_e$.

Ahora definiendo a $\beta_\perp$ como $\frac{v_\perp}{c}$ tenemos que la igualdad anterior se transforma en,

$$P = \frac{2q^2}{3c^3}\gamma^2 \left(\frac{q}{m}\right)^2 B^2 \beta_\perp^2$$

Para partículas con una distribución isotrópica de velocidad $\beta$, nosotros podemos promediar sobre el ángulo *pitch* ($\alpha$) para así derivar la potencia promedio radiada por partícula. Es decir que nos toca hallar $\langle \beta_\perp \rangle$, promedio de $\beta_\perp$ que definimos por la siguiente función en coordenadas esféricas ([7]),

$$\langle \beta_\perp^2 \rangle = \frac{\int_0^\pi \int_0^{2\pi} \int_0^r \beta_\perp^2 r^2 \sin\alpha \, dr\, d\alpha\, d\theta}{\frac{4\pi r^3}{3}} = \frac{1}{4\pi} \int_0^\pi \int_0^{2\pi} \beta_\perp^2 \sin\alpha \, d\alpha\, d\theta$$

Como $\beta_\perp = \beta \sin\alpha$, donde $\vec{B}$ es perpendicular a $\vec{\beta_\perp}$ (ver figura **4-3**).

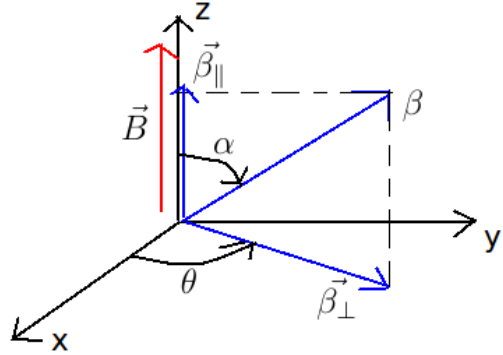

**Figura 4-3**.: Esquema cartesiano de la relación entre $\vec{\beta}$ y $\vec{B}$
.

$$\langle \beta_\perp^2 \rangle = \frac{1}{4\pi} \int_0^\pi \int_0^{2\pi} \beta^2 \sin^3\alpha \, d\alpha\, d\theta = \frac{\beta^2}{4\pi} \int_0^{2\pi} d\theta \int_0^\pi \sin^3\alpha \, d\alpha$$

$$\langle \beta_\perp^2 \rangle = \frac{\beta^2}{4\pi} 2\pi \int_0^\pi \sin\alpha(1-\cos^2\alpha) d\alpha = \frac{\beta^2}{2} \left[ -\cos\alpha\big|_0^\pi + \frac{\cos^3\alpha}{3}\big|_0^\pi \right] = \frac{\beta^2}{2}\left[ 2 - \frac{2}{3} \right]$$

O sea que el promedio buscado es,

$$\langle \beta_\perp^2 \rangle = \frac{2\beta^2}{3}$$

Entonces,

$$P_{rad} = \frac{2q^2}{3c^3} \gamma^2 \left(\frac{q}{m}\right)^2 B^2 \left(\frac{2}{3}\right)\beta^2$$

$$P_{rad} = \left(\frac{4}{9}\right)\left(\frac{q}{m}\right)^2 \frac{q^2}{c^3}\gamma^2 B^2 \beta^2 \tag{4-8}$$

se puede observar que este resultado promedio es equivalente a un ángulo *pitch* de,

$$\alpha_{equivalente} = \arcsin\sqrt{\frac{2}{3}} = 54{,}736 \text{ grados}$$

Con ayuda de la sección eficaz de Thomson ([22]),

$$\sigma_T = \frac{8\pi}{3}r_e^2 = \frac{8\pi}{3}\frac{e^4}{m^2c^4} \simeq 6{,}65 \times 10^{-25}\text{cm}^{-2} \tag{4-9}$$

en donde $e$ es la carga elemental del electrón, $r_e$ es el radio clásico del electrón, y con la definición de la densidad de energía del campo Magnético ($U_B$),

$$U_B = \frac{B^2}{8\pi} \tag{4-10}$$

vemos que la potencia total radiada promedio emitida por partícula puede ser escrita como,

$$P_{rad} = \frac{4}{3}\sigma_T c(\gamma\beta)^2 U_B \tag{4-11}$$

Ahora vamos a hacer un análisis de energía para encontrar cómo disminuye el factor $\gamma$ de Lorentz de cada electrón relativista debido a la emisión de energía de sincrotrón y a su vez hallar una expresión B como función del cambio temporal de $\gamma$.

Asumiendo conservación de la energía localmente se cumple,

$$E_i = E_f + Q$$

donde Q es la energía disipada vía emisión de sincrotrón.

$$\Rightarrow E_f - E_i = -Q$$

$$\Delta E = -Q$$

Para un intervalo menor de tiempo tenemos [175],

$$\frac{dE}{dt} = -\frac{dQ}{dt}$$

$$\frac{dE}{dt} = -P_{rad}$$

$$\frac{d(\gamma mc^2)}{dt} = -P_{rad}$$

$$\dot{\gamma} = -\frac{P_{rad}}{mc^2}$$

Utilizando la relación de la potencia radiada hallada anteriormente obtenemos que,

$$\dot{\gamma} = -\left(\frac{4}{9}\right)\left(\frac{q}{m}\right)^2 \frac{q^2}{c^3}\frac{\gamma^2 B^2 \beta^2}{mc^2}$$

$$\dot{\gamma} = -\left(\frac{4}{9}\right)\left(\frac{q}{m}\right)^3 \frac{q}{c^5}\gamma^2 B^2 \beta^2$$

$$\Rightarrow \dot{\gamma} = -\xi\gamma^2\beta^2$$

con,

$$\xi = \left(\frac{4}{9}\right)\left(\frac{q}{m}\right)^3 \frac{q}{c^5} B^2 \tag{4-12}$$

En el caso del movimiento de electrones ultrarelativistas $\gamma \gg 1 \Rightarrow \beta = 1 - \frac{1}{\gamma^2} \approx 1$

$$\dot{\gamma} = -\xi\gamma^2$$

$$\frac{d\gamma}{dt} = -\xi\gamma^2 \rightarrow \int_{\gamma_o}^{\gamma(t)} \frac{d\gamma}{\gamma^2} = \int_{t_0}^{t} \xi dt$$

$$\rightarrow -\frac{1}{\gamma}\bigg|_{\gamma_o}^{\gamma(t)} = -\xi(t - t_0)$$

$$\rightarrow \frac{1}{\gamma(t)} = \frac{1}{\gamma_o} + \xi(t - t_0)$$

$$\rightarrow \gamma(t) = \frac{1}{\frac{1}{\gamma_o} + \xi(t - t_0)}$$

$$\Rightarrow \gamma(t) = \frac{\gamma_0}{\xi\gamma_0(t - t_0) + 1}$$

$$\xi = \frac{1}{\gamma(t)(t - t_0)} - \frac{1}{\gamma_0(t - t_0)}$$

$$\xi = \frac{1}{(t - t_0)}\left(\frac{\gamma_0 - \gamma(t)}{\gamma_0\gamma(t)}\right)$$

finalmente, llevando el anterior valor de $\xi$ a la ec. (4-12), tenemos,

$$B(t) = \sqrt{\frac{c^5}{q}\left(\frac{m}{q}\right)^3 \left(\frac{3}{2}\right)^2 \frac{1}{(t - t_0)}\left(\frac{\gamma_0 - \gamma(t)}{\gamma_0\gamma(t)}\right)}, \text{con } \gamma(t) > \gamma(0)$$

Así vemos que conociendo $\gamma(t) = f(t)$ se puede obtener $B = B(t)$. En el siguiente capítulo mostraremos una forma de obtener $\gamma(t)$, a partir de un estudio hidrodinámico del movimiento de flujo de electrones ultrarelativistas.

## 4.2. Características geométricas de la emisión de sincrotrón

Para partículas cargadas no relativistas en presencia de un campo magnético transversal a su movimiento, la radiación consiste de líneas de emisión a la frecuencia fundamental $f = \frac{w_B}{2\pi}$ (en el caso no relativista tenemos que $w_B = \frac{qB}{\gamma m c} \to \frac{qB}{mc}$), con un patrón de radiación bipolar llamado "radiación de ciclotrón". Para partículas relativistas este patrón es altamente emitido en la dirección del movimiento de las partículas dentro de un cono de semi-ángulo $\theta \sim \frac{1}{\gamma}$ ([78],pp.16).

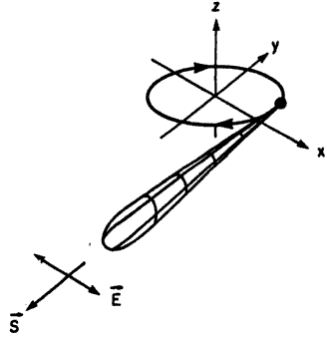

**Figura 4-4**.: Esquema pictórico del patrón de radiación de la emisión de sincrotrón emitida por un electrón ultrarelativista ([14],fig. 4.23). Observe que la luz es línealmente polarizada con el $\vec{E}$ que está vibrando en el plano de la órbita (x-y). El flujo de radiación $\vec{S}$ está confinado a un delgado cono de ancho $\frac{1}{\gamma}$ rad.

A continuación vamos a mostrar que si la emisión de fotones en un sistema $K'$ (que se encuentra en movimiento con respecto a uno $K$), es isotrópica, en el sistema $K$ esta se observará directamente en el mismo sentido de movimiento del marco de la partícula formando un ángulo $\theta \approx \frac{1}{\gamma}$ para $\gamma \gg 1$. Es decir, que la emisión detectada en el marco $K$ se observara como un cono con un semi-ángulo $\frac{1}{\gamma}$ que al aumentar la velocidad lo hará también su grado de colimación (ver figura **4-4**). Para entender esto con mas detalle tenemos que describir las ecuaciones de transformación de *Lorentz* para un ángulo $\theta$ medido en dos sistemas de referencia, en los cuales uno se esta moviendo con respecto al otro. Para comenzar tengamos en cuenta el gráfico **4-5**, donde la velocidad $u'$ es medida en el marco de referencia $K'$.

Las relaciones entre las componentes de la velocidad $u$ entre ambos sistemas de referencia se

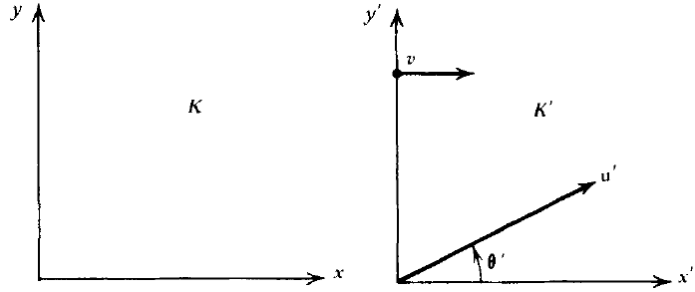

**Figura 4-5**.: Transformación de *Lorentz* de la velocidad de una partícula entre dos sistemas de referencia, uno moviéndose con respecto al otro ([142],fig. 4.2).

escriben usualmente como ([175],Cap.1).

$$u_x = \frac{dx}{dt} = \frac{u'_x + v}{1 + \frac{vu'_x}{c^2}}$$

$$u_y = \frac{u'_y}{\gamma\left(1 + \frac{vu'_x}{c^2}\right)}$$

$$u_z = \frac{u'_z}{\gamma\left(1 + \frac{vu'_x}{c^2}\right)}$$

La generalización de estas ecuaciones a una velocidad arbitraria $\vec{v}$, que no necesariamente este a lo largo del eje x, puede ser colocada en términos de las componentes de $\vec{u}$ que son paralelas ($u_\parallel$) y perpendiculares ($u_\perp$) a $\vec{v}$, ([142], pp. 109-110).

Como $u_x$ es perpendicular a $u_y$ y ambos a $u_z$ entonces una escogencia viable para $u_\perp$ y $u_\parallel$ es

$$u_\parallel = \frac{u'_\parallel + v}{1 + \frac{vu'_\parallel}{c^2}}; u_\perp = \frac{u'_\perp}{\gamma\left(1 + \frac{vu'_\parallel}{c^2}\right)}$$

Observando la figura **4-5** podemos decir que,

$$\tan\theta = \frac{u_\perp}{u_\parallel} = \frac{\frac{u'_\perp}{\gamma\left(1+\frac{vu'_\parallel}{c^2}\right)}}{\frac{u'_\parallel + v}{1+\frac{vu'_\parallel}{c^2}}} = \frac{u'_\perp}{\gamma(u'_\parallel + v)} = \frac{u'\sin\theta'}{\gamma(u'\cos\theta' + v)}$$

donde hemos considerado $u' = |\vec{u}'|$.

En el caso que $u' \equiv c$, es decir donde estamos considerando fotones, tenemos;

$$\tan\theta = \frac{c\sin\theta'}{\gamma(c\cos\theta' + v)} = \frac{c\sin\theta'}{\gamma c\left(\cos\theta' + \frac{v}{c}\right)} = \frac{\sin\theta'}{\gamma(\cos\theta' + \frac{v}{c})}$$

$$\tan^2 \theta = \frac{\sin^2 \theta'}{\gamma \left(\cos \theta' + \frac{v}{c}\right)^2}$$

la cual representa la aberración de la luz [142]. Haciendo uso de la identidad trigonométrica $\tan^2 \theta + 1 = \frac{1}{\cos^2 \theta}$ y realizando una álgebra adecuada ([175],pp.22) encontramos que,

$$\cos \theta = \frac{\cos \theta' + \left(\frac{v}{c}\right)}{1 + \left(\frac{v}{c}\right) \cos \theta'}$$

y como $\sin \theta = \tan \theta \cos \theta$, reunimos los dos resultados anteriores para obtener,

$$\Rightarrow \text{sen}\, \theta = \frac{\sin \theta'}{\gamma(\cos \theta' + \frac{v}{c})} \frac{\cos \theta' + \left(\frac{v}{c}\right)}{1 + \left(\frac{v}{c}\right) \cos \theta'}$$

$$\text{sen}\, \theta = \frac{\text{sen}\, \theta'}{\gamma \left(1 + \frac{v}{c} \cos \theta'\right)}$$

Si $\theta = \frac{\pi}{2}$, es decir si el fotón es emitido a un ángulo recto a $v$ medido desde $K'$, tenemos que: $\sin \theta = \frac{1}{\gamma}$. Si consideramos que $\gamma \gg 1$ tenemos $\sin \theta \approx 0$, es decir, que $\sin \theta \approx \theta \boxed{\Rightarrow \theta \approx \frac{1}{\gamma}}$.

Ahora si se emiten dos fotones isotrópicamente en $K'$, entonces la mitad de ellos tendrá $\theta' < \frac{\pi}{2}$ y la otra mitad $\theta' > \frac{\pi}{2}$ entonces $\theta \approx \frac{1}{\gamma}$ significará que en el sistema de referencia $K$ los fotones son concentrados en la dirección de movimiento, con la mitad de ellos estando dentro de un cono con un semi-ángulo igual a $\frac{1}{\gamma}$ (ver figura **4-6**).

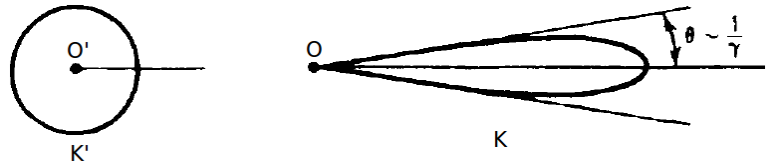

**Figura 4-6**.: Diferencia entre el patrón de radiación medido en el sistema de referencia co-móvil y en el del observador ([142],fig.4.3)

Para el caso que la propagación de las partículas cargadas relativistas (que luego emitirán radiación) se haga en forma de una onda expansiva esférica tendremos que la emisión de fotones sincrotrónicos se puede asumir isotrópica medida desde el sistema de referencia co-móvil al movimiento de los electrones y por las efectos relativistas que demostramos anteriormente vemos en el sistema de referencia del observador la detección se hace en un cono con un semi-ángulo $\theta \approx \frac{1}{\gamma}$, donde $(\gamma \gg 1)$. A medida que las partículas pierden velocidad, su factor $\gamma$ disminuye y por ende aumenta su $\theta$ medido por el observador (ver figura **4-7**).

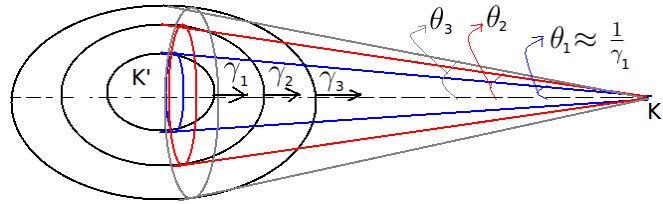

**Figura 4-7**.: Diferencia entre el patrón de radiación, proveniente de electrones ultrarelativistas, medido en el sistema de referencia co-móvil a ellos, con el medido por el observador.

## 4.3. Características temporales y en frecuencia de la detección de la radiación de sincrotrón

A partir del estudio del movimiento helicoidal que presenta el electrón relativista alrededor del campo magnético (ver figura **4-2**) podemos hallar el tiempo que dura la detección del pulso de radiación cada vez que el electrón da una vuelta en dirección hacia el punto de observación. Para comenzar a buscar este resultado debemos tener en cuenta la figura **4-8** donde se muestra una sección de arco de la trayectoria helicoidal de la partícula cargada. En este caso el radio de la trayectoria helicoidal se denota como $a$ y no se le debe confundir con el radio de la proyección circular de la trayectoria helicoidal, que está ubicado en el plano perpendicular a $\vec{B}$.

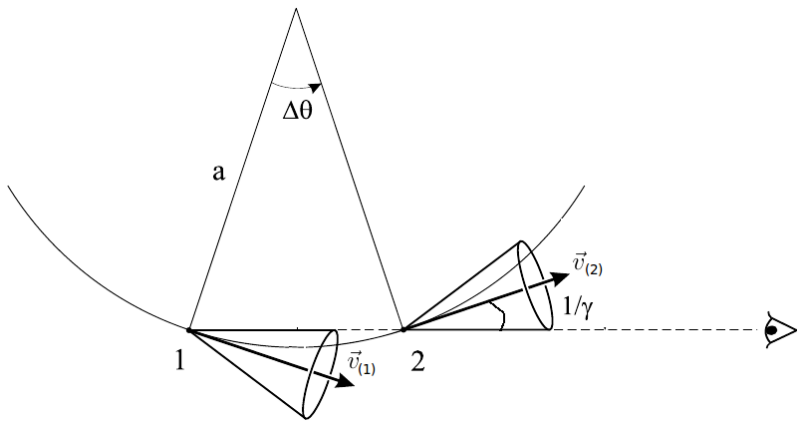

**Figura 4-8**.: Desplazamiento del cono de emisión de radiación a lo largo de los diferentes puntos de la trayectoria helicoidal con respecto a la línea de detección del observador. Modificado a partir de ([78], fig.1).

Como se puede apreciar en la figura **4-8** un observador detectará un corto (*flash*) haz de radiación cada vez que el cono de radiación esté en dirección hacia él. Teniendo en cuenta la misma figura **4-8** el *flash* comenzará cuando el cono de emisión de radiación (*beaming*) alcance la línea del observador en el punto 1 y terminará cuando este *beaming* deja esta línea de conexión hacia al observador en el punto 2. En términos angulares esta distancia entre los puntos 1 y 2 la definimos como $\Delta S$, la cual podemos hallar haciendo un minucioso análisis geométrico de la figura **4-8** en donde se encuentra que ([175],pp.75),

$$\Delta\theta = \frac{2}{\gamma} \Rightarrow \Delta S = a\Delta\theta \approx \frac{2a}{\gamma}$$

Teniendo ahora en cuenta la interacción del campo magnético con el movimiento de la partícula podemos decir que,

$$F^\alpha = \frac{dp^\alpha}{dt} \longrightarrow \gamma m \frac{\Delta \vec{v}}{\Delta t} = \frac{q}{c}\vec{v} \times \vec{B}$$

Analizando la figura **4-8** podemos darnos cuenta que,

$$|\Delta v| = v\Delta\theta$$

ocasionando que nuestra ecuación dinámica cambie a,

$$\gamma m \frac{\Delta \vec{v}}{\Delta t} = \frac{q}{c}\vec{v} \times \vec{B}$$

$$\gamma m \frac{|\Delta \vec{v}|}{\Delta t} = \frac{q}{c}|\vec{v} \times \vec{B}|$$

$$\gamma m \frac{v\Delta\theta}{\frac{\Delta S}{v}} = \frac{q}{c}vB\sin\alpha$$

$$v\frac{\Delta\theta}{\Delta S} = \frac{q}{c}\frac{B\sin\alpha}{\gamma m}$$

$$\frac{\Delta\theta}{\Delta S} = \frac{qB\sin\alpha}{\gamma mcv}$$

y como $\Delta S = a\Delta\theta \Rightarrow a = \frac{\Delta S}{\Delta \theta}$

$$\frac{1}{a} = \frac{qB\sin\alpha}{\gamma mcv}$$

$$\frac{1}{a} = \frac{w_B \sin\alpha}{v} \Rightarrow a = \frac{v}{w_B\sin\alpha} \Rightarrow \Delta S = a\Delta\theta = \frac{2a}{\gamma}$$

$$\Rightarrow \Delta S = \frac{2v}{\gamma w_B \sin\alpha}$$

Llamando $\Delta t = t_2 - t_1$ el tiempo que transcurre mientras la partícula pasa del punto 1 y al 2 (ver figura **4-8**) y como $v$ es constante en magnitud entonces,

$$\Delta S = v\Delta t \Rightarrow \Delta t = \frac{\Delta S}{v} = \frac{2v}{v\gamma w_b \sin\alpha}$$

$$\Delta t \approx \frac{2}{\gamma w_B \sin \alpha}; (\gamma \gg 1) \tag{4-13}$$

Escojamos $t_1^{ll}$ y $t_2^{ll}$ como los tiempos de llegada de la radiación en el sistema de referencia del observador, provenientes de los puntos 1 y 2 (figura **4-8**). Trabajando en el contexto Minkowskiano tenemos que.

$$\Delta t^{ll} = \Delta t - \frac{\Delta S}{c}$$

Es decir que la diferencia de tiempos: $\Delta t^{ll} \equiv \Delta t_{obs}$ que el observador percibe esta corregida por el tiempo que tarda la radiación de ir del punto 1 al punto 2.

$$\Rightarrow \Delta t_{obs} = \Delta t - \frac{v \Delta t}{c}$$

$$\Delta t_{obs} = \Delta t \left(1 - \frac{v}{c}\right)$$

Para $\gamma \gg 1$ tenemos que: $\Delta t \approx \frac{2}{\gamma w_B \sin \alpha}$.

Por su parte $\left(1 - \frac{v}{c}\right)$ para $\gamma \gg 1$ es de la forma,

$$= \left(1 - \sqrt{1 - \frac{1}{\gamma^2}}\right)$$
$$\approx 1 - \left(1 - \frac{1}{2\gamma^2}\right)$$
$$\approx \frac{1}{2\gamma^2}$$

entonces uniendo los dos últimos resultados, tenemos que para $\gamma \gg 1$ se satisface que,

$$\Delta t_{obs} = \frac{2}{\gamma w_B \sin \alpha} \frac{1}{2\gamma^2}$$

$$\Delta t_{obs} = \frac{1}{\gamma^3 w_B \sin \alpha}; \qquad \gamma \gg 1 \tag{4-14}$$

El resultado encontrado (ec. 4-14) muestra que a medida que la partícula disminuya su velocidad, el tiempo de detección del pulso de radiación sera mayor.

En conclusión vimos como la partícula emite radiación que ilumina al detector del observador durante un intervalo $\Delta t_{obs}$ cada cierto periodo de tiempo $T$ detectando así un pulso de radiación periódico (ver figura **4-9**).

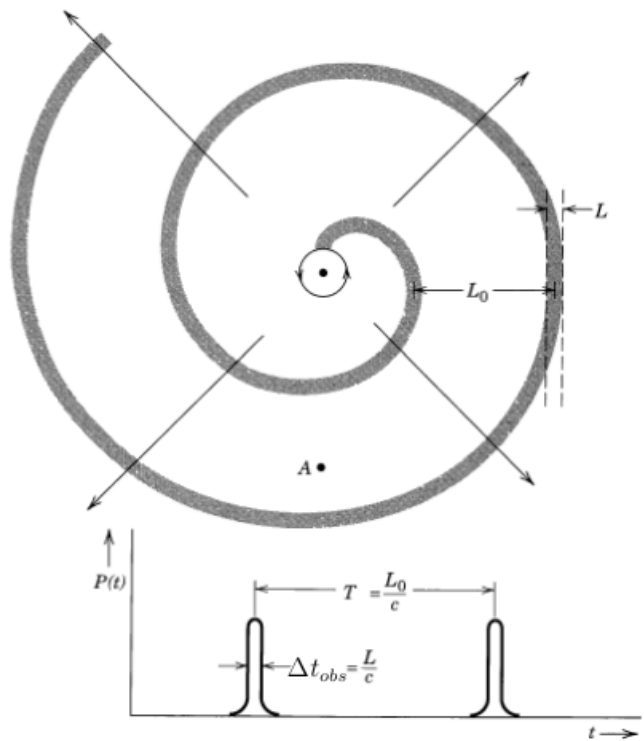

**Figura 4-9**.: Una partícula relativista en movimiento periódico emite un patrón espiral de radiación que un observador en el punto A detecta como cortas ráfagas (*bursts*) de radiación de duración de tiempo $\Delta t_{Obs}$, ocurriendo a intervalos regulares de tiempo T. Modificado de ([83], fig 14.7).

Es decir que la potencia radiada la puedo expresar como una distribución de potencias en el tiempo o en la frecuencia teniendo en cuenta la teoría de *Fourier*, es decir que para cada pulso tengo,

$$\Delta t_{obs} \Delta w \geq 1$$

El espectro de potencias contiene un conjunto de frecuencias hasta un máximo denominado $w_c \sim (\Delta t)^{-1}$ (figura **4-10**).

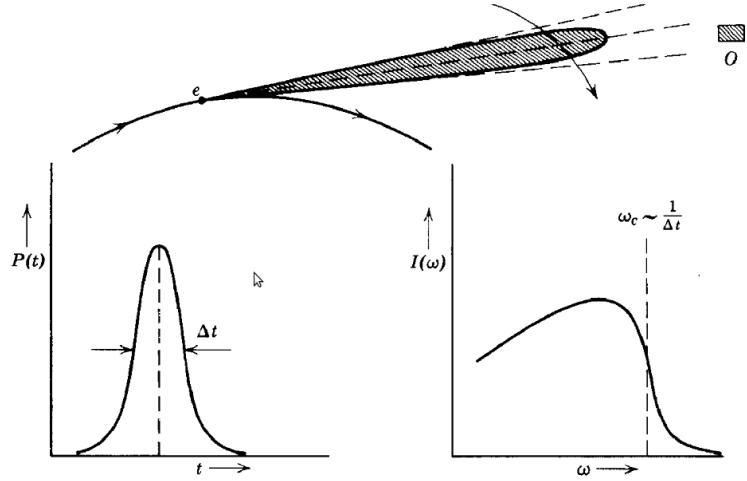

**Figura 4-10**.: La partícula ilumina la detector en un intervalo de tiempo $\Delta_{t_{Obs}}$ que se traduce en un $\Delta w$ [82].

Como $\Delta t_{obs} = \frac{1}{\gamma^3 w_B \sin \alpha}$ y utilizando que $\Delta t_{obs} \Delta w > 1$ entonces se puede decir que,

$$\Delta w \geq \frac{1}{\Delta t_{obs}}$$

para obtener,

$$\Delta w \geq \gamma^3 w_B \sin \alpha \equiv 2\pi \underbrace{\left(\frac{1}{2\pi} \gamma^3 w_B \sin \alpha\right)}_{\Delta f^*}$$

es decir que se emite radiación a frecuencias,

$$\Delta f^* \gg f = \frac{2\pi}{w_B}; \qquad \gamma \gg 1$$

Esto demuestra el porqué se pueden explicar las detecciones de radiación electromagnética desde frecuencias altas (por ejemplo rayos x), hasta frecuencias bajas (ondas de radio) en los *afterglows* de GRBs vía mecanismos de emisión de sincrotrón. También pudimos ver (teniendo en cuenta la teoría de *Fourier* relacionada con señales) que el pulso de radiación que se emitió en un corto tiempo corresponde a un ancho de frecuencia grande y viceversa.

Para un electrón ultrarelativista ($\gamma \gg 1$) tenemos que,

$$\Delta t \approx \frac{1}{\gamma^3 w_B \sin \alpha}$$
$$\Rightarrow \Delta w \approx \gamma^3 w_B \sin \alpha$$

asumiendo como $w_c$ el máximo de frecuencias de la muestra contenida en el ancho de banda $\Delta w$ una buena escogencia para hallar el valor de $w_c$ sería evaluar $\Delta w$ en el caso que $\sin \alpha$ sea máximo, es decir para $\alpha = \frac{\pi}{2}$, resultando que,

$w_c = \gamma^3 w_B$

Si $\gamma_{max} \approx 10^3 \Rightarrow w_v \approx 10^{18}$

$w_c \approx 10^9 \left(\dfrac{qB}{\gamma mc}\right)$

$w_c \approx 10^9 \left(\dfrac{qB}{10^3 mc}\right) \approx 10^6 \left(\dfrac{q}{mc}\right) B$

con $v_c = \frac{w_c}{2\pi}$

Si ahora decimos por ejemplo que el electrón ultrarelativista tendrá asociado un $\gamma \approx 10^3$ y emitirá un rayo x con $\approx 10^{16} - 10^{18} Hz$, entonces en principio de podría calcular los $B$ generados por los primeros electrones,

$B \approx \left(\dfrac{m_e c}{q}\right) \dfrac{10^{17}}{10^6}(2\pi) = 3{,}57 T$

El calculo del campo magnético generado por electrones ultrarelativistas ($\gamma \approx 10^3$) utilizando la relación de la frecuencia de corte $w_c$, obteniendo para una radiación emitida de $\frac{w_c}{2\pi} \approx 10^{17} Hz$(rayos x), un $B \cong 3{,}57 T \approx 3{,}57 \times 10^4 Gauss$, es cercana a la suposición hecha por *G. Beuefi* y *A. Barret* de un $B = 0{,}8 \times 10^4 Gauss$ transversal al movimiento de unos electrones acelerados a una energía de $E = 3 \times 10^9 eV$, en donde se obtiene una frecuencia máxima de emisión $w_c \approx 4{,}9 \times 10^{18} Hz$. [14].

Otros autores definen convenientemente una frecuencia de corte de la forma [142, 78],

$w_c = \dfrac{3}{2}\gamma^3 w_B \sin \alpha$

## 4.4. Dependencia de la radiación de sincrotrón con la frecuencia

El ánimo de esta sección es mostrar con detalle la deducción de la potencia radiada por unidad de frecuencia vía emisión de sincrotrón proveniente de electrones relativistas en presencia de campos magnéticos transversales. Para ello el autor de la presente tesis consultó en primera medida varias fuentes al tratarse de un tema estándar de electrodinámica clásica [155, 30, 101, 14, 82, 97, 151, 42, 142]. Se escogió como base el desarrollo realizado en el tema por *P. J. Duke* [42], debido a su gran detalle en pasos de cálculo intermedio; sin embargo en algunos apartes se utilizó conocimiento albergado en las demás fuentes citadas.

Nosotros podemos expresar la derivada de la potencia radiada con respecto al ángulo sólido como [151]:

$$\frac{dP_{rad}}{d\Omega} = \frac{c}{4\pi}|\vec{E}|^2|\vec{r}-\vec{r'}|^2$$

Donde $\vec{E} \equiv \vec{E}(\vec{r},t)$ corresponde a la llamada componente de radiación del campo electromagnético medido en el punto de observación a una distancia $|\vec{r}-\vec{r'}|$ de la carga acelerada, donde $\vec{r'}$ es la posición de la partícula en un tiempo t' al cual el campo electromagnético empieza a propagarse. Este tiempo de emisión o retardado está definido por: $t' = t - \frac{|\vec{r}-\vec{r'}|}{c}$ [42].

Llamando,

$$\vec{G}(t) = \sqrt{\frac{c}{4\pi}}|\vec{E}||\vec{r}-\vec{r'}|$$

Entonces,

$$\frac{dP_{rad}}{d\Omega} = |\vec{G}(t)|^2$$

$$\frac{d^2U_{rad}}{d\Omega dt} = |\vec{G}(t)|^2; U_{rad} \equiv \text{Energía radiada}$$

Tal que el observador estacionario (por ejemplo *Swift* en un instante dado) quien está detectando la radiación emitida dentro de un ángulo solido $d\Omega$, mide la energía total recibida en el $\frac{J}{sr}$ durante un intervalo de tiempo $dt$, es decir que,

$$\frac{dU_{rad}}{d\Omega} = \int_{t_0}^{t_1} |\vec{G}(t)|^2 dt$$

Debido a que el pulso de radiación es emitido en un intervalo de tiempo $t_1 - t_0 = \Delta t > 0$, entonces podemos escribir la siguiente expresión,

$$\frac{dU_{rad}}{d\Omega} = \int_{-\infty}^{t_0} |\vec{G}(t)|^2 dt + \int_{t_0}^{t_1} |\vec{G}(t)|^2 dt + \int_{t_1}^{\infty} |\vec{G}(t)|^2 dt$$

$$\frac{dU_{rad}}{d\Omega} = \int_{-\infty}^{\infty} |\vec{G}(t)|^2 dt$$

Utilizando el formalismo de la transformada de *Fourier* para expresar la energía radiada en términos de la frecuencia, para ello expresamos la función $\vec{G}(t)$, que está en el dominio del tiempo, en términos de una función $\vec{G}(w)$ (en el dominio de la frecuencia),

$$\vec{G}(t) = \frac{1}{\sqrt{2\pi}} \int_{-\infty}^{\infty} \vec{G}(w) e^{-iwt} dw$$

cuya transformada inversa es de la forma,

$$\vec{G}(w) = \frac{1}{\sqrt{2\pi}} \int_{-\infty}^{\infty} \vec{G}(t) e^{iwt} dt$$

La anterior expresión nos permite expresar $U_{rad} \equiv U_{rad}(w)$ tal que,

$$\frac{dU_{rad}}{d\Omega} = \int_{-\infty}^{\infty} |\vec{G}(w)| dw$$

Introduciendo el término $I_w$ como la intensidad de la radiación emitida a cierta frecuencia podemos expresar la energía radiada total por ángulo sólido de la siguiente manera,

$$\frac{dU_{rad}}{d\Omega} = \int_{-\infty}^{\infty} \frac{d^2 I_w}{d\Omega dw} dw$$

Pero hablar de frecuencia negativas no tendría mucho sentido observacional. Entonces,

$$\frac{U_{rad}}{d\Omega} = \int_{0}^{\infty} \frac{d^2 I_w}{d\Omega dw} dw$$

Así que en el dominio de las frecuencias se satisface,

$$\frac{dU_{rad}}{d\Omega} = \int_{-\infty}^{\infty} |\vec{G}(w)|^2 dw$$

Asumiendo que $w$ debe ser positiva entonces,

$$\int_{-\infty}^{\infty} |\vec{G}(w)|^2 dw = \int_{-\infty}^{0} |\vec{G}(w)|^2 dw + \int_{0}^{\infty} |\vec{G}(w)|^2 dw$$

$$\int_{-\infty}^{\infty} |\vec{G}(w)|^2 dw = \int_{0}^{\infty} |\vec{G}(-w)|^2 dw + \int_{0}^{\infty} |\vec{G}(w)|^2 dw$$

$$\int_{-\infty}^{\infty} |\vec{G}(w)|^2 dw = \int_{0}^{\infty} [|\vec{G}(w)|^2 + |\vec{G}(-w)|^2] dw$$

$$\frac{dU_{rad}}{d\Omega} = \int_{0}^{\infty} \frac{d^2 I_w}{d\Omega dw} dw = \int_{0}^{\infty} |\vec{G}(w)|^2 + |\vec{G}(-w)|^2 dw$$

$$\frac{d^2 I_w}{d\Omega dw} = |\vec{G}(w)|^2 + |\vec{G}(-w)|^2$$

por la definición que hicimos anteriormente de $\vec{G}(w)$, entonces $\vec{G}(-w)$ queda determinada por,

$$\vec{G}(-w) = \frac{1}{\sqrt{2\pi}} \int_{-\infty}^{\infty} \vec{G}(t) e^{-iwt} dt$$

es decir que $\vec{G}(-w) = \vec{G}^*(-w)$ el cual es el complejo conjugado de $\vec{G}(w)$, tal que,

$$|\vec{G}^*(w)|^2 = |\vec{G}(w)|^2$$

$$\frac{d^2 I_w}{d\Omega dw} = 2|\vec{G}(w)|^2$$

entonces,

$$\frac{dU_{rad}}{d\Omega} = \int_0^\infty 2|\vec{G}(w)|^2$$

Recordando que habíamos denotado $\vec{G}(t)$ como,

$$\vec{G}(t) = \sqrt{\frac{c}{4\pi}} |\vec{E}||\vec{r}-\vec{r'}|$$

donde $\vec{E} = \vec{E}(\vec{r},t)$ corresponde a la componente de radiación es, válido escribir [4]:

$$\vec{G}(t) = \sqrt{\frac{c}{4\pi}} \left( \frac{q}{c} \left[ \frac{\hat{n} \times \left[ \left(\hat{n} - \frac{\vec{v}(t')}{c}\right) \times \frac{\vec{a}(t')}{c} \right]}{\left(1 - \hat{n}\cdot\frac{\vec{v}(t')}{c}\right)^3 |\vec{r}-\vec{r'}|} \right] |\vec{r}-\vec{r'}| \right)$$

$$\vec{G}(t) = \frac{q}{\sqrt{4\pi c}} \left( \left[ \frac{\hat{n} \times \left[ \left(\hat{n} - \frac{\vec{v}(t')}{c}\right) \times \frac{\vec{a}(t')}{c} \right]}{\left(1 - \hat{n}\cdot\frac{\vec{v}(t')}{c}\right)^3} \right] \right)$$

donde $t' = t - \frac{|\vec{r}-\vec{r'}|}{c} = t - \frac{|\vec{R}|}{c}$

$$\vec{G}(w) = \frac{q}{\sqrt{8\pi^2 c}} \int_{-\infty}^{\infty} \frac{\hat{n} \times \left[ \left(\hat{n} - \frac{\vec{v}(t')}{c}\right) \times \frac{\vec{a}(t')}{c} \right]}{\left(1 - \hat{n}\cdot\frac{\vec{v}(t')}{c}\right)^3} e^{iwt} dt$$

Observando la anterior integral nos damos cuenta que tenemos variables en función de $t'$ y la integración es con respecto a $t$, así que vamos a utilizar que $t' = t - \frac{|\vec{R}|}{c}$ tal que $t = t' + \frac{|\vec{R}|}{c}$ para hallar $dt$ en función de $dt'$ y así rescribir la anterior integral en función de $t'$. Para ello asumamos en primera medida que el punto de observación de la radiación emitida está muy lejos de la región donde la carga se está acelerando.

En el siguiente gráfico se muestra la situación estudiada, donde $x$ es la distancia del origen al punto de observación $P$ y $\vec{r}(t')$ es la posición de la partícula con respecto a "O"

---

[4] Para mirar los términos completos del Campo eléctrico y Magnético retardados, además de un debido análisis de ellos recomendamos remitirse a ([42], Cap. 5) y a ([151], Cap. 14).

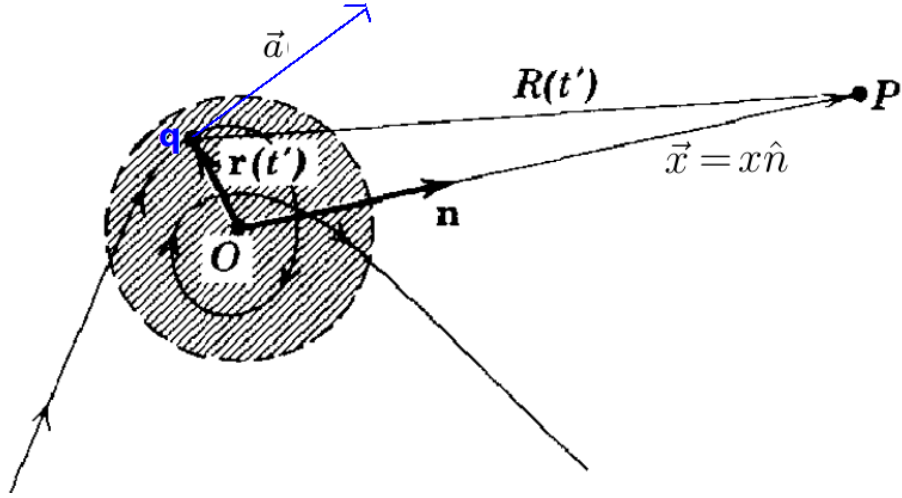

**Figura 4-11**.: Movimiento espiral del electrón medido por un observador que se encuentra muy lejos de la carga. Gráfico modificado de ([82], Fig.155).

De la figura anterior podemos inferir que,

$$\vec{r}(t) + \vec{R} = \vec{x}$$
$$\vec{R} = \vec{x} - \vec{r}(t)$$
$$\vec{R} \cdot \vec{R} = \vec{R} \cdot (\vec{x} - \vec{r}(t))$$
$$= (\vec{x} - \vec{r}(t)) \cdot (\vec{x} - \vec{r}(t))$$
$$= x^2 + |\vec{r}(t)|^2 - 2\vec{r}(t') \cdot \vec{x}$$
$$|R|^2 = x^2 + |\vec{r}(t)|^2 - 2\vec{r}(t') \cdot x\hat{n}$$
$$\Rightarrow |R(t')|^2 = x^2 \left(1 + \frac{|\vec{r}(t)|^2}{x^2} - \frac{2\vec{r}(t') \cdot x\hat{n}}{x^2}\right)$$

Como estamos muy alejados de la carga acelerada entonces $|\vec{r}(t)|^2 \ll x$ así que $\left|\frac{|\vec{r}(t)|}{x}\right| \ll 1$ y la anterior relación se aproxima a:

$$\Rightarrow |R(t')|^2 \simeq x^2 \left(1 - \frac{2\vec{r}(t') \cdot \hat{n}}{x}\right)$$

$$|R(t')| \simeq x \left(1 - \frac{2\vec{r}(t') \cdot \hat{n}}{x}\right)^{\frac{1}{2}}$$

Entonces a primer orden tenemos,

$$|R(t')| \simeq x \left(1 - \frac{1}{\cancel{2}} \frac{\cancel{2}\vec{r}(t') \cdot \hat{n}}{x}\right)$$

$$|R(t')| \simeq x \left(x - \vec{r}(t') \cdot \hat{n}\right)$$

Así podemos expresar $e^{iwt}$ como,

$$e^{iwt} = e^{iw\left(t' + \frac{|R(t')|}{c}\right)} = e^{iw\left(t' + \frac{x}{c} - \frac{\vec{r}(t')\cdot\hat{n}}{c}\right)}$$

Considerando que [42],ec. 4.45,

$$\frac{dt}{dt'} = \left(1 - \hat{n}\cdot\frac{\vec{v}(t')}{c}\right)$$

Entonces a $dt$ lo podemos escribir como: $dt = \left(1 - \hat{n}\cdot\frac{\vec{v}(t')}{c}\right)dt'$ y reemplazando los últimos resultados encontrados para $e^{iwt}$ y para $dt$ tenemos que $\vec{G}(w)$ queda escrita como,

$$\vec{G}(w) = \frac{q}{\sqrt{8\pi^2 c}} \int_{-\infty}^{\infty} \frac{\hat{n}\times\left[\left(\hat{n}-\frac{\vec{v}(t')}{c}\right)\times\frac{\vec{a}(t')}{c}\right]}{\left(1-\hat{n}\cdot\frac{\vec{v}(t')}{c}\right)^{\cancel{3}2}} e^{iw\left(t'+\frac{x}{c}-\hat{n}\cdot\frac{\vec{r}(t')}{c}\right)} \cancel{\left[1-\hat{n}\cdot\frac{\vec{v}(t')}{c}\right]} dt'$$

$$\vec{G}(w) = \frac{q}{\sqrt{8\pi^2 c}} \int_{-\infty}^{\infty} \frac{\hat{n}\times\left[\left(\hat{n}-\frac{\vec{v}(t')}{c}\right)\times\frac{\vec{a}(t')}{c}\right]}{\left(1-\hat{n}\cdot\frac{\vec{v}(t')}{c}\right)^2} e^{iw\left(t'+\frac{x}{c}-\frac{\vec{r}(t')\cdot\hat{n}}{c}\right)} dt'$$

$$\vec{G}(w) = \frac{q e^{iw\frac{x}{c}}}{\sqrt{8\pi^2 c}} \int_{-\infty}^{\infty} \frac{\hat{n}\times\left[\left(\hat{n}-\frac{\vec{v}(t')}{c}\right)\times\frac{\vec{a}(t')}{c}\right]}{\left(1-\hat{n}\cdot\frac{\vec{v}(t')}{c}\right)^2} e^{iw\left(t'-\frac{\vec{r}(t')\cdot\hat{n}}{c}\right)} dt'$$

Denotando que $\zeta = e^{iw\left(t'-\frac{\vec{r}(t')}{c}\cdot\hat{n}\right)}$ y haciendo que,

$$d\eta = \frac{\hat{n}\times\left[\left(\hat{n}-\frac{\vec{v}(t')}{c}\right)\times\frac{\vec{a}(t')}{c}\right]}{\left(1-\hat{n}\cdot\frac{\vec{v}(t')}{c}\right)^2}$$

Podemos evaluar por partes la anterior integral,

$$\int_{-\infty}^{\infty} \zeta d\eta = \zeta\eta\Big|_{-\infty}^{\infty} - \int_{-\infty}^{\infty} \eta d\zeta$$

Así, para hacerlo necesitamos hallar la respectiva representación de $\eta$ y de $d\zeta$.

$$\frac{d\zeta}{dt'} = iw\left(1 - \left(\frac{\vec{r}(t')\cdot\cancel{\frac{d\hat{n}}{dt'}}^0}{c} + \frac{\hat{n}\cdot\frac{d\vec{r}(t')}{dt'}}{c}\right)\right) e^{iw\left(t'-\frac{\vec{r}(t)\cdot\hat{n}}{c}\right)}$$

$$d\zeta = iw\left(1 - \frac{\hat{n}\cdot\vec{v}(t')}{c}\right) e^{iw\left(t'-\frac{\vec{r}(t)\cdot\hat{n}}{c}\right)} dt'$$

Por otra parte [5],

$$\frac{\hat{n} \times \left[\left(\hat{n} - \frac{\vec{v}(t')}{c}\right) \times \frac{\vec{a}(t')}{c}\right]}{\left(1 - \hat{n} \cdot \frac{\vec{v}(t')}{c}\right)^2} = \frac{d}{dt'}\left[\frac{\hat{n} \times \left(\hat{n} \times \frac{\vec{v}(t')}{c}\right)}{\left(1 - \hat{n} \cdot \frac{\vec{v}(t')}{c}\right)}\right]$$

Así que nuestro $\eta$ es de la forma,

$$\eta = \frac{\hat{n} \times \left(\hat{n} \times \frac{\vec{v}(t')}{c}\right)}{1 - \left(\hat{n} - \frac{\vec{v}(t')}{c}\right)}$$

Completando los términos la integral queda representada por,

$$\int_{-\infty}^{\infty} \eta d\zeta = e^{iw\left(t' - \frac{\vec{r}(t) \cdot \hat{n}}{c}\right)} \frac{\hat{n} \times \left(\hat{n} \times \frac{\vec{v}(t')}{c}\right)}{1 - \left(\hat{n} \cdot \frac{\vec{v}(t')}{c}\right)}\bigg|_{t' \to -\infty}^{t' \to \infty} - iw \int_{-\infty}^{\infty} \frac{\hat{n} \times \left(\hat{n} \times \frac{\vec{v}(t')}{c}\right)}{1 - \left(\hat{n} \cdot \frac{\vec{v}(t')}{c}\right)} \cancel{1 - \left(\hat{n} \cdot \frac{\vec{v}(t')}{c}\right)} e^{iw\left(t' - \frac{\vec{r}(t) \cdot \hat{n}}{c}\right)} dt'$$

El primer término cuando $t' \to -\infty$ a causa de la exponencial tiende a cero y cuando $t' \to \infty$ también desaparece pues la radiación (que estaría relacionada con la norma al cuadrado de este término) emitida por la carga acelerada está confinada a un cono (cercano a $t' = 0$). Por las anteriores razones,

$$\int_{-\infty}^{\infty} \zeta d\eta = -iw \int_{-\infty}^{\infty} \hat{n} \times \left(\hat{n} \times \frac{\vec{v}(t')}{c}\right) e^{iw\left(t' - \frac{\vec{r}(t) \cdot \hat{n}}{c}\right)} dt'$$

Recordemos que,

$$\vec{G}(w) = \frac{qe^{iw\frac{x}{c}}}{\sqrt{8\pi^2 c}} \int_{-\infty}^{\infty} \zeta d\eta$$

así que

$$\vec{G}(w) = \frac{qe^{iw\frac{x}{c}}}{\sqrt{8\pi^2 c}} \int_{-\infty}^{\infty} -iw(\hat{n} \times \left(\hat{n} \times \frac{\vec{v}(t')}{c}\right)) e^{iw\left(t' - \frac{\vec{r}(t') \cdot \hat{n}}{c}\right)} dt'$$

omo vimos anteriormente,

$$\frac{d^2 I_w}{d\Omega dw} = 2|\vec{G}(w)|^2$$

Entonces,

$$\frac{d^2 I_w}{d\Omega dw} = 2\left|\frac{-qiwe^{iw\frac{x}{c}}}{\sqrt{8\pi^2 c}}\right|^2 \left|\int_{-\infty}^{\infty} \hat{n} \times \left(\hat{n} \times \frac{\vec{v}(t')}{c}\right) e^{iw\left(t' - \frac{\vec{r}(t') \cdot \hat{n}}{c}\right)} dt'\right|^2$$

---

[5]demostrada con lujo de detalles en el apéndice C de [175]

Como $e^{iw\frac{x}{c}}$ lo puedo representar en el plano complejo como $z = cos(\frac{wx}{c}) + isen(\frac{wx}{c})$ tal que $|z|^2 = zz* = 1$, que para nuestro caso lleva a que $|e^{iw\frac{x}{c}}|^2 = 1$ y además tenemos que $|iw|^2 = w^2$. Así finalmente obtenemos la siguiente expresión:

$$\frac{d^2 I_w}{d\Omega dw} = \frac{qw^2}{4\pi^2 c} \left| \int_{-\infty}^{\infty} \hat{n} \times \left( \hat{n} \times \frac{\vec{v}(t')}{c} \right) e^{iw\left(t' - \frac{\vec{r}(t')\cdot \hat{n}}{c}\right)} dt' \right|^2$$

## 4.5. Radiación emitida por un electrón acelerado que se mueve en un arco circular (distribución espectral)

Considerando el movimiento de un electrón relativista en un arco de círculo de radio $r$ (ver figura **4-12**), vamos a encontrar el espectro de frecuencias de la radiación emitida por el electrón.

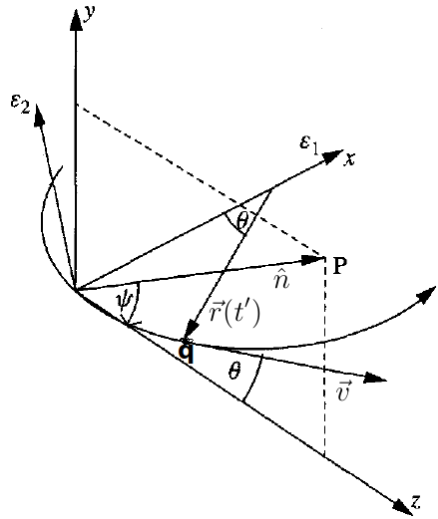

**Figura 4-12**.: Descripción vectorial del movimiento del electrón relativista en una sección de su trayectoria. Gráfica adaptada de (fig.5.10,[42])

En la anterior figura el observador se encuentra localizado en el punto $P$ (en el plano y-z). La línea que conecta $P$ y el origen de coordenadas, definida por el vector unitario $\hat{n}$, forma un ángulo $\psi$ con el eje $z$.

En el tiempo $t'$ el radio vector $\vec{r}$ hace un ángulo $\theta$ con el eje $x$, que es igual a $w_0 t'$, tomando a un $w_0$ cono la frecuencia angular del electrón en su órbita circular, es decir que,

$$w_0 = \frac{|\vec{v}|}{|\vec{r}|} \equiv \frac{v}{r}$$

.

Teniendo en cuenta nuestra anterior figura (**4-12**), podemos escribir $\hat{n}$, $\vec{r}(t')$ y $\hat{v}$ en términos de los ángulos mostrados en el sistema cartesiano $x,y,z$.

$$\hat{n} = (0, \sin\psi, \cos\psi) \qquad (4\text{-}15)$$
$$\vec{r}(t') = |\vec{r}|(\cos(w_0 t'), 0, \text{sen}(w_0 t')) \qquad (4\text{-}16)$$
$$\vec{v}(t') = |\vec{v}|(\text{sen}(w_0 t'), 0, \cos(w_0 t'))^6 \qquad (4\text{-}17)$$

tanto $\vec{r}(t')$ y $\vec{v}(t')$ están en el plano x-z, es decir que el intervalo de trayectoria está en el plano x-y con un radio de curvatura $|\vec{r}|$. En el esquema anterior también hemos definido dos vectores polarización ($\varepsilon_1$ y $\varepsilon_2$), que indican dos direcciones posibles del campo eléctrico. Estas direcciones deben formar ángulos rectos con la dirección de observación.

Llamando a $\hat{\varepsilon}_1$ como vector polarización horizontal (i.e. $\varepsilon_\parallel$) y a $\hat{\varepsilon}_2$ como la dirección vertical de polarización ($\varepsilon_\perp$) entonces observando el anterior gráfico es válido afirmar que,

$\hat{\varepsilon}_2 = \hat{n} \times \hat{\varepsilon}_1$

Ahora expresaremos $\vec{v}$ en términos de los vectores unitarios $\hat{\varepsilon}_1, \hat{\varepsilon}_2, \hat{n}$, para ello debemos dibujar a $\vec{v}$ en un sistema cartesiano dominado por $\hat{\varepsilon}_1, \hat{\varepsilon}_2, \hat{n}$.

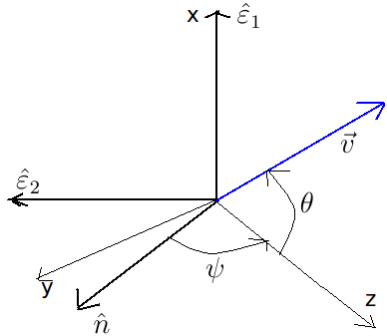

**Figura 4-13**.: Esquema pictórico de la relación vectorial entre la velocidad de la partícula y los vectores de polarización

El orden de los ejes se escogió de tal forma que tuviera en cuenta la figura (**4-12**) y que preservara que $\hat{\varepsilon}_2 = \hat{n} \times \hat{\varepsilon}_1$.

Así nuestro vector $\vec{v}$ queda representado por,

$\vec{v} = |\vec{v}|\left[\cos\psi\cos\theta\hat{n} - \text{sen}\,\psi\cos\theta\hat{\varepsilon}_2 + \text{sen}\,\theta\hat{\varepsilon}_1\right]$

Reemplazando $\theta$ por $w_0 t'$ tenemos que,

$\vec{v} = |\vec{v}|\left[\cos w_0 t'(\cos\psi\hat{n} - \text{sen}\,\psi\hat{\varepsilon}_2) + \text{sen}\,w_0 t'\hat{\varepsilon}_1\right]$

Para calcular el valor de $\frac{d^2 I_w}{d\Omega dw}$ debemos calcular una integral de una función que presenta el producto triple $\hat{n} \times \left(\hat{n} \times \frac{\vec{V}}{c}\right)$ y al producto punto $\hat{n} \cdot \vec{r}(t')$, así que para poder hallar la integral debemos encontrar estos productos para describir la potencia radiada como función de la frecuencia y ángulo sólido del electrón moviéndose en un arco circular.

En primera medida calculamos $\hat{n} \times \vec{v}$,

$$\hat{n} \times \vec{v} = |\vec{v}| \left[\cos w_0 t' \left[\cos\psi \underbrace{(\hat{n} \times \hat{n})}_{0} - \text{sen}\,\psi(\hat{n} \times \hat{\varepsilon}_2)\right] + \text{sen}\,w_0 t' (\hat{n} \times \hat{\varepsilon}_1)\right]$$

Como $\hat{\varepsilon}_2 = \hat{n} \times \hat{\varepsilon}_1 \Rightarrow \hat{n} \times \hat{\varepsilon}_2 = -\hat{\varepsilon}_1$

$\hat{n} \times \hat{v}) = |\vec{v}|(\text{sen}\,w_0 t' \varepsilon_2 + \cos w_0 t' \,\text{sen}\,\psi \varepsilon_1)$

$\hat{n} \times (\hat{n} \times \hat{V}) = |\vec{v}|(-\text{sen}\,w_0 t' \varepsilon_1 + \cos w_0 t' \,\text{sen}\,\psi \varepsilon_2)$

Por otra parte utilizamos las definiciones de $\hat{n}$ y de $\vec{r}$ en el sistema cartesiano $x, y, z$

$\hat{n} = \text{sen}\,\psi \hat{j} + \cos\psi \hat{k}$

$\vec{r} = |\vec{r}|\cos w_0 t' \hat{i} + |\vec{r}|\text{sen}\,w_0 t' \hat{k}$

$\hat{n} \cdot \vec{r} = |\vec{r}|\cos\psi \,\text{sen}\,w_0 t'$

Debido a que estamos haciendo nuestro análisis para $t'$ cercanos a cero [175], esto nos provoca tanto que $\theta \equiv w_0 t'$ como $\psi$ (asociado con el semi-ángulo del cono de radiación) sean muy pequeños. Esto nos motiva a expandir hasta segundo orden las funciones trigonométricas $\cos\psi$ y $\text{sen}\,w_0 t'$,

$$\cos\psi \approx 1 - \frac{\psi^2}{2}$$

$$\text{sen}\,w_0 t' \approx w_0 t' - \frac{w_0^3 t'^3}{6}$$

Finalmente nuestra expresión para $t' - \frac{\vec{r}(t') \cdot \hat{n}}{c}$ queda expresada como,

$$t' - \frac{r}{c}\cos\psi \,\text{sen}\,w_0 t'; \text{ donde}|\vec{r}| \equiv r$$

y utilizando las anteriores expansiones de la funciones trigonométricas tenemos que se satisface,

$$t' - \frac{r}{c}\left(1 - \frac{\psi^2}{2}\right)\left(w_0 t' - \frac{w_0^3 t'^3}{6}\right)$$

Reemplazando $w_0$ por $\frac{v}{r}$ (condición valida en un arco de círculo, donde $w_0$ es la velocidad angular a $t = 0$)

$$t' - \frac{r}{c}\left(1 - \frac{\psi^2}{2}\right)\left(\frac{v}{r}t' - \frac{\frac{v^3}{r^3}t'^3}{6}\right) =$$

$$t' - \frac{v}{c}\left(1 - \frac{\psi^2}{2}\right)t' + \left(1 - \frac{\psi^2}{2}\right)\frac{v^3 t'^3}{6cr^2} =$$

$$t' - \frac{v}{c}t' + \frac{v}{c}\frac{\psi^2}{2}t' + \frac{v^3 t'^3}{6cr^2}\frac{c^3}{c^3} - \frac{\psi^2}{2}\frac{v^3 t'^3}{6cr^2} =$$

$$t'\left(1 - \frac{v}{c}\right) + \frac{v}{c}\frac{\psi^2}{2}t' + \frac{t'^3 c^2}{6r^2}\left(\frac{v^3}{c^3}\right)\left[1 - \frac{\psi^2}{2}\right] =$$

Llamando $\frac{v}{c} = \beta$, en donde $\gamma = \frac{1}{\sqrt{1-\beta^2}}$ tenemos que si la partícula presenta una velocidad cercana a la de la luz, es decir $v \approx c; (\gamma \gg 1) \Rightarrow \beta \to 1$

Sí $\gamma = \frac{1}{\sqrt{1-\beta^2}} \Rightarrow 1 - \beta^2 = \frac{1}{\gamma^2}$ es decir que $(1 - \beta)(1 + \beta) = \frac{1}{\gamma^2}$

Con $\beta \to 1 \Rightarrow (1 - \beta) \to \frac{1}{2\gamma^2}$, o lo mismo que decir que: $(1 - \frac{v}{c}) \approx \frac{1}{2\gamma^2}$

Entonces podemos escribir que para el caso del electrón ultra-relativista, $(\beta \to 1)$.

$$t' - \frac{r}{c}\cos\psi\,\text{sen}\,(w_0 t') \approx \frac{t'}{2\gamma^2}\left(1 + \gamma^2\psi^2\right) + \frac{t'^3 c^2}{6r^2}\left(1 - \frac{\psi^2}{2}\right)$$

Como $\psi$ es un ángulo pequeño entonces podemos aproximar $\left(1 - \frac{\psi^2}{2}\right) \approx 1$.[7] Entonces para este caso particular de estudio tenemos que,

$$t' - \frac{\hat{n}\cdot\hat{r}(t')}{c} \approx \frac{t'}{2\gamma^2}(1 + \gamma^2\psi^2) + \frac{t'^3 c^2}{6r^2}$$

Reemplazando la expresiones encontradas de: $t' - \frac{\hat{n}\cdot\hat{r}(t')}{c}$ y de $\hat{n} \times \left(\hat{n} \times \frac{\vec{v}}{c}\right)$ en la relación hallada anteriormente para $\vec{G}(w)$ tenemos ahora que,

$$\vec{G}(w) = \frac{qe^{iw\frac{x}{c}}}{\sqrt{8\pi^2 c}}\int_{-\infty}^{\infty} -iw\left[\hat{n} \times \left(\hat{n} \times \frac{\vec{v}(t')}{c}\right)e^{iw\left(t' - \frac{\vec{r}(t')\cdot\hat{n}}{c}\right)}\right]dt'$$

$$\vec{G}(w) = \frac{qe^{iw\frac{x}{c}}}{\sqrt{8\pi^2 c}}\int_{-\infty}^{\infty} -iw\left[\frac{v}{c}\left(-\text{sen}\,(w_0 t')\hat{\varepsilon}_1 + \cos(w_0 t')\,\text{sen}\,\psi\hat{\varepsilon}_2\right)\right]e^{iw\left(\frac{t'}{2\gamma^2}(1+\gamma^2\psi^2) + \frac{t'^3 c^2}{6r^2}\right)}dt'$$

Es decir que podemos escribir,

$$\vec{G}(w) = \vec{G}_1(w) + \vec{G}_2(w)$$
$$\vec{G}(w) = G_1(w)\hat{\varepsilon}_1 + G_2(w)\hat{\varepsilon}_2$$
$$2|\vec{G}(w)|^2 = 2(G_1^2(w) + G_2^2(w))$$

donde,

$$G_1(w) = iw\frac{q}{\sqrt{8\pi^2 c}}\int_{-\infty}^{\infty}\left[\frac{v}{c}\text{sen}(w_0 t')\right]e^{iw\left(\frac{t'}{2\gamma^2}(1+\gamma^2\psi^2) + \frac{t'^3 c^2}{6r^2}\right)}dt'$$

---
[7] $\psi$ dado en radianes

$$G_2(w) = -iw\frac{q}{\sqrt{8\pi^2 c}} \int_{-\infty}^{\infty} \left[\frac{v}{c}\cos(w_0 t')\sin\psi\right] e^{iw\left(\frac{t'}{2\gamma^2}(1+\gamma^2\psi^2)+\frac{t'^3 c^2}{6r^2}\right)} dt'$$

Asumiendo que el movimiento del electrón es ultra-relativista ($\gamma \gg 1$, por ejemplo $\gamma = [10^2 - 10^3]$) tenemos que $\beta = \frac{v}{c} \approx 1$ y como estamos haciendo el análisis en una vecindad temporal alrededor de $t' = 0$ entonces es válido afirmar que, (a primer orden)

$\sin(w_0 t') \approx w_0 t'$

$\cos(w_0 t') \approx 1$

$\sin\varepsilon \approx \varepsilon$

Transformando nuestras expresiones halladas anteriormente para $G_1(w)$ y $G_2(w)$ a las siguientes,

$$G_1(w) = iw\frac{qe^{w\frac{x}{c}}}{\sqrt{8\pi^2 c}} \int_{-\infty}^{\infty} w_0 t' e^{iw\left(\frac{t'}{2\gamma^2}(1+\gamma^2\psi^2)+\frac{t'^3 c^2}{6r^2}\right)} dt'$$

$$G_2(w) = -iw\frac{qe^{w\frac{x}{c}}}{\sqrt{8\pi^2 c}} \int_{-\infty}^{\infty} \psi e^{iw\left(\frac{t'}{2\gamma^2}(1+\gamma^2\psi^2)+\frac{t'^3 c^2}{6r^2}\right)} dt'$$

Observando la expresión obtenida para $G_2$ nos damos cuenta que ella desaparece cuando $\psi = 0$, es decir cuando el observador está en el plano de la trayectoria del electrón ver (**4-12**). Además no tendríamos componente vertical de la radiación ($\hat{\varepsilon}_2 \equiv \hat{\varepsilon}_\perp$) y el observador solo detectaría radiación polarizada horizontalmente correspondiente a la componente horizontal ($\hat{\varepsilon}_1 \equiv \hat{\varepsilon}_\parallel$),

El siguiente paso es resolver las integrales involucradas en $G_1(w)$ y $G_2(w)$. Para ello hacemos el siguiente cambio de variable,([42], pp. 95),

$$x = \frac{\gamma w_0}{(1+\gamma^2\psi^2)^{\frac{1}{2}}} t' \text{ y } \xi = \frac{1}{2}\frac{w}{w_c}(1+\gamma^2\psi^2)^{\frac{3}{2}}$$

donde utilizamos que $\boxed{w_c = \frac{3}{2}\gamma^3 w_0}$. Es decir que en realidad $\xi$ es de la forma,

$$\xi = \frac{1}{3}\frac{w}{\gamma^3 w_0}(1+\gamma^2\psi^2)^{\frac{3}{2}}$$

Cuando llevamos a cabo la cabo este cambio de variable obtenemos una nueva escritura de nuestras funciones $G_1(w)$ y $G_2(w)$, tal que.

$$G_1(w) = \frac{iwqe^{iw\frac{x}{c}}}{\sqrt{8\pi^2 c}} \int_{-\infty}^{\infty} \frac{x}{\gamma}(1+\gamma^2\psi^2)^{\frac{1}{2}} e^{iw\left(\frac{x(1+\gamma^2\psi^2)^{\frac{3}{2}}}{2\gamma^3 w_0} + \frac{x^3(1+\gamma^2\psi^2)^{\frac{3}{2}} c^2}{\gamma^3 w_0^3 6 r^2}\right)} \frac{(1+\gamma^2\psi^2)^{\frac{1}{2}}}{\gamma w_0} dx$$

pues si $x = \frac{\gamma w_0}{(1+\gamma^2\psi^2)^{\frac{1}{2}}} t' \Rightarrow dx = \frac{\gamma w_0}{(1+\gamma^2\psi^2)^{\frac{1}{2}}} dt'$; así $dt' = \frac{(1+\gamma^2\psi^2)^{\frac{1}{2}}}{\gamma w_0} dt'$

Juntando términos obtenemos que,

$$G_1(w) = \frac{iwqe^{iw\frac{x}{c}}}{\sqrt{8\pi^2 c}} \int_{-\infty}^{\infty} \frac{x(1+\gamma^2\psi^2)^{\frac{1}{2}}}{\gamma^2 w_0} e^{iw\left((1+\gamma^2\psi^2)^{\frac{3}{2}}\left[\frac{x}{2\gamma^3 w_0} + \frac{x^3 c^2}{\gamma^3 w_0^3 6 r^2}\right]\right)} dx$$

Analizando el térmíno que se encuentra dentro de la exponencial tenemos que asumiendo $v = rw_0 \Rightarrow v^2 = r^2 w_0^2$ y entonces $\frac{c^2}{v^2} = \frac{1}{\beta^2}$ si $(\gamma \gg 1)$ entonces $\beta \approx 1$ y entonces $\frac{1}{\beta^2} \approx 1$ así que, (caso ultra-relativista)

$$\frac{(1+\gamma^2\psi^2)^{\frac{3}{2}}}{\gamma^3 w_0}\left[\frac{x}{2} + \frac{x^3 c^2}{6 w_0^2 r^2}\right] \approx \frac{(1+\gamma^2\psi^2)^{\frac{3}{2}}}{\gamma^3 w_0}\left[\frac{x}{2} + \frac{x^3}{6}\right]$$

Como $\xi = \frac{1}{3}\frac{w}{\gamma^3 w_0}(1+\gamma^2\psi^2)^{\frac{3}{2}} \Rightarrow \frac{w(1+\gamma^2\psi^2)^{\frac{3}{2}}}{\gamma^3 w_0} = 3\xi$, así,

$$\frac{(1+\gamma^2\psi^2)^{\frac{3}{2}}}{\gamma^3 w_0}\left[\frac{x}{2} + \frac{x^3}{6}\right] \approx 3\xi\left[\frac{x}{2} + \frac{x^3}{6}\right]$$

Reemplazando en $G_1(w)$ obtenemos,

$$G_1(w) = \frac{iwqe^{iw\frac{x}{c}}(1+\gamma^2\psi^2)}{\gamma^2 w_0 \sqrt{8\pi^2 c}} \int_{-\infty}^{\infty} xe^{i\frac{3}{2}\xi\left[x+\frac{x^3}{3}\right]} dx$$

Por su parte haciendo un análisis similar y utilizando el mismo cambio de variable obtenemos que $G_2(w)$ es de la forma,[42],

$$G_2(w) = \frac{-iwqe^{iw\frac{x}{c}}\psi(1+\gamma^2\psi^2)^{\frac{1}{2}}}{\gamma w_0 \sqrt{8\pi^2 c}} \int_{-\infty}^{\infty} e^{i\frac{3}{2}\xi\left[x+\frac{x^3}{3}\right]} dx$$

Así nos podemos dar cuenta que nuestro asunto sera resolver las integrales involucradas. Para $G_1(w)$ tenemos que resolver,

$$\int_{-\infty}^{\infty} xe^{i\frac{3}{2}\xi\left[x+\frac{x^3}{3}\right]} = I_{p1} \tag{4-18}$$

y para $G_1(w)$ tenemos que resolver.

$$\int_{-\infty}^{\infty} e^{i\frac{3}{2}\xi\left[x+\frac{x^3}{3}\right]} = I_{p2} \tag{4-19}$$

Al representar $e^{i\frac{3}{2}\xi\left(x+\frac{x^3}{3}\right)}$ en su parte real e imaginaria,

$$I_{p1} = \int_{-\infty}^{\infty} x\left[\cos\left(\frac{3}{2}\xi\left(x+\frac{1}{3}x^3\right)\right) + i\,\text{sen}\left(\frac{3}{2}\xi\left(x+\frac{1}{3}x^3\right)\right)\right]$$

$$I_{p1} = \int_{-\infty}^{\infty} x \cos\left(\frac{3}{2}\xi\left(x + \frac{x^3}{3}\right)\right)dx + i\int_{-\infty}^{\infty} x \sen\left(\frac{3}{2}\xi\left(x + \frac{x^3}{3}\right)\right)dx$$

Por su parte,

$$I_{p1} = \int_{-\infty}^{\infty} \cos\left(\frac{3}{2}\xi\left(x + \frac{x^3}{3}\right)\right)dx + i\int_{-\infty}^{\infty} \sen\left(\frac{3}{2}\xi\left(x + \frac{x^3}{3}\right)\right)dx$$

Teniendo en cuenta que la integral de una función impar entre $-a$ y $a$ (intervalo simétrico) es cero y la integral de una función par entre $-a$ y $a$ es el doble de la integral de la función desde 0 hasta $a$ es válido afirmar que (sin perdida de generalidad),

$$\int_{-\infty}^{\infty} x \cos\left(\frac{3}{2}\xi\left(x + \frac{x^3}{3}\right)\right)dx = 0$$

Pues el producto de una función impar $(x)$ por una función par $\cos\left(\frac{3}{2}\xi\left(x + \frac{x^3}{3}\right)\right)$ es impar.

$$\int_{-\infty}^{\infty} x \sen\left(\frac{3}{2}\xi\left(x + \frac{x^3}{3}\right)\right)dx = 2\int_{0}^{\infty} x \sen\left(\frac{3}{2}\xi\left(x + \frac{x^3}{3}\right)\right)dx$$

Pues el producto de una función impar $(x)$ por una función impar $\cos\left(\frac{3}{2}\xi\left(x + \frac{x^3}{3}\right)\right)$ es una función par[8].

Por argumentos similares entonces

$$\int_{-\infty}^{\infty} \cos\left(\frac{3}{2}\xi\left(x + \frac{x^3}{3}\right)\right)dx = 2\int_{0}^{\infty} \cos\left(\frac{3}{2}\xi\left(x + \frac{x^3}{3}\right)\right)dx$$

Por su parte

$$\int_{-\infty}^{\infty} \sen\left(\frac{3}{2}\xi\left(x + \frac{x^3}{3}\right)\right)dx = 0$$

Reemplazando los respectivos valores de loas integrales en $I_{p1}$ y $I_{p2}$, para posteriormente realizar un reemplazo en $G_1(w)$ y $G_2(w)$ hallamos que,

$$G_1(w) = \frac{iwqe^{iw\frac{x}{c}}(1+\gamma^2\psi^2)}{\gamma^2 w_0 \sqrt{8\pi^2 c}}\left[2i\int_{0}^{\infty} x \sen\left(\frac{3}{2}\xi\left(x + \frac{x^3}{3}\right)\right)dx\right]$$

$$G_1(w) = -\frac{qe^{iw\frac{x}{c}}}{\sqrt{8\pi^2 c}}\frac{w(1+\gamma^2\psi^2)}{\gamma^2 w_0}\left[2\int_{0}^{\infty} x \sen\left(\frac{3}{2}\xi\left(x + \frac{x^3}{3}\right)\right)dx\right]$$

Por su parte,

$$G_2(w) = -\frac{iqe^{iw\frac{x}{c}}\psi}{\sqrt{8\pi^2 c}}\frac{w(1+\gamma^2\psi^2)^{\frac{1}{2}}}{\gamma w_0}\left[2\int_{0}^{\infty} x \cos\left(\frac{3}{2}\xi\left(x + \frac{x^3}{3}\right)\right)dx\right]$$

---

[8]Nota: las funciones son bien comportadas

Las integrales mostradas en $G_1(w)$ y $G_2(w)$ pueden ser escritas en términos de las funciones de Bessel modificada de segunda clase $K_{\frac{2}{3}}$ y $K_{\frac{1}{3}}$ (([81], Pg. 1918, ecs. II.10 y II.12),([94], pg. 277)).

$$\int_0^\infty \cos\left(\frac{3}{2}\xi\left(x+\frac{x^3}{3}\right)\right)dx = \frac{1}{\sqrt{3}}K_{\frac{1}{3}}(\xi)$$

$$\int_0^\infty x\,\text{sen}\left(\frac{3}{2}\xi\left(x+\frac{x^3}{3}\right)\right)dx = \frac{1}{\sqrt{3}}K_{\frac{2}{3}}(\xi)$$

Recordemos que estas funciones $K_n(x)$ (en este caso $n=\frac{1}{3}$ y $n=\frac{2}{3}$) son soluciones de la ecuación modificada de Bessel ([42] cap. 6, Pg. 101).

$$x^2\frac{d^2y}{dx^2}+x\frac{dy}{dx}-(x^2+n^2)y=0$$

Las primeras soluciones que se encuentran a esta ecuación usando el método de "Frobenios" son conocidas como las funciones modificadas de Bessel de primera clase, denotadas como

$$I_{+n}(x)=\sum_{k=0}^\infty \frac{1}{k!\Gamma(k+n+1)}\left(\frac{x}{2}\right)^{2k+n}$$

$$I_{-n}(x)=\sum_{k=0}^\infty \frac{1}{k!\Gamma(k-n+1)}\left(\frac{x}{2}\right)^{2k-n}$$

donde estas funciones $I_n(x)$ están relacionadas con las funciones de Bessel $J_n(x)$ como

$$I_n(x)=i^{-n}J(ix)$$

Así nuestras funciones $K_n(x)$ están relacionadas con estas funciones $I_n(x)$ de la siguiente forma;

$$K_n(x)=\frac{1}{2}\frac{\pi}{\text{sen}(n\pi)}(I_{-n}(x)-I_{+n}(x))$$

Finalmente haciendo los debidos reemplazos obtenemos una nueva re-escritura de las funciones $G_1(w)$ y $G_2(w)$,

$$G_1(w)=-\frac{qe^{iw\frac{x}{c}}}{\sqrt{8\pi^2c}}\frac{w(1+\gamma^2\psi^2)}{\gamma^2 w_0}\frac{2}{\sqrt{3}}K_{\frac{2}{3}}(\xi)$$

$$G_2(w)=-\frac{iqe^{iw\frac{x}{c}}\psi}{\sqrt{8\pi^2c}}\frac{w(1+\gamma^2\psi^2)^{\frac{1}{2}}}{\gamma w_0}\frac{2}{\sqrt{3}}K_{\frac{1}{3}}(\xi)$$

Recordando que,

$$\frac{d^2I_w}{d\Omega dw}=2|\vec{G}(w)|^2=2\left(G_1^2(w)+G_2^2(w)\right)$$

Puesto que $\vec{G}(w) = G_1(w)\hat{\varepsilon}_1 + G_2(w)\hat{\varepsilon}_2$, además anteriormente habíamos denotado a $\hat{\varepsilon}_1$ como polarización horizontal y a $\hat{\varepsilon}_2$ como polarización vertical. Entonces también podemos notar,([175] Pg. 86)

$$\frac{d^2 I_w}{d\Omega dw} = \frac{d^2 I_\parallel}{d\Omega dw} + \frac{d^2 I_\perp}{d\Omega dw}$$

Al reemplazar las ultimas funciones encontradas para $G_1(w)$ y $G_2(w)$ en la potencia radiada por ángulo solido y por frecuencia tenemos que,

$$\frac{d^2 I_w}{d\Omega dw} = \frac{2q^2 w^2}{8\pi^8 c}\frac{(1+\gamma^2\psi^2)^2}{\gamma^4 w_0^2}\frac{4}{3}K_{\frac{2}{3}}^2(\xi) + 2\frac{q^2\psi^2}{8\pi^2 c}\frac{w^2(1+\gamma^2\psi^2)}{\gamma^2 w_0^2}\frac{4}{3}K_{\frac{1}{2}}^2(\xi)$$

Entonces obtenemos,

$$\frac{d^2 I_w}{d\Omega dw} = \frac{1}{3}q^2\frac{w^2(1+\gamma^2\psi^2)^2}{\pi^2\gamma^4 w_0^2}\left[K_{\frac{2}{3}}^2(\xi) + \frac{\gamma^2\psi^2}{(1+\gamma^2\psi^2)}K_{\frac{1}{3}}^2(\xi)\right]$$

finalmente,

$$\frac{d^2 I_w}{d\Omega dw} = \frac{3}{4}q^2\frac{(1+\gamma^2\psi^2)^2}{\pi^2}\left(\frac{w}{w_c}\right)^2\left[K_{\frac{2}{3}}^2(\xi) + \frac{\gamma^2\psi^2}{(1+\gamma^2\psi^2)}K_{\frac{1}{3}}^2(\xi)\right] \qquad (4\text{-}20)$$

Con $\xi = \frac{1}{2}\frac{w}{w_c}(1+\gamma^2\psi^2)^{\frac{3}{2}}$ donde hemos usado que $w_c = \frac{3}{2}\gamma^3 w_0$, tal que $w_c^2 = \frac{9}{4}\gamma^6 w_0^2$. Esta última ecuación muestra el espectro de radiación en frecuencia proveniente de un electrón que sigue una trayectoria en un arco de círculo, la cual es medida en $\frac{J}{Sr}$ en un intervalo de frecuencia. Esta energía radiada esta localizada a un ángulo $\varepsilon$ sobre el plano de la órbita del electrón(ver angulo que forma vector $\vec{n}$ con el plano orbital del electrón, en dirección hacia punto de observación $P$).

A partir de integrar sobre el ángulo sólido la ecuación (4-20) podemos hallar la energía radiada por rango de frecuencia $(w, w+dw)$ por el electrón por órbita completa en el plano normal proyectado (por eso la escogencia de los vectores $\vec{r}$ y $\vec{v}$ en la figura **4-12**). Es decir que es válido afirmar,

$$\frac{dI}{dw} = \int \frac{d^2 I}{d\Omega dw} d\Omega$$

Durante cada órbita la radiación emitida es mayormente confinada a un ángulo solido como se muestra en la figura (**4-14**),

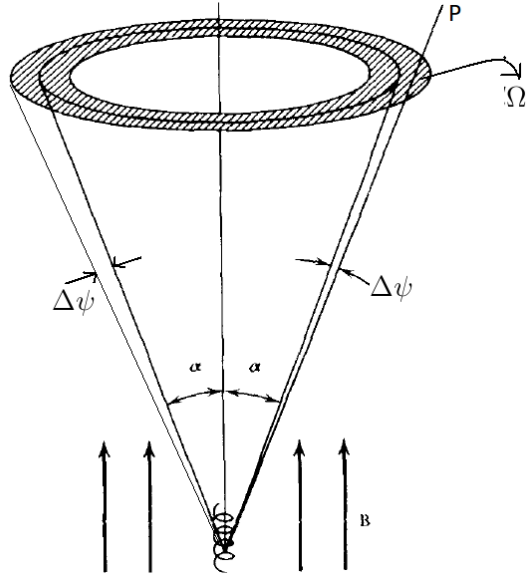

**Figura 4-14**.: Area efectiva donde podemos detectar la radiación de sincrotrón. Adaptado de (fig. 6.5 [142])

Entonces observando el anterior gráfico podemos deducir que,

$$d\Omega = 2\pi \operatorname{sen}\alpha \, d\psi$$

donde en este caso (figura **4-14**) $\Delta\psi \approx \frac{1}{\gamma}$, pues sabemos que la radiación se emitió en un delgado cono de semiángulo $\frac{1}{\gamma}$ (como lo vimos anteriormente en el comienzo del capítulo) comprendido entre la dirección del vector velocidad y la dirección del punto de observación a cada instante de tiempo.
Teniendo en cuenta lo anterior podemos escribir,

$$\frac{dI}{dw} = \int \frac{d^2I}{d\Omega dw} d\Omega = 2\pi \operatorname{sen}\alpha \int_{-\infty}^{\infty} \frac{d^2I}{d\Omega dw} d\psi \qquad (4\text{-}21)$$

donde $\alpha$ es el ángulo *pitch* comprendido por la dirección del campo magnético y la velocidad de la partícula.

Más adelante vamos a utilizar el caso extremo donde $\alpha = \frac{\pi}{2}$ así que $\operatorname{sen}\alpha = 1$. De otra parte los límites hacia el infinito en el integrando de la relación anterior se justifican debido a que $\Delta\psi \approx \frac{1}{\gamma}$.
Reemplazando (4-20) en (4-21) obtenemos,

$$\frac{dI}{dw} = \frac{3}{4}\frac{q^2}{\pi^2}\left(\frac{w}{w_c}\right)^2 2\pi \operatorname{sen}\alpha \int_{-\infty}^{\infty} \left[(1+\gamma^2\psi^2)^2 K_{\frac{2}{3}}^2(\xi) + \gamma^2\psi^2(1+\gamma^2\psi^2)K_{\frac{1}{3}}^2(\xi)\right] d\psi$$

Después de hacer un proceso de fino cálculo, que incluye utilizar las funciones de Airy, ver detalles específicos en ([42], Pg. 115-122, Cap. 7). Por su parte en [168] se hace un tratamiento alternativo que lleva al mismo resultado, en definitiva,

$$\frac{dI}{dw} = \frac{\sqrt{3}}{2}\gamma\left(\frac{q^2}{c}\right)\left(\frac{w}{w_c}\right)\operatorname{sen}\alpha\left[\left(K_{\frac{2}{3}}\left(\frac{w}{w_c}\right) + \int_{\frac{w}{c}}^{\infty} K_{\frac{5}{3}}(y)dy\right) + \left(-K_{\frac{2}{3}}\left(\frac{w}{w_c}\right) + \int_{\frac{w}{c}}^{\infty} K_{\frac{5}{3}}(y)dy\right)\right]$$

$$F(x) = x\int_x^{\infty} K_{\frac{5}{3}}(y)dy$$

$$G(x) = xK_{\frac{2}{3}}, \text{ con } x \equiv \frac{w}{w_c}$$

Tenemos que la energía radiada por el intervalo de frecuencia se reescribe como,

$$\frac{dI}{dw} = \frac{\sqrt{3}}{2}\gamma\left(\frac{q^2}{c}\right)\operatorname{sen}\alpha\left[(G(x) + F(x)) + (-G(x) + F(x))\right]$$

Como

$$\frac{dI}{dw} = \frac{dI_\perp}{dw} + \frac{dI_\parallel}{dw}$$

Entonces comparando término a término, podemos decir que,

$$\frac{dI_\perp}{dw} = \operatorname{sen}\alpha\frac{\sqrt{3}}{2}\gamma\left(\frac{q^2}{c}\right)[F(x) + G(x)]$$

$$\frac{dI_\parallel}{dw} = \operatorname{sen}\alpha\frac{\sqrt{3}}{2}\gamma\left(\frac{q^2}{c}\right)[F(x) - G(x)]$$

En conclusión es válido afirmar que,

$$\frac{dI}{dw} = \frac{\sqrt{3}}{2}\gamma\left(\frac{q^2}{c}\right)\operatorname{sen}\alpha F(x)$$

Finalmente para hallar la potencia radiada por unidad de frecuencia, dividimos la anterior ecuación por el periodo orbital de la carga, es decir por $T = \frac{2\pi}{w_B}$ con $w_B = \frac{qB}{\gamma mc}$,

$$\frac{dP(w)}{dw} = \frac{q^3 B \operatorname{sen}\alpha\sqrt{3}}{2\pi mc^2}F(x)$$

Si queremos expresar el anterior resultado en términos de la frecuencia de los fotones emitidos ($\nu$) tenemos que ($w = 2\pi\nu$), $\frac{dP(\nu)}{d\nu} = \frac{q^3 B \operatorname{sen}\alpha\sqrt{3}}{mc^2}F(\frac{\nu}{\nu_c})$ donde $\nu_c = \frac{w_c}{2\pi} = \frac{1}{2\pi}\left(\frac{3}{2}\gamma^2 w_B\right) = \frac{3\gamma^2 B}{4\pi mc}$

La función $F(x)$ ha sido graficada para tener un mayor entendimiento cualitativo del resultado encontrado,

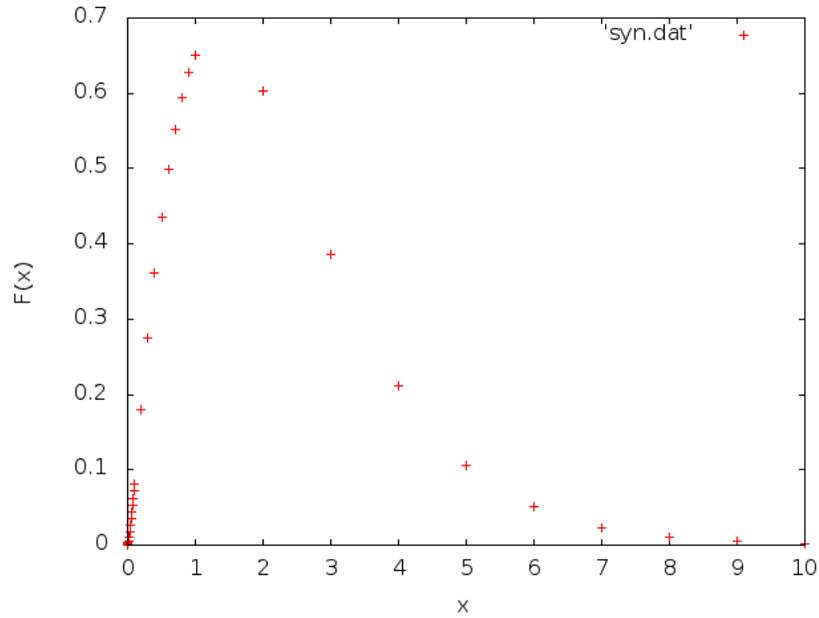

**Figura 4-15**.: Representación gráfica de la función $F(x) = x \int_x^\infty K_{\frac{5}{3}}(y)dy$. Esta gráfica se genero a partir de los datos de la tabla 7.1, [42].

Otros autores han analizado el comportamiento de esta función $F(x)$ y han propuesto las siguientes aproximaciones,([142] Pg. 179)

$$F(x) \approx \frac{4\pi}{\sqrt{3}\Gamma(\frac{1}{3})}\left(\frac{x}{2}\right)^{\frac{1}{3}}, x \ll 1$$

$$F(x) \approx \left(\frac{\pi}{2}\right)^{\frac{1}{2}} e^{-x} x^{\frac{1}{2}}, x \gg 1$$

$$F(x) = cx^{\frac{1}{3}}e^{-x}$$

Con $c \simeq 1,85$, en donde $0,1 \leq x \leq 10$ [56].

# 5. Modelo teórico de la evolución temporal del Afterglow de un Gamma-Ray Burst

En este capítulo se muestra el estudio teórico que se hizo de la evolución temporal del *afterglow* en rayos x de un GRB detectado por la sonda espacial *Swift Gamma-Ray Burst Mission*, específicamente el GRB 050525 que ocurrió el 25 de Mayo del 2005 [23]. Para ello se estudio el problema desde dos puntos de vista complementarios: hidrodinámico y radiativo, los cuales componen el llamado modelo del *fireball*.

En el aspecto hidrodinámico se estudió la variación temporal del factor gamma de *Lorentz* de un fluido adiabático ultra-relativista mediante el estudio de la onda de choque relativista producida por la interacción de este fluido (compuesto por electrones) con el Medio interestelar-*ISM*. Se encontró la forma funcional del crecimiento en el tiempo del radio de la onda expansiva que contiene la flujo de electrones, además se obtuvo que el factor gamma de *Lorentz* va decayendo a medida que pasa el tiempo

En el contexto radiativo se tuvo en cuenta que los electrones por su movimiento ultra-relativista generan campos magnéticos transversales al movimiento de los electrones que vienen detrás de ellos y se produce emisión de radiación de sincrotrón, que nos permite explicar la evolución temporal del flujo de radiación detectado por los sensores de *Swift* en la banda de los 2-10 keV.

## 5.1. Modelo del Fireball (Bola de Fuego)

En búsqueda de entender los procesos físicos involucrados con las detecciones de *Gamma-Ray Bursts*, a mediados de los años ochenta se realizaron modelos iniciales donde la producción de pares fue la clave inicial para explicar cómo se detectaba tanta energía en tan corto tiempo [27]. En estos modelos se consideraba la emisión de una onda expansiva que contenía un flujo denso de fotones [1] ultra-energéticos al cual llamaron *fireball* ( con $\tau_{\gamma\gamma} \gg 1$ [2] ).

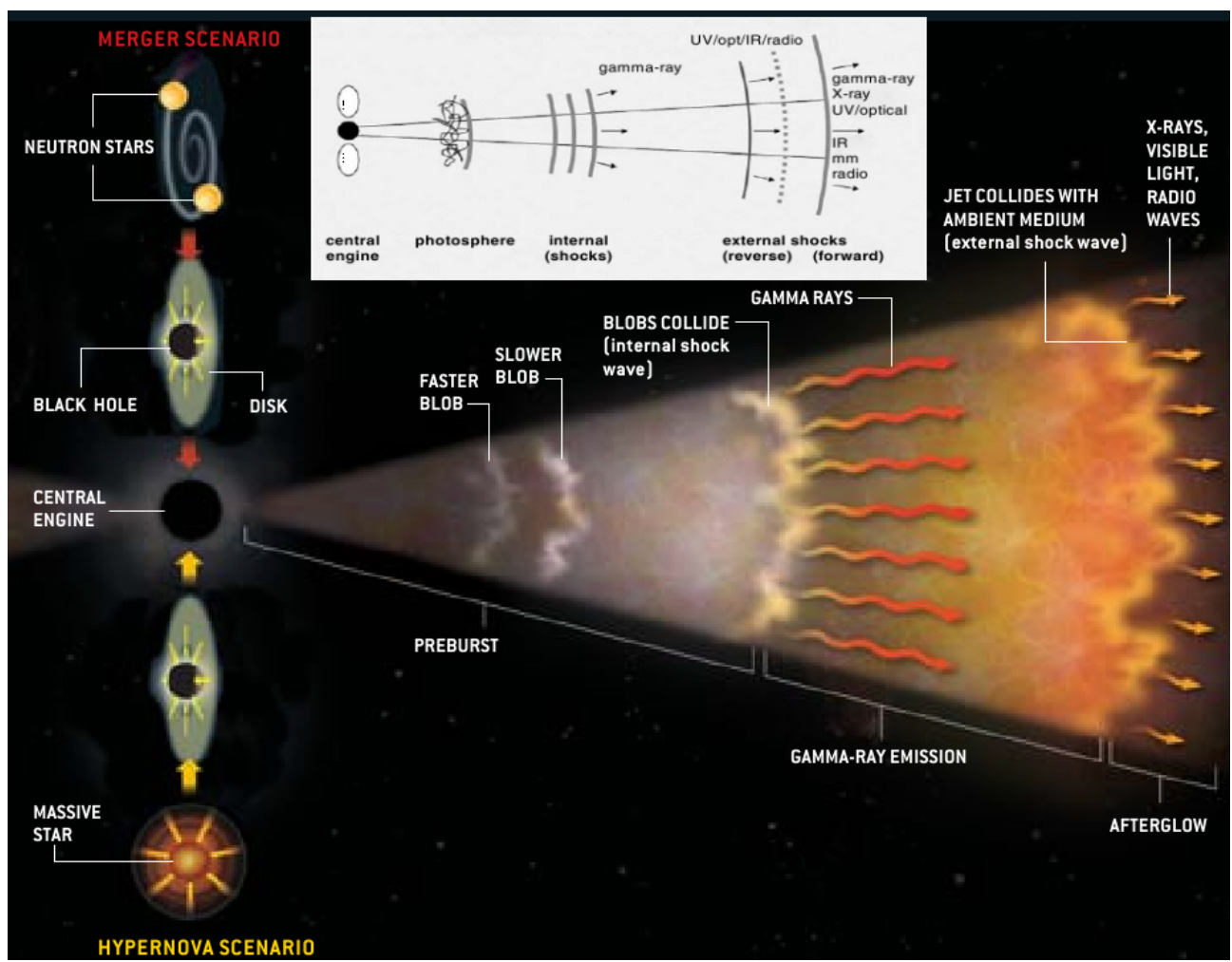

**Figura 5-1**.: Esquema pictórico que muestra las etapas que componen el modelo estándar del *fireball*. Imagen compuesta a partir de: [116, 59]

Estos fotones ultra-energéticos son capaces de producir, mediante la reacción: $\gamma + \gamma \longrightarrow e^+ + e^-$ pares electrón-positrón que a medida que pasa el tiempo van haciendo parte del flujo de partículas que contiene la onda expansiva, es decir que ahora ese flujo se compone de fotones menos energéticos, electrones y positrones. Al cabo de un tiempo se llega a que los fotones pierden aún más energía (en esos instantes el flujo presenta un $\tau_{\gamma\gamma} \approx 1$), se liberan los primeros GRBs [153, 63, 129, 128], y el flujo queda compuesto en su mayoría por electrones ultra-relativistas, que pueden explicar la emisión de radiación en las otras bandas.

---

[1] Es decir que esta onda expansiva contiene una gran cantidad de energía en una pequeña región de volumen generando así una "fotosfera".

[2] Recordemos que un medio se dice que es ópticamente opaco cuando la profundidad óptica $\tau_\nu$ es mayor que uno. En el caso que $\tau_\nu < 1$ el medio se considera ópticamente transparente [142]. En el caso del *fireball*, los primeros GRBs que se emiten no dejan detectar que eventos se esconden detrás de ellos y por eso se asume un $\tau_{\gamma\gamma} \gg 1$ para el flujo inicial de fotones que salen acelerados por el mecanismo central.

Con el paso de los años este modelo inicial se fue refinando hasta llegar a un modelo estándar del *fireball* que está compuesto de varias etapas (ver figura **5-1**), aunque hay que aclarar que aunque no es un modelo completo que explique todas las características observacionales de los GRBs (mostradas en el capítulo 2), sí explica la mayoría de ellas.

Observando la figura **5-1** nos damos cuenta que actualmente el modelo presenta 4 escenarios [116]:

**i** *Central engine*: Mecanismo central de la emisión de las primeras ondas expansivas.

**ii** *Preburst*: Fase en la que consideramos la dinámica de las primeras ondas expansivas provenientes del motor central, que transportan al *fireball*.

**iii** *Gamma-Ray emission*: Etapa en la cual se liberan los primeros rayos gamma, que son los detectados como Gamma-Ray Bursts.

**iv** *Afterglow*: Fase en la cual se liberan los primeros rayos x, uv, visible, infrarojo y ondas de radio.

En la fase (i) el Mecanismo central es el encargado de generar la emisión de las primeras ondas expansivas; actualmente se encuentran con mayor aceptación por parte de la comunidad dos escenarios: Colapso de Hipernovas y Supernovas (por ejemplo ver: [71]) para la familia de los *long* GRBs detectados y el *merge* de un sistema binario de estrellas de neutrones (por ejemplo ver: [91]) para la familia de los *short* GRBs detectados. Este tema se considera todavía un campo abierto de investigación, pues hay una decena de modelos de posibles mecanismos centrales de emisión de *fireballs* (ver referencias citadas en el capítulo 2).

En la etapa (ii) consideramos la dinámica de ondas expansivas (con densidad de fotones, electrones y positrones) que se mueven a diferentes velocidades, así que a un radio: $R_0 \sim 10^{13} cm$ [115] una de ellas que haya viajado mas rápido que la otra la puede alcanzar y chocar, llevando al tercer proceso (iii) donde se libera una porción de su energía (puede ser la mitad [27] ) en emisión de los primeros rayos gamma y la otra parte se transforma energía cinética de las partículas, generalmente protones del medio interestelar-ISM que van formar parte de una nueva onda expansiva que esta vez va a estar conformada de electrones ultarrelativistas contaminado de bariones del ISM.

Debido a que el flujo esta compuesto por una alta densidad de electrones ultra-relativistas, entonces podemos considerarlo como un flujo ideal adiabático con coeficiente adiabático de 4/3 (dominado por radiación), así que se le puede considerar como un nuevo *fireball* aunque estrictamente no lo sea pues no va a tener contenido de fotones. Al ser estos electrones relativistas van a generar campos magnéticos transversales a su movimiento al de los que vienen detrás de ellos, esto va generar en principio una emisión de radiación de sincrotrón, como explicamos con detalle en el capítulo 4 de la presente tesis. Esta radiación que ocurre hacia los $10^{16}$ cm con respecto al motor central [115] representa la fase del *afterglow* (iv) que presenta el modelo.

En la presente tesis se tuvo en cuenta este modelo del *fireball* para estudiar la evolución temporal del *afterglow* en rayos x (2-10keV) del GRB 050525.

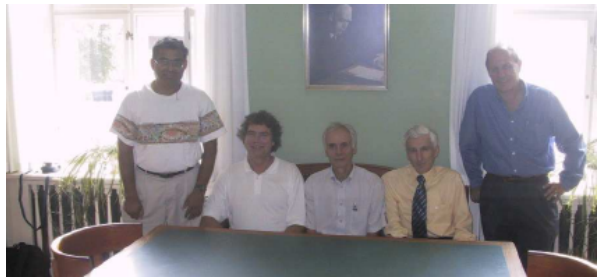

**Figura 5-2**.: Los Mayores representantes del modelo estándar del *fireball*. De izquierda a derecha: *S. Kulkarni, S. Woosley, P. Meszaros, M. Rees* y *T. Piran*. http://www.slac.stanford.edu/econf/C010815/proceedings.html

## 5.2. Modelo hidrodinámico empleado

Con el fin de estudiar la evolución temporal de un *afterglow* de un GRB es necesario describir tanto el cambio temporal del flujo de partículas que componen el *fireball* como el de la onda expansiva que los envuelve grupalmente. Para ello se va a mostrar a continuación el modelo hidrodinámico realizado [3] que nos dé razón del cambio temporal del radio de la onda expansiva con respecto al mecanismo central *central engine* y la evolución temporal del factor $\gamma$ de *Lorentz* de las partículas que componen el *fireball*; este dato es fundamental para hacer una buena descripción radiativa de las curvas de luz de los afterglows de GRBs. En la figura **5-3** se muestra la situación hidrodinámica de estudio.

---

[3]Teniendo en mente por supuesto modelos realizados por otros autores [113, 144, 36, 77, 77, 92, 76, 75].

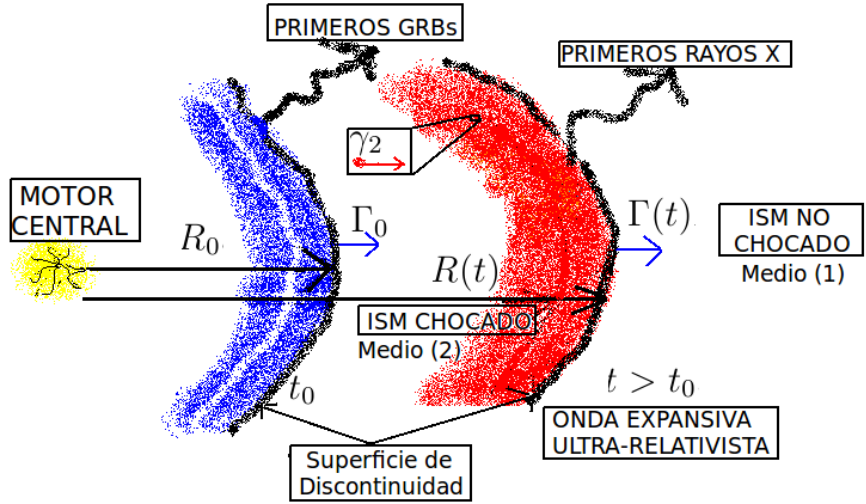

**Figura 5-3**.: Esquema pictórico de la evolución temporal de un *fireball* tenido en cuenta en el modelo hidrodinámico propuesto.

En la figura **5-3** podemos mirar que se desarrollan tres etapas principalmente: La fase que ocurre antes de la emisión de rayos gamma, *Pre-Burst*, el proceso que ocurre después de la emisión de rayos gamma, *Gamma-Ray emission*, que es la misma que la de antes de emisión en rayos x, y la fase posterior de la emisión de rayos x.

En la primera de ellas (parte izquierda de la figura **5-3** ) se muestran dos ondas expansivas que llevan electrones ultra-relativistas y fotones, y que están cerca de colisionar debido a que la de atrás (más cercana al motor central) viaja más rápido que la de adelante, lo que implica que a un tiempo $t_0$, cuando ellas se encuentren a una distancia $R_0$ con respecto al motor central que las impulsó inicialmente, van a colisionar de tal forma que parte de su energía se emitirá en los primeros rayos gamma y la otra parte generará una onda expansiva que llevara consigo electrones relativistas contaminado de bariones del medio interestelar (ISM) que ha sido chocado (medio 2). Finalmente la evolución temporal de ese plasma bariónico ultra-relativista [4] producirá la emisión en principio de rayos x vía emisión de sincrotrón.

En la dinámica propuesta para la onda expansiva que contiene al plasma bariónico ultra-relativista tenemos que por conservación de energía,

$$E_s = E_e(t) + Q(t)$$

Donde $E_s$ es la energía total proporcionada al fluido al inicio de su movimiento, es decir la proporción de energía que no se emitió en los rayos gamma iniciales. Por su parte $E_e(t)$ es

---

[4]Se llama bariónico por la contaminación de protones que se produjo al flujo inicial de electrones ultra-relativistas en la fase del choque con el medio interestelar

la energía total contenida por el fluido en expansión a un $t > t_0$ y $Q$ es la energía perdida por el fluido a causa de procesos internos, que más adelante extenderemos un poco.

Considerando como esférica la propagación de la onda expansiva que lleva consigo al fluido adiabático ultra-relativista nosotros tenemos que la energía que lleva esta onda a un tiempo determinado $t$ es ([20], Sec. III),

$$E_e(R_0, R_t, t) = \int_0^{2\pi} \int_0^{\pi} \int_{R_0}^{R(t)} e_2 r^2 \sin\phi \, dr \, d\phi \, d\theta$$

donde $e_2$ es la densidad de energía que es medida desde un sistema de referencia co-móvil al fluido en el medio 2 (ISM chocado), que está detrás de la superficie de discontinuidad en la figura **5-3**.

Por las condiciones de choque relativistas que se presentan cuando interacciona el fluido ultra-relativista, que lleva consigo el ISM chocado (medio 2), con el ISM no chocado (medio 1) tenemos que se cumple [20],

$$e_2 = 2\Gamma^2 w_1 \qquad n_2' = 2\Gamma^2 n_1 \qquad \gamma_2^2 = \frac{1}{2}\Gamma^2 \qquad (5\text{-}1)$$

Donde $\Gamma$ es el factor de *Lorentz* del frente de la onda expansiva, $\gamma_2$ es el factor gamma de *Lorentz* de las partículas al interior del fluido ultra-relativista que componen el ISM chocado, $w_1$ es la entalpía por unidad de volumen del fluido que esta delante de la superficie de discontinuidad en la figura **5-3**, $n_2'$ es el número de partículas por unidad de volumen del medio 2 medido en el marco de referencia del fluido que no ha sido chocado (medio 1) que presenta una densidad de partículas por unidad de volumen $n_1$. Cabe aclarar que como vemos en la figura **5-3** esta superficie de discontinuidad separa el ISM chocado (medio 2) del ISM que no ha sido chocado (medio 1), debido a que los valores de las cantidades termodinámicas y la velocidad de las partículas del medio 2 cambian drásticamente al pasar al medio 1, por eso es necesario tener en mente en este análisis el formalismo conceptual de las ondas de choque que vimos en el capítulo 3. Las ecuaciones mostradas en (5-1) se llaman las *jump conditions* que son una adaptación al contexto astrofísico que hicieron *R. D. Blandford* y *C. F. Mckee* en 1976 [20] de las ecuaciones originales de choque relativistas desarrolladas por *A. H. Taub* en 1949 [157] y que mostramos su deducción en el capítulo 3. En este caso específico del caso ultra-relativista hemos considerado que el coeficiente adiabático de nuestro gas de electrones $\hat{\gamma} = \frac{4}{3}$ y $\Gamma >> 1$.

Debido a que asumimos que el ISM no chocado es frío (por ejemplo 10 K) entonces para esta condición particular se encuentra que $w_1 = \rho_1 = n_1 m_p c^2$, donde $\rho_1$ es la densidad por unidad de volumen del medio 1, $m_p$ es la masa de protones del medio 1 (ISM no chocado) y c es la velocidad de la luz. Reemplazando esta condición en la ec. (5-1) obtenemos,

$$e_2 = 4\gamma_2^2 n_1 m_p c^2$$

$$E_e(R_0, R_t, t) = \int_{R_0}^{R(t)} 4\gamma_2^2 n_1 m_p c^2 (4\pi r^2) dr = \frac{16\pi \gamma_2^2(t) n_1 m_p c^2}{3}(R^3(t) - R_0^3)$$

$$\Rightarrow E_e = \frac{8\pi \Gamma^2(t) n_1 m_p c^2}{3}(R^3(t) - R_0^3)$$

Llamando $\beta_\Gamma = \frac{\dot{R}}{c} = \sqrt{1 - \frac{1}{\Gamma^2}}$, donde $\dot{R} = \Delta R/\Delta t_b$ es la tasa de cambio del radio del frente de onda expansiva medido desde el sistema de referencia del *burster* (desde donde se emitió el primer *burst* de rayos gamma, es decir a $R_0$ del motor central). Si ahora consideramos que la onda expansiva es ultra-relativista, entonces al ser $\Gamma \gg 1$ es válido escribir que,

$$E_e = \frac{8\pi \Gamma^2 n_1 m_p c^2}{3}(R^3(t) - R_0^3)(1 - \frac{1}{\Gamma^2})$$

$$E_e = \frac{8\pi \Gamma^2 n_1 m_p c^2}{3}(R^3(t) - R_0^3)\beta_\Gamma^2 \quad (5\text{-}2)$$

De otro modo teniendo en cuenta la definición de $\beta_\Gamma$ podemos decir que,

$$\Delta R = \sqrt{1 - \frac{1}{\Gamma^2}} c \Delta t_b$$

$$\Delta R = \sqrt{\frac{\Gamma^2 - 1}{\Gamma}} c \Delta t_b$$

Donde $\Delta t_b$ es el intervalo de tiempo medido desde el sistema de referencia del *Burster*.

Los fotones observados (en la cercanía de la Tierra) a un intervalo de tiempo $\Delta t$ realmente son emitidos en un intervalo de tiempo $\Delta t_b$ ([97],pg. 217),

$$\Delta t_b = \frac{\Delta t}{1 - \frac{v}{c}}$$

esto sucede puesto que tenemos que tener en cuenta el tiempo que tarda el fotón en recorrer una distancia de $\frac{\Delta r}{c}$ tal que el observador en Tierra detectara los fotones en un intervalo de tiempo definido por,

$$\Delta t = \Delta t - \frac{\Delta r}{c} = \Delta t - \frac{v_2 \Delta t}{c} = \Delta t(1 - \frac{v_2}{c})$$

$$\Rightarrow \Delta t_b = \frac{\Delta t}{1 - \frac{v}{c}} = \frac{\Delta t}{1 - \frac{\sqrt{\gamma_2^2 - 1}}{\gamma_2}} = \frac{\gamma_2 \Delta t}{\gamma_2 - \sqrt{\gamma_2^2 - 1}}$$

Para hacer la ultima igualdad utilizamos que $\frac{v_2}{c} = \frac{\sqrt{\gamma_2^2-1}}{\gamma_2}$ que es válido puesto que en general,

$$\sqrt{\gamma^2 - 1} = \sqrt{\frac{1}{1 - \frac{v^2}{c^2}} - 1} = \frac{v/c}{\sqrt{1 - \frac{v^2}{c^2}}} = \frac{v}{c}\gamma$$

Reemplazando la expresión que obtuvimos para $\Delta t_b$ en la que teníamos de para $\Delta R$ nos da como resultado que,

$$\Delta R = \sqrt{\frac{\Gamma^2 - 1}{\Gamma}} c \left(\frac{\gamma_2 \Delta t}{\gamma_2 - \sqrt{\gamma_2^2 - 1}}\right)$$

Para cambios infinitesimales es válido afirmar que,

$$\dot{R} = \sqrt{\frac{\Gamma^2 - 1}{\Gamma}} \left(\frac{\gamma_2}{\gamma_2 - \sqrt{\gamma_2^2 - 1}}\right) c$$

$$\beta_\Gamma = \frac{\dot{R}}{c} = \sqrt{\frac{\Gamma^2 - 1}{\Gamma}} \left(\frac{\gamma_2}{\gamma_2 - \sqrt{\gamma_2^2 - 1}}\right) \tag{5-3}$$

A continuación vamos a hacer una pequeña discusión acerca de la energía suministrada $E_s$ en el tiempo $t_0$ a la onda expansiva ultra-relativista.

Cuando describimos la figura **5-1**, mencionamos que la energía contenida en el *fireball* en $t_0$, impulsado inicialmente por un motor central, radiaba parte de su energía en emisión de una ráfaga de rayos gamma y la otra parte se suministraba para darle energía a una onda expansiva ultra-relativista que contenía un flujo de electrones, contaminado de una densidad de protones provenientes del medio interestelar. Por conservación de energía y llamando a $E_0$ la energía liberada en rayos gamma tenemos que la situación anteriormente descrita se puede expresar como,

$$E_{Fireball}(t_0) = E_0 + E_s$$

Si consideramos que $E_{Fireball} = fE_0$ entonces $E_s = fE_0 - E_0 \Rightarrow E_s = E_0(f - 1)$, donde $1 < f < 2$. Por ejemplo si $f = 3/2$ obtenemos que: $E_s = \frac{E_0}{2}$, es decir que la mitad de la energía contenida en el *fireball* inicial se libera en los primeros rayos gamma.

Volviendo al análisis realizado para la descripción de la evolución temporal de la onda expansiva ultra-relativista teníamos que $E_s = E_e(t) + Q(t)$. Así, después de remplazar los términos de $E_s$ y de $E_e$ obtenemos nuestro conjunto de ecuaciones para describir la evolución temporal hidrodinámica de nuestro plasma bariónico ultra-relativista,

$$E_0(f-1) = \frac{8\pi \Gamma^2 n_1 m_p c^2}{3}(R^3(t) - R_0^3)\beta_\Gamma^2 + Q(t) \tag{5-4}$$

$$\beta_\Gamma = \frac{\dot{R}}{c} = \sqrt{\frac{\Gamma^2 - 1}{\Gamma}}\left(\frac{\gamma_2}{\gamma_2 - \sqrt{\gamma_2^2 - 1}}\right) \tag{5-5}$$

$$Con \quad \Gamma^2 = 2\gamma_2^2 \tag{5-6}$$

Algunos investigadores proponen que $Q(t)$ puede ser de la forma [144],

$$Q(t) = E_0\left(\frac{t}{t_0}\right)^{\frac{-17\epsilon_e}{16}}$$

Donde $\epsilon_e$ es la fracción de energía interna que es radiada y perdida del sistema, de alrededor del 10 %. En nuestros cálculos numéricos consideraremos que $Q \approx 0$.

Ahora para incluir en nuestro análisis el caso que se presenta cuando la onda expansiva entra en régimen no relativista, $\beta_\Gamma << 1$, debemos utilizar una expresión general de $E_s$ que contemple los casos: no relativista, medianamente relativista y ultra-relativista. Esta relación fue hallada por primera vez por *R. D. Blandford* y *C. F. McKee* y a continuación las mostramos ([20], Sec. V),

$$E_e = \sigma w_1 \Gamma^2 \beta_\Gamma^2 \frac{4\pi}{3} R^3(t)$$

Donde $\sigma$ nos determina cada caso: relativista y no relativista, dependiendo exclusivamente del valor del coeficiente adiabático $\widehat{\gamma}$ que presente el flujo de partículas. Para los diferentes casos no relativistas donde se ha supuesto que tenemos un gas adiabático de partículas que se mueven por un impulso inicial *R. D. Blandford* y *C. F. Mckee* calcularon los diferentes valores de $\sigma$ [20]. A continuación se muestra un gráfico que construimos con esos datos para hallar una forma funcional de esa relación que podamos incluir en el modelo hidrodinámico para analizar el caso no-relativista con mayor detalle,

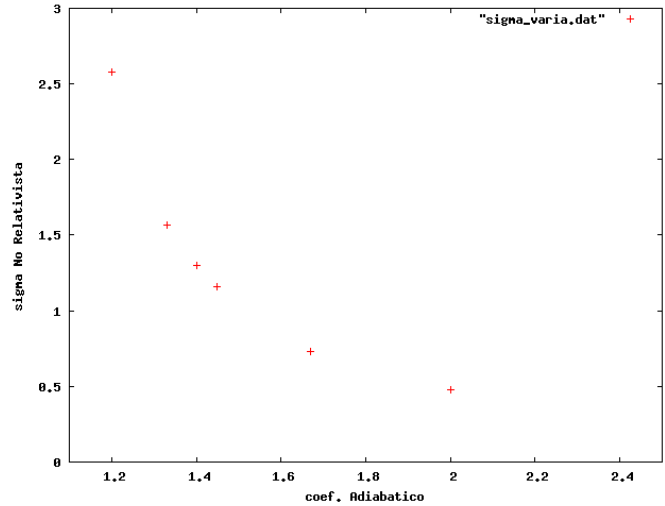

**Figura 5-4**.: Relación teórica entre el coeficiente adiabático y el $\sigma$ no relativista.

Buscando una relación analítica entre $\sigma_{No\ \ Relativista}$ y $\widehat{\gamma}$, entonces usamos *GnuPlot* para ajustar una función a los datos de origen teórico que mostramos en la figura **5-4**. El resultado fue el siguiente:

$$\sigma_{No\ \ Relativista} = \frac{a}{b - \widehat{\gamma}} = \frac{-0{,}509}{1{,}003 - \widehat{\gamma}}$$

Con un error de ajuste numérico de: $2.293\,\%$ para el valor de $a$, y $0.557\,\%$ para el valor de $b$.

Con la ayuda de los mismos datos calculados en [20] otros investigadores han propuesto una forma general de $\sigma$ que abarque tanto el caso relativista como el no relativista. A continuación mostramos una de estas propuestas [76],

$$\sigma = 0{,}73 - 0{,}38\beta$$

la cual se construye a partir de los casos extremos, es decir cuando $\beta \longrightarrow 0$; entonces $\sigma \longrightarrow 0{,}73$ que es el valor de $\sigma$ no relativista asociado con un gas ideal monoatómico $\widehat{\gamma} = \frac{5}{3}$ y cuando $\beta \longrightarrow 1$ obtenemos que $\sigma \longrightarrow 0{,}35$ que es el valor de $\sigma$ para el caso ultra-relativista donde $\widehat{\gamma} = \frac{4}{3}$.

Finalmente si queremos construir un modelo hidrodinámico general que abarque no solo el caso ultra-relativista, sino también al relativista debemos también hacer consideraciones sobre el factor gamma de *Lorentz* del frente de onda expansiva, pues en su versión relativista (recordemos que hemos estado trabajando en su versión ultra-relativista) es de la forma [20],

$$\Gamma^2 = \frac{(\gamma_2+1)(\widehat{\gamma}(\gamma_2-1)+1)^2}{\widehat{\gamma}(2-\widehat{\gamma})(\gamma_2-1)+2}$$

## 5.3. Modelo radiativo utilizado

En esta sección vamos a mostrar las ecuaciones que permiten estudiar al evolución temporal del flujo de radiación x proveniente del plasma bariónico ultra-relativista mostrado en la figura **5-3**.

En este estudio radiativo nos vamos a concentrar en la emisión de sincrotrón, pues este mecanismo ha sido aceptado por la comunidad como que origina el flujo de radiación detectado en los *afterglows* de GRBs [136, 103, 146]. Como vimos en el capitulo anterior esta radiación proviene del movimiento helicoidal de electrones alrededor del campo magnético generado por electrones que se encuentran moviéndose adelante de ellos en este fluido ultra-relativista.

A continuación vamos a hacer una extensión de la emisión de sincrotrón en términos de la frecuencia para un electrón que vimos en el capítulo 4, al de un número $n$ de electrones por unidad de volumen.

Cuando se liberan los primeros GRBs también se genera una onda expansiva ultra-relativista que arrastra electrones a velocidades relativistas con un factor gamma de *Lorentz*, $\gamma_2$, que varía entre $10^2$ y $10^3$ [135], es decir que estos electrones son ultra-relativistas. Teniendo en cuenta las ecuaciones de choque relativista en su versión astrofísica, mejor llamadas *jump conditions* ( ecs. 5-1) [20], justo cuando la onda expansiva sale impulsada (a $R_0$ del motor central) tenemos que es válido considerar que,

$$n_2 = (4\gamma_2 + 3)n_1$$

donde hemos considerado que $n_2$ y $n_1$ son el número de partículas por unidad de volumen (*number density*) del ISM chocado y del ISM no chocado respectivamente.

Si nosotros derivamos la anterior relación con respecto a $\gamma_2$ obtenemos que,

$$\frac{dn_2}{d\gamma_2} = 4n_1$$

que significa el cambio infinitesimal en la vecindad de $R_0$ que sufriría la población de partículas del ISM chocado ($n_2$), pues las ecuaciones de choque relativistas presentan su mayor validez en la vecindad de la superficie de discontinuidad (ver esta región en la figura **5-3**). Esta suposición fue fundamental para encontrar las ecuaciones de choque relativista en el capítulo 3.

Si queremos observar la evolución temporal de $n_2$ para $R > R_0$ entonces es usual utilizar una ley de potencia en $\gamma_2$ [36], pues a medida que el pasa el tiempo la onda expansiva se va frenando pues va ganado más número de partículas del ISM no chocado. Con estas ideas en mente planteamos la evolución temporal (por su dependencia en función de $\gamma_2 = \gamma_2(t)$) de la población $n_2$ de la siguiente forma,

$$\frac{dn_2}{d\gamma_2} = 4n_1(1+\gamma_2^{-p}): \quad p > 1$$

Donde se considera que $p$ puede estar entre $1,4 < p < 2,8$, pues ha habido un buen ajuste teórico en este rango de valores al espectro observados de afterglows de GRBs [174, 131].

En el sistema de referencia co-móvil, la potencia radiada de sincrotrón por un electrón del ISM chocado en términos de su frecuencia mostramos en el capítulo 4 que es de la forma,

$$\frac{dP'(\nu')}{d\nu'} = \frac{q_e^3 B' \sqrt{3}}{m_e c^2} F\left(\frac{\nu'}{\nu_c'}\right)$$

Con,

$$F\left(\frac{\nu'}{\nu_c'}\right) = \frac{\nu'}{\nu_c'} \int_{\frac{\nu'}{\nu_c'}}^{\infty} K_{\frac{5}{3}}(y) dy$$

donde $\nu_c' = \frac{w_c'}{2\pi} = \frac{1}{2\pi}\left(\frac{3}{2}\gamma_2^3 w_{B'}\right) = \frac{3\gamma^2 q_e B'}{4\pi m_e c}$ y $K_{5/3}$ es la función modificada de *Bessel* de segundo orden.

Ahora teniendo en cuenta lo anterior para una distribución de electrones del ISM chocado tenemos que su potencia radiada va a ser de la forma [36],

$$\frac{dP'(\nu')}{d\nu'} = \frac{q_e^3 B' \sqrt{3}}{m_e c^2} \int_{\gamma_{2,min}}^{\gamma_{2,max}} \left(\frac{dn_2'}{d\gamma_2}\right) F\left(\frac{\nu'}{\nu_c'}\right) d\gamma_2$$

Si asumimos que la potencia radiada se emite de forma isotópica tenemos que [77],

$$\frac{dP'_{\nu'}}{d\nu' d\Omega'} = \frac{1}{4\pi} \frac{dP'(\nu')}{d\nu'}$$

De otra parte desde el sistema de referencia del observador tenemos que la potencia radiada medida es de la forma ([142] pp. 140-141),

$$\frac{dP_\nu}{d\nu d\Omega} = \frac{1}{\gamma_2^3 (1 - \beta_2 Cos(\theta))} \frac{dP'_{\nu'}}{d\nu' d\Omega'}$$

$$\Rightarrow \frac{dP_\nu}{d\nu d\Omega} = \frac{1}{\gamma_2^3 (1 - \beta_2 Cos(\theta))} \frac{1}{4\pi} \frac{dP'(\nu')}{d\nu'}$$

donde al tener en cuenta el efecto *Doppler* relativista ([142],pp. 111) tenemos que,

$$\nu = \frac{\nu'}{(1 - \beta_2 Cos(\theta))}$$

Por su parte recordemos que $\theta$ es el ángulo que se forma entre la velocidad de los electrones y la línea de señal hacia el observador.

La onda expansiva que lleva electrones relativistas al expandirse produce una densidad de flujo observada con geometría elipsoidal [139] $S_\nu$ cuya expresión es [75],

$$S_\nu = \frac{\pi (\gamma c t)^2 \frac{dP_\nu}{d\nu d\Omega}}{D_L^2}$$

Donde $D_L$ es la distancia desde la fuente hacia la tierra.

Por último, el flujo $F$ de radiación observado por un detector de rayos x es la integral de $S_\nu$ en el rango de frecuencias $(\nu_{menor}, \nu_{mayor})$ de rayos x,

$$F_{obs}(t) = \int_{\nu_{menor}}^{\nu_{mayor}} S_\nu d\nu$$

Finalmente en el caso de la radiación emitida vía emisión de sincrotrón por un fluido adiabático ultra-relativista se ha encontrado que hay dos posibles pendientes que expliquen la disminución temporal de flujo de radiación en la curva de luz de los *afterglows*. Escribiendo $F_\nu \sim t^{-\beta}$, los dos casos son: $\beta = \frac{3}{4}(p-1)$ y $\beta = \frac{3p}{4} - \frac{1}{2}$ [146].

## 5.4. Análisis del GRB 050525

En esta sección vamos a estudiar numéricamente la evolución temporal del *afterglow* del GRB 050525 utilizando los resultados generales mostrados en el modelo hidrodinámico y radiativo. Este Gamma Ray Burst fue detectado por el instrumento BAT (*Burst Alert Telescope*) a bordo de la sonda espacial *Swift* a las 00:02:53.26 UT del 25 de mayo de 2005 y tuvo una duracion de $T_{90} = 8,8 \pm 0,5$ $s$ [23]. La medición de su emisión en rayos x asociada no se hizo esperar y 75 segundos después el instrumento XRT (*X-ray telescope*) también a bordo de *Swift*, empezó a registrar la disminución del flujo de radiación en la banda de los 2keV a los 10keV a medida que el tiempo pasaba (ver figura **5-5**).

El primer análisis que se realizó fue resolver numéricamente el sistema de ecuaciones hidrodinámicas (ecs. (5-4,5-5, 5-6)) con las condiciones iniciales asociadas al GRB 050525, las cuales fueron: $E_0 = 2,3 \times 10^{45}$ $J$, $R_0 = 3,53 \times 10^{13}$ $m$ y $\gamma_2 = 4542$.

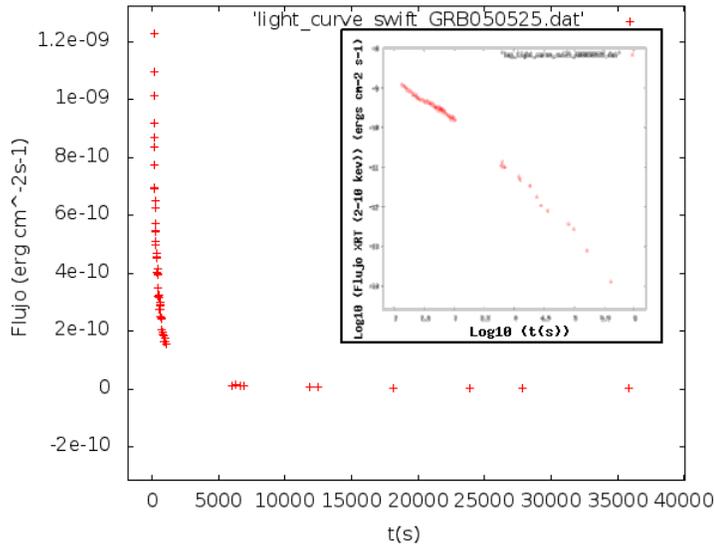

**Figura 5-5**.: Medición de la curva de luz del *afterglow* en rayos x del GRB 050525 por el instrumento XRT a bordo de *Swift*, el recuadro muestra la gráfica de $Log_{10}F$ vs $Log_{10}t$.

Usando *Sage math* que es un software libre de matemáticas hecho en *Python* [5][111], en donde implementamos un *Runge-Kutta* de orden 4 para resolver la ec. (5-4) asumiendo: $f = 3/2$,

---

[5]Con el que se pueden realizar cálculos numéricos de álgebra, cálculo diferencial, cálculo integral, ecuaciones diferenciales, teoría de números, entre otros . Sage se creó y se utiliza como una alternativa libre a programas como Maple, Magma, Mathematica y Matlab. La sintaxis y velocidad de Sage lo han hecho un real competidor a los demás programas de matemáticas y es utilizado en diferentes universidades alrededor del mundo.[46]

$Q \approx 0$, teniendo un paso de 10s y un intervalo temporal de 6000s. En la figura **5-6** se muestra el resultado obtenido.

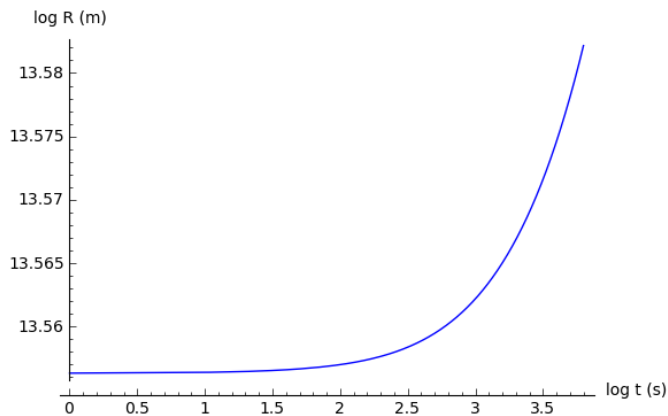

**Figura 5-6**.: Variación del radio de la onda expansiva relativista con el paso del tiempo.

La gráfica **5-6** se realizó con parejas ordenadas iteradas (t,R) así que las graficamos por aparte y así pudimos hallar usando *GnuPlot* una función analítica de R, la cual describo a continuación: $R(t) = (1{,}977 \times 10^{13})t^{1/4} + (3{,}55 \times 10^{12})t^{1/2}$m. Podemos ver que cuando $R >> R_0$ entonces $R \sim t^{1/4}$, que es un resultado estándar en la evolución temporal de un fluido adiabático ultarelativista [144].

Finalmente como nuestro caso de estudio fue la evolución temporal de un fluido adiabático ultrarelativista, entonces es válido ajustar la curva de luz del *afterglow* en rayos x del GRB 050525 con la relación del flujo definida en la sección anterior por: $F_\nu \sim t^{-\beta}$, los dos casos son: $\beta = \frac{3}{4}(p-1)$ y $\beta = \frac{3p}{4} - \frac{1}{2}$ [146]. Los ajustes se hicieron usando *GnuPlot* y se muestran en las figuras **5-7**, **5-8**.

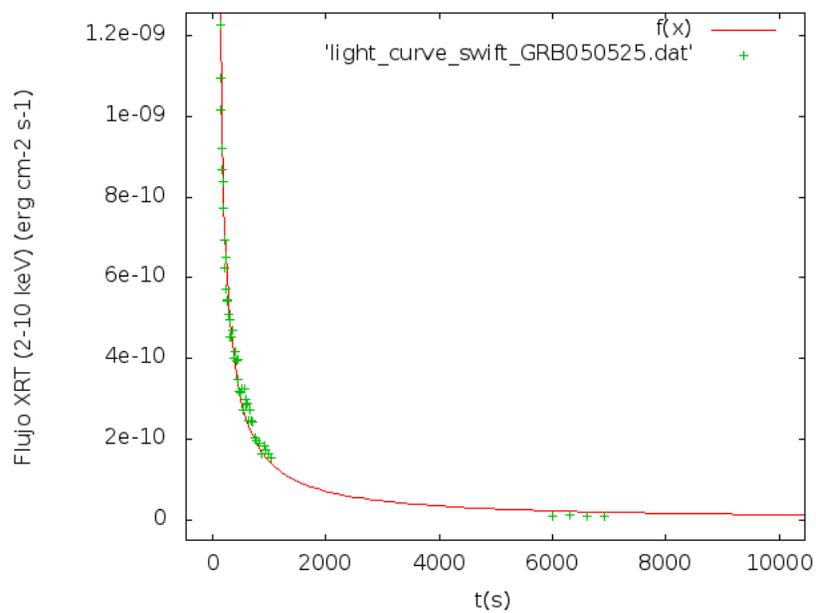

**Figura 5-7**.: Ajuste teórico de la curva de luz del *afterglow* del GRB 050525 con índice espectral $p = 2{,}35 \pm 0{,}02(1{,}041\,\%)$

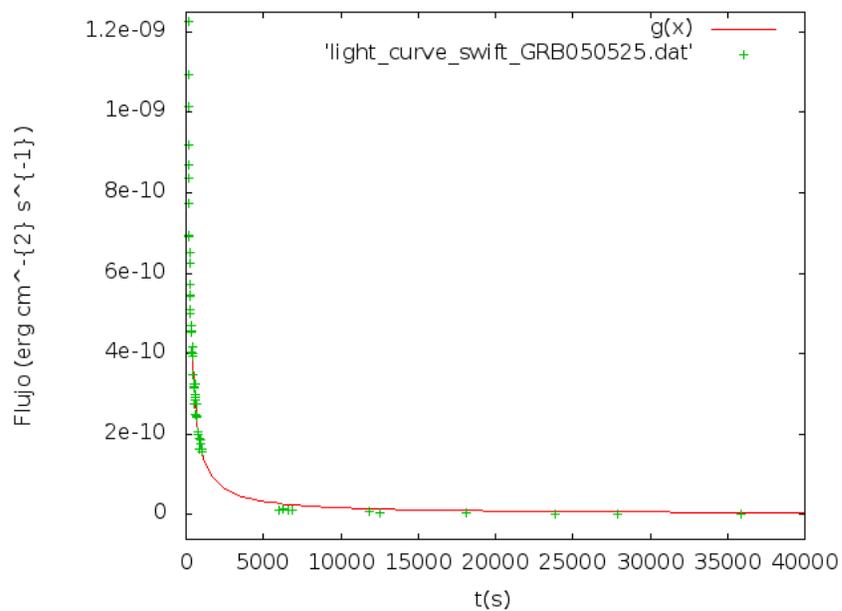

**Figura 5-8**.: Ajuste teórico de la curva de luz del *afterglow* del GRB 050525 con índice espectral $p = 1{,}97 \pm 0{,}02(1{,}186\,\%)$

# 6. Conclusiones

En el presente trabajo se hizo una revisión exhaustiva de la fenomenología de los *Gamma Ray Bursts* con mas de un centenar de referencias, la mayoría de los cuales son artículos provenientes de revistas internacionales de prestigio en el campo de la física, la astrofísica y la astronomía. Que se remontan desde el tratamiento relativista de los choques, inicialmente de manera general por *H. Taub*, seguida de su implementación en el contexto astrofísico de las ondas expansivas relativistas por *Blandford* y *Mckee* hasta que se aplican en el contexto de un modelo que busca explicar características observacionales de los GRBs, implementado por *Meszaros*, *T. Piran*, *M. Rees*, *S. Kulkarni*, *S. Woosley*, para citar a los pioneros en la construcción del modelo del *fireball*.

Encontramos de otro lado, de la parte observacional unas propiedades comunes a todos los GRBs que los divide en distintas etapas de emisión de energía enlazados en una secuencia, donde inicialmente es un gran *Burst* de rayos gamma, seguido de una radiación $x$ y así sucesivamente hasta llegar a tener emisiones en el óptico e incluso en radio. De otro lado también observacionalmente se han clasificado en dos grandes familias: las de corta y larga duración.

En la presente tesis nos limitamos entonces a seguir el modelo del *fireball* y para ello necesitábamos hacer dos descripciones físicas para darle soporte teórico al presente trabajo. Así, desarrollamos un capítulo sobre choques relativistas, haciendo énfasis en el caso ultra-relativista y otro capítulo sobre radiación sincrotrónica. Nosotros nos concentramos en el presente trabajo, no en lo que se denomina el motor central (*central engine*) sino en la fase posterior denominada el rescoldo o *afterglow* básicamente por dos razones. La primera de ellas, debido a que no tiene confrontación observacional directa, es materia de mucho debate, y podemos decir que contiene un grado relativamente alto de especulación. La segunda razón, es que porque la gran mayoría de los GRBs detectados por las diferentes misiones están en el rango del *afterglow*, lo cual hace que un modelo pueda ser contrastado con dichas observaciones. La evolución temporal de prácticamente todos los GRBs detectados a la fecha tienen dos zonas de comportamiento evolutivo, una primera parte que decae fuertemente, el cual ha sido distinguido con el calificativo de enfriamiento rápido, que corresponde a la emisión de la ráfaga en rayos x, y una segunda fase de decaimiento mas lento que corresponde al choque con el medio interestelar. Nosotros logramos reproducir teóricamente ese comportamiento radiativo combinando las descripciones de choque relativista y el de radiación sincrotrónica y lo aplicamos a un caso en particular como lo fue con el GRB 050525 el cual aconteció el 25 de Mayo de 2005. Los ajustes de la evolución temporal de la señal detectada del GRB 050525, se expresaron como es usual y como fue corroborado en el presente trabajo: $F_\nu \propto t^{-\beta}$.

En este caso para el GRB 050525, los mejores ajustes tomando solamente el canal de 2-10keV del instrumento XRT a bordo del *Swift* fueron: $\beta = \frac{3}{4}(p-1)$ con $p = 2{,}35 \pm 0{,}02 (1{,}041\,\%)$ para la zona de decaimiento fuerte (enfriamiento rápido), y $\beta = \frac{3p}{4} - \frac{1}{2}$ con $p = 1{,}97 \pm 0{,}02 (1{,}186\,\%)$ para la segunda fase del decaimiento. El error en el ajuste global fue menor del orden de 1,2 %.

Podemos inferir con el presente trabajo que el comportamiento temporal del decaimiento conocido como *afterglow* o rescoldo de los GRBs se describe muy bien con el modelo de la bola de fuego o *Fireball* el cual se distingue por tener dos comportamientos fuertemente demarcados como lo son: un decaimiento fuerte y uno mas lento. Nosotros lo pudimos corroborar con los datos pertinentes del GRB 050525, con los exponentes arriba citados, pero sin embargo creemos que los diferentes *afterglows* de los GRBs que se presentan se pueden modelar de la manera aquí descrita. Recordemos que en el trabajo de *P. Kumar* y *A. Panateisco* [131] se había establecido con base en el estudio de 8 GRBs que el factor p esta en el rango de valores comprendido entre $1{,}4 < p < 2{,}8$. Nuestro trabajo va en la dirección de confirmar este resultado, y creemos que el análisis de diversos GRBs podrían seguir confirmando este rango de valores, el cual podría seguir como una extensión del presente trabajo hacia el futuro para una muestra lo suficientemente alta de GRBs, con lo cual reforzaríamos la idea también ventilada en este trabajo de que el modelo del *fireball* definitivamente es el modelo apropiado para la descripción de todos los GRBs. Sin embargo la puerta queda abierta a describir el motor central de los GRBs.

# A. Anexo: Descripción matemática del flujo de energía del fluido ideal

En la siguiente demostración [98] se podrá determinar analíticamente como varía con el tiempo la energía del fluido ideal. Para comenzar sabemos que la energía de un elemento de volumen arbitrario del fluido es de la forma[1],

$$E = \frac{1}{2}\rho\nu^2 + \rho\epsilon \tag{A-1}$$

$$\Rightarrow \frac{\partial E}{\partial t} = \frac{\partial(\frac{1}{2}\rho\nu^2 + \rho\epsilon)}{\partial t} \tag{A-2}$$

Recordando que,
$\nu^2 = \vec{\nu} \cdot \vec{\nu}$,
$\frac{\partial \rho}{\partial t} = -\vec{\nabla} \cdot (\rho\vec{\nu})$ (Continuidad), y
$\frac{\partial \vec{\nu}}{\partial t} = -(\vec{\nu} \cdot \vec{\nabla})\vec{\nu} - \frac{1}{\rho}\vec{\nabla}P = -\frac{1}{2}\vec{\nabla}(\nu^2) + (\vec{\nu} \times (\vec{\nabla} \times \vec{\nu})) - \frac{1}{\rho}\vec{\nabla}P$ ( Euler [2] )
tenemos que el primer termino de la anterior ecuación es de la forma,

$$\Rightarrow \frac{\partial(\frac{1}{2}\rho\nu^2 + \rho\epsilon)}{\partial t} = \frac{1}{2}\nu^2\frac{\partial \rho}{\partial t} + \rho\vec{\nu} \cdot \frac{\partial \vec{\nu}}{\partial t} \tag{A-3}$$

$$\Rightarrow \frac{\partial(\frac{1}{2}\rho\nu^2 + \rho\epsilon)}{\partial t} = -\frac{1}{2}\nu^2(\vec{\nabla} \cdot (\rho\vec{\nu})) - \rho\vec{\nu} \cdot \frac{1}{2}\vec{\nabla}(\nu^2) + \rho\vec{\nu} \cdot (\vec{\nu} \times \vec{\nabla} \times \vec{\nu}) - \nu \cdot \vec{\nabla}P \tag{A-4}$$

Como $d\omega = Tds + \frac{1}{\rho}dP \Rightarrow \vec{\nabla}P = \rho\vec{\nabla}\omega - \rho T\vec{\nabla}s$, donde $\omega$ es la entalpía por unidad de masa.

$$\Rightarrow \frac{\partial(\frac{1}{2}\rho\nu^2 + \rho\epsilon)}{\partial t} = -\frac{1}{2}\nu^2(\vec{\nabla} \cdot (\rho\vec{\nu})) - \rho\vec{\nu} \cdot \vec{\nabla}(\frac{1}{2}\nu^2 + \omega) + \rho T\vec{\nu} \cdot \vec{\nabla}s \tag{A-5}$$

---
[1]En donde el primer termino es la energía cinética y el segundo es la energía interna, siendo $\epsilon$ la energía interna por unidad de masa

[2]Recordando que $\vec{\nabla}(\vec{A} \cdot \vec{B}) = (\vec{A} \cdot \vec{\nabla})\vec{B} + (\vec{B} \cdot \vec{\nabla})\vec{A} + \vec{A} \times (\vec{\nabla} \times \vec{B}) + \vec{B} \times (\vec{\nabla} \times \vec{A})$ entonces
$\vec{\nabla}(\vec{\nu} \cdot \vec{\nu}) = (\vec{\nu} \cdot \vec{\nabla})\vec{\nu} + (\vec{\nu} \cdot \vec{\nabla})\vec{\nu} + \vec{\nu} \times (\vec{\nabla} \times \vec{\nu}) + \vec{\nu} \times (\vec{\nabla} \times \vec{\nu}) \Rightarrow \vec{\nabla}(\vec{\nu} \cdot \vec{\nu}) = 2(\vec{\nu} \cdot \vec{\nabla})\vec{\nu} + 2\vec{\nu} \times (\vec{\nabla} \times \vec{\nu})$,
$\Rightarrow \frac{1}{2}\vec{\nabla}(\nu^2) = (\vec{\nu} \cdot \vec{\nabla})\vec{\nu} + \vec{\nu} \times (\vec{\nabla} \times \vec{\nu})$

El siguiente paso de nuestra demostración es encontrar a que es igual $\frac{\partial(\rho\epsilon)}{\partial t}$.

Partiendo de la siguientes relaciones termodinámicas:
$d\omega = Tds + \frac{1}{\rho}dP$,
$d\epsilon = Tds - pdV = Tds + (\frac{P}{\rho^2})$ Donde utilizamos que $V = \frac{1}{\rho} \Rightarrow dV = -\frac{d\rho}{\rho^2}$
$\Rightarrow Tds = d\omega - \frac{1}{\rho}dP \Rightarrow d\epsilon = d\omega - \frac{1}{\rho}dP - \frac{P}{\rho^2} \Rightarrow d\omega = d\epsilon + d(\frac{P}{\rho}) \Rightarrow \omega = \epsilon + \frac{P}{\rho} \Rightarrow PV = \omega - \epsilon$
$\Rightarrow d(\rho\epsilon) = \epsilon d\rho + \rho d\epsilon = \epsilon d\rho + \rho Tds + \frac{P}{\rho}d\rho \Rightarrow d(\rho\epsilon) = \epsilon d\rho + \rho d\epsilon = \epsilon d\rho + \rho Tds + \omega d\rho - \epsilon d\rho \Rightarrow$
$d(\rho\epsilon) = \omega d\rho + \rho Tds$

$$\Rightarrow \frac{\partial(\rho\epsilon)}{\partial t} = \omega\frac{\partial\rho}{\partial t} + \rho T\frac{\partial s}{\partial t} \tag{A-6}$$

Asumiendo que el fluido es isoentropico, es decir que $\frac{Ds}{Dt} = 0 \Rightarrow \frac{\partial s}{\partial t} + \vec{\nu}\cdot\vec{\nabla}s = 0$ obtenemos,

$$\frac{\partial(\rho\epsilon)}{\partial t} = -\omega\vec{\nabla}\cdot(\rho\vec{\nu}) - \rho T\vec{\nu}\cdot\vec{\nabla}s \tag{A-7}$$

Finalmente sumando (36) con (38) tenemos que la variación de energía es,

$$\frac{\partial(\frac{1}{2}\rho\nu^2 + \rho\epsilon)}{\partial t} = -(\frac{1}{2}\nu^2 + \omega)\vec{\nabla}\cdot(\rho\vec{\nu}) - \rho\vec{\nu}\cdot\vec{\nabla}(\frac{1}{2}\nu^2 + \omega) \tag{A-8}$$

utilizado de forma conveniente la identidad: $\vec{\nabla}\cdot(\psi\vec{A}) = (\vec{\nabla}\psi)\cdot\vec{A} + \psi(\vec{\nabla}\cdot\vec{A})$, la anterior ecuación se nos convierte en,

$$\frac{\partial(\frac{1}{2}\rho\nu^2 + \rho\epsilon)}{\partial t} = -\vec{\nabla}\cdot(\rho\vec{\nu}(\frac{1}{2}\nu^2 + \omega)) \tag{A-9}$$

Ahora integrando la anterior ecuación en un volumen determinado y utilizando el teorema de Gauss (o de la divergencia) [3],

$$\frac{\partial[\int(\frac{1}{2}\rho\nu^2 + \rho\epsilon)dV]}{\partial t} = -\int\vec{\nabla}\cdot(\rho\vec{\nu}(\frac{1}{2}\nu^2 + \omega))dV \tag{A-10}$$

$$\Rightarrow \frac{\partial[\int(\frac{1}{2}\rho\nu^2 + \rho\epsilon)dV]}{\partial t} = -\oint[\vec{\nabla}\cdot(\rho\vec{\nu}(\frac{1}{2}\nu^2 + \omega))]\cdot d\vec{a} \tag{A-11}$$

---

[3] Cabe recordar que este teorema relaciona el flujo de un campo vectorial a través de una superficie cerrada con la integral de su divergencia en el volumen delimitado por dicha superficie.

Analizando los términos de la anterior ecuación tenemos que el de la izquierda es la variación por unidad de tiempo de la energía del fluido en un volumen determinado, y el de la derecha es la cantidad de energía que fluye hacía el exterior de este volumen en la unidad de tiempo. Así es muy claro notar que el termino $\rho\vec{\nu}(\frac{1}{2}\nu^2 + \omega)$ corresponde al vector densidad de flujo de energía, donde su magnitud es la cantidad de energía que pasa por unidad de tiempo y de área en la dirección perpendicular de la superficie.

# B. Anexo: Descripción matemática del flujo de momento del fluido ideal

En la siguiente demostración [98] se podrá determinar analíticamente como varía con el tiempo el momento del fluido ideal.

El momento por unidad de volumen es $\rho\vec{\nu}$ y nos interesa saber a que es igual el flujo de momento. Para ello primero vamos a determinar $\frac{\partial(\rho\vec{\nu})}{\partial t}$ utilizando la notación tensorial [1], de tal forma que podemos expresar $\vec{A}\cdot\vec{B}$ de la forma $A_i B_i$.

$$\Rightarrow \frac{\partial(\rho\nu_i)}{\partial t} = \rho\frac{\partial\nu_i}{\partial t} + \nu_i\frac{\partial\rho}{\partial t} \tag{B-1}$$

En esta notación la ecuaciones de continuidad y de Euler quedan de la siguiente forma,

$$\frac{\partial\rho}{\partial t} = -\frac{\partial(\rho\nu_k)}{\partial x_k} \tag{B-2}$$

$$\frac{\partial\nu_i}{\partial t} = -\nu_k\frac{\partial\nu_i}{\partial x_k} - \frac{1}{\rho}\frac{\partial P}{\partial x_i} \tag{B-3}$$

Remplazando (44) y (45) en (43) obtenemos,

$$\Rightarrow \frac{\partial(\rho\nu_i)}{\partial t} = -\rho\nu_k\frac{\partial\nu_i}{\partial x_k} - \rho\frac{1}{\rho}\frac{\partial P}{\partial x_i} + -\nu_i\frac{\partial(\rho\nu_k)}{\partial x_k} \tag{B-4}$$

Reuniendo términos [2] tenemos que se satisface la siguiente relación,

$$\Rightarrow \frac{\partial P}{\partial x_i} = \delta_{ik}\frac{\partial P}{\partial x_k} \Rightarrow \frac{\partial(\rho\nu_i)}{\partial t} = -\delta_{ik}\frac{\partial P}{\partial x_k} - \frac{\partial\rho\nu_i\nu_k}{\partial x_k} \tag{B-5}$$

---

[1]Donde los sufijos $i$, $k$,..., toman los valores 1,2,3 correspondientes a las componentes de los vectores y tensores a lo largo de los ejes $x$, $y$ y $z$ correspondientemente. Cabe notar que según lo anterior se satisface que $\vec{A}\cdot\vec{B} = A_1 B_1 + A_2 B_2 + A_3 B_3 = \sum_{i=0}^{3} A_i B_i$

[2]También debemos recordar que dados un $A_i$ y un $A_k$, representaciones de vectores en $R^3$ es valida la siguiente relación: $A_i = \delta_{ik} A_k$

$$\frac{\partial(\rho\nu_i)}{\partial t} = -\left(\frac{\partial(\delta_{ik}P + \rho\nu_i\nu_k)}{\partial x_k}\right) \tag{B-6}$$

$$\frac{\partial(\rho\nu_i)}{\partial t} = -\left(\frac{\partial \Pi_{ik}}{\partial x_k}\right) \tag{B-7}$$

Donde $\Pi_{ik}$ es el tensor de de densidad de flujo de momento de nuestro fluido ideal, y lo denotamos como $\Pi_{ik} = \delta_{ik}P + \rho\nu_i\nu_k$. Este tensor representa una transferencia de momento completamente reversible debida simplemente al transporte mecánico de las distintas partículas de un lugar a otro y a las fuerzas de presión que actúan en dicho fluido

Para ver el significado del tensor $\Pi_{ik}$ integremos la anterior ecuación con respecto a un volumen determinado,

$$\frac{\partial[\int(\rho\nu_i)dV]}{\partial t} = -\int\left(\frac{\partial \Pi_{ik}}{\partial x_k}\right)dV \tag{B-8}$$

La integral de la derecha se transforma en una integral de superficie utilizando para ello el teorema de Gauss en el espacio [3],

$$\frac{\partial[\int(\rho\nu_i)dV]}{\partial t} = -\oint(\Pi_{ik}df_k)dV \tag{B-9}$$

Observado la anterior ecuación nos damos cuenta que el término de la izquierda es la variación respecto al tiempo de la componente $i$ del momento contenido en el volumen considerado. La integral de superficie del término de la derecha representa la cantidad de momento que fluye (por unidad de tiempo) hacia fuera a través de la superficie límite. Así $\Pi_{ik}df_k$ es la componente $i$ que fluye a través del elemento superficial $df$. Ahora sí escribimos $df_k$ como $n_k df$, donde $df$ es el área del elemento superficial y $\vec{n}$ un vector unitario a lo largo de la normal y dirigido hacia fuera del volumen en cuestión, veremos que $\Pi_{ik}n_k$ es el flujo de la componente $i$ del momento a través del área superficial unidad. Finalmente la expresión $\Pi_{ik}n_k = Pn_i + \rho\nu_i\nu_k n_k$ se puede escribir en forma vectorial como: $P\vec{n} + \rho\vec{\nu}(\vec{\nu}\cdot\vec{n})$ y es claro notar que nuestro tensor de densidad de flujo de momento $\Pi_{ik}$ es la componente $i$ de la cantidad de momento que fluye a través del elemento superficial, el cual es perpendicular al eje $x_k$.

Finalmente el vector $P\vec{n} + \rho\vec{\nu}(\vec{\nu}\cdot\vec{n})$ nos proporciona el flujo de momento en la dirección del vector unitario $\vec{n}$, el cual es normal a la superficie. En particular tomando a $\vec{n}$ en una dirección paralela a la velocidad del fluido vemos que solo la componente longitudinal del momento se ve transportada en dicha dirección y su densidad de flujo es $P + \rho\nu^2$

---

[3] $\int_S \vec{B}\cdot\vec{n}dS = \int_V \vec{\nabla}\cdot\vec{B}dV$

# Bibliografía